\documentclass[a4paper,10pt,fleqn]{article}
\usepackage{jheppub}
\usepackage[T1]{fontenc}
\usepackage{setspace}
\usepackage{enumitem}
\usepackage{amsmath}
\usepackage{amsthm}
\usepackage{bm}
\usepackage[normalem]{ulem}
\usepackage{mathtools}
\usepackage{caption}
\usepackage{slashed}
\usepackage{subcaption}
\usepackage{tikz}

\definecolor{jlab_red}{RGB}{192,39,45}
\definecolor{jlab_orange}{RGB}{249,102,0}
\definecolor{jlab_blue}{RGB}{47,122,121}
\definecolor{jlab_green}{RGB}{65,125,10}

\hypersetup{%
pdftitle = {t-channel-scattering},
pdfsubject = {QCD},
pdfauthor = {Baiao-Raposo and Hansen},
colorlinks = {true},
filecolor = {black},
linkcolor = {jlab_blue},
menucolor = {black},
citecolor = {jlab_green},
urlcolor = {jlab_green},
}{}

\renewcommand{\>}{\rangle}
\newcommand{\<}{\langle}
\renewcommand{\Tilde}{\widetilde}

\newcommand{\onShellSingularity}[0]{Luscher:1985dn,Luscher:1986pf,Kim:2005gf}
\newcommand{\TOPT}[0]{Hansen:2014eka,Blanton:2020gha}
\newcommand{\LHCbTcc}[0]{LHCb:2021vvq,LHCb:2021auc}

\newcommand{\ThreeBodyRFT}[0]{Hansen:2014eka}
\newcommand{\threeNonDeg}[0]{Blanton:2020gmf}
\newcommand{\luscher}[0]{Luscher:1986pf}
\newcommand{\spinPapers}[0]{Briceno:2014oea,Briceno:2015csa}
\newcommand{\dfInspiration}[0]{Hansen:2014eka,Briceno:2015tza}
\newcommand{\intEquationStrats}[0]{Dawid:2023kxu}
\newcommand{\twoScatter}[0]{Rummukainen:1995vs,He:2005ey,Christ:2005gi,Kim:2005gf,Lage:2009zv,Bernard:2010fp,Fu:2011xz,Briceno:2012yi,Hansen:2012tf,Guo:2012hv,Briceno:2014oea}
\newcommand{\threeScatter}[0]{Polejaeva:2012ut,Hansen:2014eka,Hansen:2015zga,Briceno:2017tce,Guo:2017ism,Hammer:2017uqm,Hammer:2017kms,Mai:2017bge,Doring:2018xxx,Briceno:2018mlh,Klos:2018sen,Briceno:2018aml,Guo:2018ibd,Blanton:2019igq,Pang:2019dfe,Romero-Lopez:2019qrt,Hansen:2020zhy,Blanton:2020gha,Blanton:2020jnm,Guo:2020spn,Pang:2020pkl,Romero-Lopez:2020rdq,Blanton:2020gmf,Muller:2020vtt,Muller:2020wjo,Hansen:2021ofl,Blanton:2021mih,Muller:2021uur,Blanton:2021eyf,Muller:2022oyw,Jackura:2022xml,Severt:2022jtg,Baeza-Ballesteros:2023ljl,Draper:2023xvu,Bubna:2023oxo}
\newcommand{\threeForTChannel}[0]{Hansen:2014eka,Hansen:2015zga}
\newcommand{\numericalThreeBody}[0]{Beane:2007es,Detmold:2008fn,Woss:2018irj,Romero-Lopez:2018rcb,Mai:2018djl,Horz:2019rrn,Blanton:2019vdk,Mai:2019fba,Culver:2019vvu,Fischer:2020jzp,Alexandru:2020xqf,Hansen:2020otl,Brett:2021wyd,Blanton:2021llb,Mai:2021nul,Garofalo:2022pux,Baeza-Ballesteros:2022bsn,Draper:2023boj}
\newcommand{\threeBodyIntegralEquations}[0]{Dawid:2020uhn,Jackura:2020bsk,Jackura:2022gib,Dawid:2023jrj,Dawid:2023kxu}
\newcommand{\rescaleBarrier}[0]{Morningstar:2017spu}
\newcommand{\cutFormalismNeeded}[0]{Green:2021qol}
\newcommand{\TccCollection}[0]{Padmanath:2022cvl,Du:2023hlu}
\newcommand{\polesInCL}[0]{Luscher:1986pf,Kim:2005gf}
\newcommand{\groupTheory}[0]{Luscher:1986pf,Dudek:2010ew}

\newcommand{\SpinBases}[0]{Briceno:2014oea,Briceno:2015csa}

\title{Finite-volume scattering on the left-hand cut}

\author[a]{A.~Bai\~ao Raposo}
\author[a]{and M.~T.~Hansen}
\affiliation[a]{Higgs Centre for Theoretical Physics, School of Physics and Astronomy, The University of Edinburgh, Edinburgh EH9 3FD, UK}
\emailAdd{a.baiao-raposo@sms.ed.ac.uk}
\emailAdd{maxwell.hansen@ed.ac.uk}

\abstract{The two-particle finite-volume scattering formalism derived by L{\"u}scher and generalized in many subsequent works does not hold for energies far enough below the two-particle threshold to reach the nearest left-hand cut. The breakdown of the formalism is signaled by the fact that a real scattering amplitude is predicted in a regime where it should be complex. In this work, we address this limitation by deriving an extended formalism that includes the nearest branch cut, arising from single particle exchange. We focus on two-nucleon ($NN \to NN$) scattering, for which the cut arises from pion exchange, but give expressions for any system with a single channel of identical particles. The new result takes the form of a modified quantization condition that can be used to constrain an intermediate K-matrix in which the cut is removed. In a second step, integral equations, also derived in this work, must be used to convert the K-matrix to the physical scattering amplitude. We also show how the new formalism reduces to the standard approach when the $N \to N \pi$ coupling is set to zero.
}

\begin{document}
\maketitle
\flushbottom
\abovedisplayskip 11pt
\belowdisplayskip 11pt
\clearpage

\section{Introduction}

A powerful method for reliably predicting properties of quantum chromodynamics (QCD), is the application of Monte Carlo importance sampling to numerically evaluate the imaginary-time, discretized, finite-volume QCD path integral. This approach, called lattice QCD, delivers estimates of imaginary-time, discretized, finite-volume correlation functions, and various theoretical frameworks are then applied to relate this data to physical observables.

One example application of this general approach is the extraction of finite-volume energies in a given spatial volume, with periodicity $L$, defined with a particular set of internal quantum numbers and a specified total momentum $\boldsymbol P$ in the finite-volume frame. Generally speaking, the numerical values of such finite-volume energies depend on the interaction strength of the hadrons in the channel. Following the seminal work of L{\"u}scher \cite{\luscher} and subsequent generalizations \cite{\twoScatter}, this can be used to constrain the infinite-volume hadronic scattering amplitudes.

Such relations between energies and amplitudes necessarily come with kinematic restrictions. Denoting the $n^{\text{th}}$ finite-volume energy for a given periodicity and momentum by $E_n(L, \boldsymbol P)$, the original work of L\"uscher and the extension to non-zero $\boldsymbol P$ of refs.~\cite{Rummukainen:1995vs,Kim:2005gf,Christ:2005gi} apply only for two-pion elastic scattering, i.e.~for energies satisfying
\begin{equation}
0 < E_n(L, \boldsymbol P)^2 - \boldsymbol{P}^2 < (4 M_\pi)^2 \,,
\label{eq:pion_energy_range}
\end{equation}
where $M_\pi$ is the physical, infinite-volume pion mass. Because $G$-parity prevents the coupling between even- and odd-number multi-pion states, the lowest inelastic threshold is $4 M_\pi$, as indicated. The lower bound of eq.~\eqref{eq:pion_energy_range} is completely irrelevant in practical calculations, because the lowest finite-volume energy is generically either near or above $2 M_\pi$ or else near some bound state mass that, while perhaps well below $2 M_\pi$, is still well above zero.

For this work it is nevertheless instructive to recall the origin of the lower bound. Two reasons can be given (see also ref.~\cite{Kim:2005gf}): One is that the boost matrices to the center-of-mass (CM) frame become ill-defined, or at the very least require a subtle analytic continuation, if the four-vector $ P^\mu = (E, \boldsymbol P)$ becomes either light-like or space-like. The other reason is that the two-to-two amplitude has a left-hand branch cut, with a branch point at Mandelstam $s = P^2 = E^2 - \boldsymbol P^2 = 0$ and this is not taken into account in the derivation.

In the extension to $NN$ systems \cite{Briceno:2014oea}, the analogous restriction is
\begin{equation}
(2 M_N)^2 - M_\pi^2 < E_n(L, \boldsymbol P)^2 - \boldsymbol{P}^2 < (2 M_N+M_\pi)^2 \,,
\label{eq:two_nuc_restriction}
\end{equation}
where $M_N$ is the nucleon mass. The upper bound here is simply the lowest-lying inelastic threshold, that of $NN\pi$ production. The lower bound is due to a left-hand cut from single-pion exchange and is the focus of this work.

\begin{figure}
\begin{center}
\includegraphics[width=.8\textwidth]{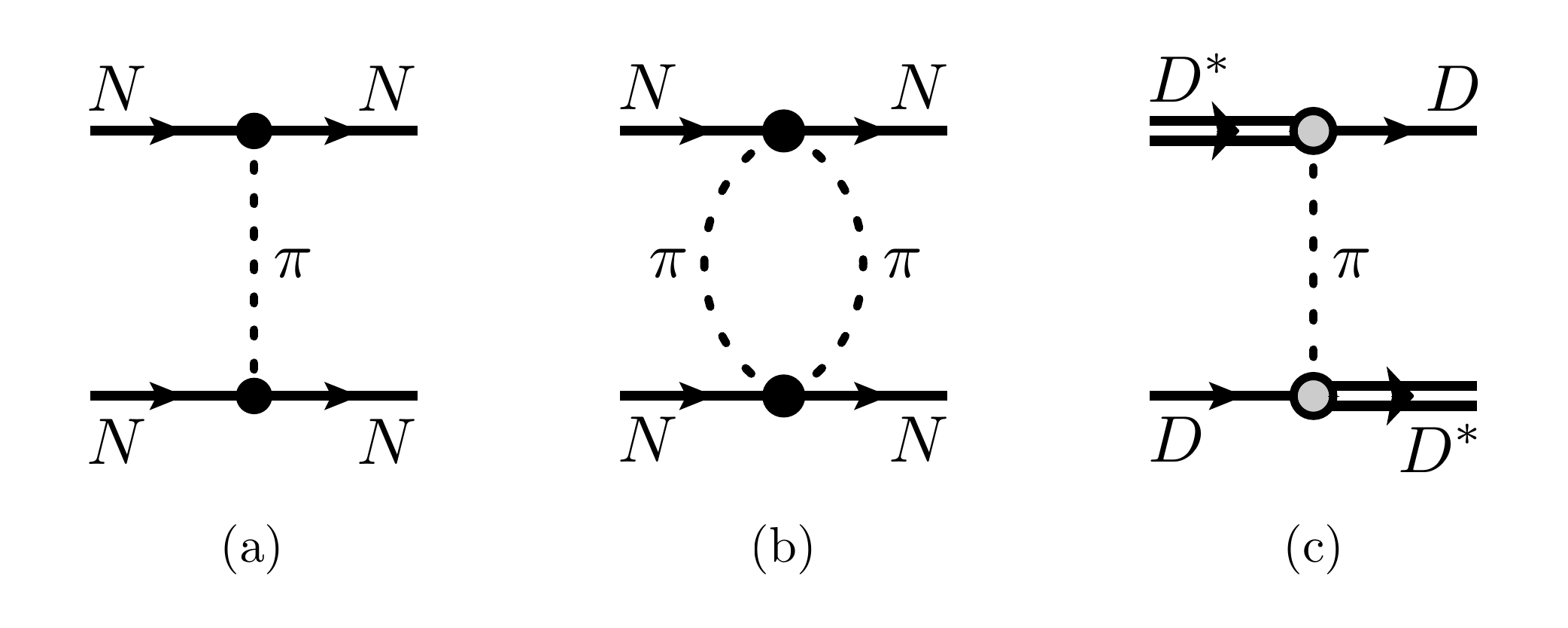}
\caption{Examples of pion exchanges leading to left-hand cuts for various processes: (a) single-pion exchange in $NN \to NN$, leading to the branch cut in angular-momentum projected $N N \to NN$ amplitudes treated in this work, (b) two-pion exchange, leading to a branch cut starting at $s = 4 M_N^2 - 4 M_\pi^2$, \emph{not} treated in this work, (c) single-pion exchange in $D D^* \to D D^*$. \label{fig:various_examples}}
\end{center}
\end{figure}

Over the last decade, extending the range of validity for finite-volume scattering formulae has received much attention. One aspect of this is the generalization to three-particle amplitudes \cite{\threeScatter}. This has progressed rapidly in recent years and a theoretical framework is now in place, along with first numerical lattice QCD determinations of three-to-three scattering amplitudes \cite{\numericalThreeBody}. With respect to \eqref{eq:two_nuc_restriction}, this can be understood as extending beyond the upper cutoff of $(2 M_N + M_\pi)^2$.%
\footnote{For this particular system, the generalization is still outstanding. Given the recent work treating non-identical and non-degenerate particles \cite{\threeNonDeg} and particles with intrinsic spin \cite{Draper:2023xvu}, no fundamental issues are expected in deriving the relevant formalism.}
In this article, we are concerned with extending the regime of validity below the lower cutoff, which arises due to the diagram shown in figure~\ref{fig:various_examples}(a). This generates a pole in the amplitude with fixed external momenta and when the amplitudes latter is projected to definite angular momentum, the pole becomes a branch cut with a branch point at $s = (2 M_N)^2 - M_\pi^2$.

Remarkably, recent lattice calculations have extracted finite-volume energies violating the lower bound of eq.~\eqref{eq:two_nuc_restriction}, and it is currently unclear how to use these energies to constrain the scattering amplitude. As already discussed in ref.~\cite{Green:2021qol}, the simplest way to see that the standard formalism breaks down is to note that the latter always predicts a real-valued scattering amplitude for $s < (2 M_N)^2$ whereas, for $s < (2 M_N)^2 - M_\pi^2$, it is known that the angular-momentum-projected amplitude must be complex.

The realness of the scattering amplitude is a symptom of the problem, but does not directly reveal the underlying reason. As we discuss in detail in section~\ref{sec:FVformalism}, the fundamental issue is subtle. In short, the relation between finite-volume energies and infinite-volume scattering amplitudes is derived by studying an all orders diagrammatic expansion of a finite-volume correlation function. At various steps, subdiagrams with four-external legs are identified and, because these are inside a larger diagram, the momenta they carry take on all values and do not satisfy the on-shell condition $k^2 = M_N^2$. However, at a particular step in the derivation, one can show that the replacement $k^2 \to M_N^2$ is valid and only leads to neglected $L$-dependence of the form $e^{- \mu(s) L}$ for some (possibly energy-dependent) characteristic scale $\mu(s)$.

For $s < (2 M_N)^2 - M_\pi^2$, however, this on-shell replacement is invalid. In this work we resolve the issue by separating out the pion exchanges in all diagrams and avoiding the problematic step for these contributions. The result is a generalized formalism that relates finite-volume energies to scattering amplitudes in an extended region:
\begin{equation}
(2 M_N)^2 - (2 M_\pi)^2 < E_n(L, \boldsymbol P)^2 - \boldsymbol{P}^2 < (2 M_N+M_\pi)^2 \,,
\label{eq:two_nuc_extended}
\end{equation}
where the new lower limit arises from two-pion exchanges, such as that shown in figure~\ref{fig:various_examples}(b). These are not treated in our generalization.

It should be emphasized that neither L\"uscher nor subsequent authors claimed that the scattering relations would hold on the cut. However, it was only recently stressed in ref.~\cite{\cutFormalismNeeded} that the subthreshold analytic continuation is both invalid as one approaches the cut and also needed due to lattice-determined energies in this regime.

We further comment that the inequality~\eqref{eq:two_nuc_restriction} does not define strict boundaries of validity. As one approaches the inelastic threshold at $s = (2 M_N + M_\pi)^2$ from below, $\mu (s)$ tends to zero so that the neglected $e^{- \mu(s) L}$ terms become enhanced. In this work, we find that the same breakdown occurs as one approaches the lower limit of $s = (2 M_N)^2 - M_\pi^2$ from above. Thus, it is prudent to apply the formulas derived here, even for finite-volume energies above the branch point, to study the effect of potential enhancement of neglected exponentials. An analogous issue is that, for infinite-volume scattering amplitudes, the convergence of the partial-wave expansion becomes arbitrarily bad as one approaches $s = (2 M_N)^2 - M_\pi^2$. This is also addressed with the formalism presented in this work, since partial-wave projection is only required for an intermediate infinite-volume quantity in which the offending cut is absent.

The basic workflow resulting from our generalized scattering formalism closely mimics that used in three-particle finite-volume scattering calculations. In a first step, the finite-volume energies are used to constrain the $N N \pi$ coupling together with an intermediate K-matrix, denoted $\overline {\mathcal K}^{\sf os}$, where the bar indicates a non-standard quantity and the label ${\sf os}$ stands for on shell, emphasizing the physical kinematic dependence. Then, in a second step, integral equations are solved to relate the $N N \pi$ coupling and $\overline {\mathcal K}^{\sf os}$ to the physical two-to-two scattering amplitude. Still, the results presented here are not just some version of the three-particle formalism in disguise. For one thing, the present method is manifestly invalid for $s > (2 M_N + M_\pi)^2$.

At the same time, it has been recently understood that the three-particle formalism of refs.~\cite{\threeForTChannel} can also provide a solution to the left-hand cut problem. If one of the scatterers in the target two-particle channel can be represented as a pole in a two-to-two subprocess amplitude, then it is possible to analytically continue the three-to-three relations below threshold and infer a two-to-two formalism in which the cut is properly treated.
This was first identified in ref.~\cite{Dawid:2023jrj}, in the context of a toy system of three identical spin-zero particles. In this important work, the authors predict both the two-to-two scattering amplitude and the finite-volume spectrum from the underlying three-particle formalism. They then show numerically that the two are clearly related by standard two-particle formalism for energies away from the left-hand cut, but that the relation breaks as expected on the cut.

While the present article focuses on $NN \to NN$, it is applicable for any single-channel two-to-two system with identical scattering particles. In particular, the relations are provided for any value of intrinsic spin. The extension to non-identical and non-degenerate particles will be considered in future work, and is needed to include systems such as $D D^* \to D D^*$, i.e.~the vector-pseudoscalar channel with the internal quantum numbers of two charm quarks. This sector is of great interest due to the experimental discovery of a doubly charmed tetraquark $T_{cc}(3875)^+$ by LHCb, as presented in refs.~\cite{\LHCbTcc}.

In fact, a lattice QCD calculation of the relevant finite-volume spectrum for $D D^* \to D D^*$ was already performed and presented in ref.~\cite{Padmanath:2022cvl}, including the prediction of virtual bound state at the heavier than physical pion mass of $M_\pi \approx 280 \, \text{MeV}$, for which the $D^*$ is a stable particle. The authors argue that the virtual bound state is likely associated with the $T_{cc}^+$. Subsequently, ref.~\cite{Du:2023hlu} argued that the left-hand cut, resulting from the pion exchange shown in figure~\ref{fig:various_examples}(c), could qualitatively change the result. In light of this, it would clearly be valuable to make use of the finite-volume energies that overlap the branch cut. One formal method to address this is to extend the three-particle formalism to $D D \pi$; this is underway. The alternative method of extending the work here to non-identical and non-degenerate particles, and the relation of the latter to the three-particle approach, will also be considered in a future publication.

Two other publications, of relevance for the left-hand cut in the finite-volume context, have come to our attention during the completion of this work. These are ref.~\cite{Sato:2007ms}, concerning exponentially suppressed corrections to finite-volume scattering formula for $NN \to NN$, as well as ref.~\cite{Meng:2021uhz}, which makes use of a plane-wave basis for the quantization condition. We compare our approach to that used in these earlier works in section~\ref{sec:previous_work}.

The remainder of this manuscript is organized as follows: In the next section, we review properties and relations for the infinite-volume $NN \to NN$ scattering amplitude in the region given by the inequality \eqref{eq:two_nuc_extended}. Then, in section~\ref{sec:FVformalism}, we turn to the finite-volume system. Over numerous subsections, we detail the breakdown in the usual derivation and provide the resolution, summarized in section \ref{sec:result}. As we explain there, this depends on an intermediate quantity that is related to the scattering amplitude via integral equations in section~\ref{sec:int_eqs}. Finally, in section~\ref{sec:explore}, we explore the result in various ways, e.g.~by recovering the standard formalism in the limiting case when the $N N \pi$ coupling vanishes. In section~\ref{sec:conclusions} we briefly conclude. We also include three appendices collecting technical details of the derivation.

\section{Infinite-volume scattering}

Here we review aspects of infinite-volume scattering theory that are relevant to this work. In the following subsection, we recall a derivation of the relation between the scattering amplitude and the K-matrix. We give the relation using all-orders diagrammatic arguments, and this serves as a warm-up for the finite-volume derivation presented in section~\ref{sec:FVformalism}. In section~\ref{sec:inf_vol_cut}, we discuss the analytic continuation of the amplitude below threshold and the origin of the left-hand cut. To avoid clutter of notation, we first give the discussion for spin-zero nucleons. Section~\ref{sec:inf_vol_spin} then provides the details to extend all results to generic spin, including the physical spin-half case.

\subsection{Scattering amplitude, K-matrix, and phase space} \label{sec:scattering_K_phase}

Consider a generic relativistic quantum field theory that has one heavy spin-zero particle, with mass $M_N$, and one light spin-zero particle, with mass $M_\pi$, such that $M_\pi \ll M_N$. Refer to the light particle as a pion (with associated quantities labelled $\pi$) and the heavy particle as a nucleon (with associated quantities labelled $N$). Further require that the nucleon is charged under a $U(1)$ symmetry such that the single-particle states can be written $\vert N, \boldsymbol p, \pm \rangle$ where the second label is spatial momentum and the third is charge. In QCD, baryon number plays the role of this conserved charge. In this set-up we consider a single flavor of pion, corresponding to the $\pi^0$ in nature. The field is neutral under the $U(1)$ charge, i.e.~has baryon number zero.

The infinite-volume single-particle states are normalized as
\begin{align}
\label{eq:N_scalaR_Normalization} \langle N, \boldsymbol p', q' \vert N, \boldsymbol p, q \rangle & = \delta_{q'q} 2 \omega_N(\boldsymbol p) (2 \pi)^3 \delta^3(\boldsymbol p' - \boldsymbol p) \,,
\\
\langle \pi, \boldsymbol p' \vert \pi, \boldsymbol p \rangle & = 2 \omega_\pi(\boldsymbol p) (2 \pi)^3 \delta^3(\boldsymbol p' - \boldsymbol p) \,,
\end{align}
where
\begin{align}
\omega_N(\boldsymbol p) = \sqrt{\boldsymbol{p}^2 + M_N^2} \,, \qquad \omega_\pi(\boldsymbol p) = \sqrt{\boldsymbol{p}^2 + M_\pi^2} \,.
\end{align}
We allow the theory to have generic interactions respecting the conservation of baryon number. In particular, we include $N \to N \pi$ couplings of the form
\begin{equation}
\mathcal L_{N^\dagger \! N \pi}(x) = N^\dagger(x) N(x) \mathcal F \big ( \pi(x) \big ) \,, \label{eq:lagrangian_density}
\end{equation}
where $\mathcal F$ is a generic polynomial in the pion field, starting with a linear dependence on $\pi(x)$.

For both the pion and the scalar nucleon, the momentum-space dressed scalar propagator takes the form
\begin{equation}
\label{eq:Delta_full_def}
\Delta_{x, i \epsilon}(k) = \frac{i}{k^2 - M_x^2 - \Pi_x(k^2)+ i \epsilon} \,,
\end{equation}
for $x \in \{\pi, N\}$. We require the self-energy $\Pi_x(k^2)$ to satisfy $\Pi_x(M_x^2) = \Pi'_x(M_x^2) = 0$ where the prime indicates a derivative. This implies that the residue of the single-particle pole has magnitude one and that $M_x$ is the physical (pole) mass. Throughout this work, we use Minkowski signature four-vectors satisfying $k^2 = (k^0)^2 - \boldsymbol k^2$ where $\boldsymbol k$ is a spatial three-vector. The use of Minkowski signature may seem surprising, in particular in the finite-volume analysis of section \ref{sec:FVformalism}, since the purpose of this derivation is to interpret Euclidean-signature lattice calculations. However, as the finite-volume energies are metric-independent, this detail of the numerical calculation is irrelevant here.

\begin{figure}
\begin{center}
\includegraphics[width=\textwidth]{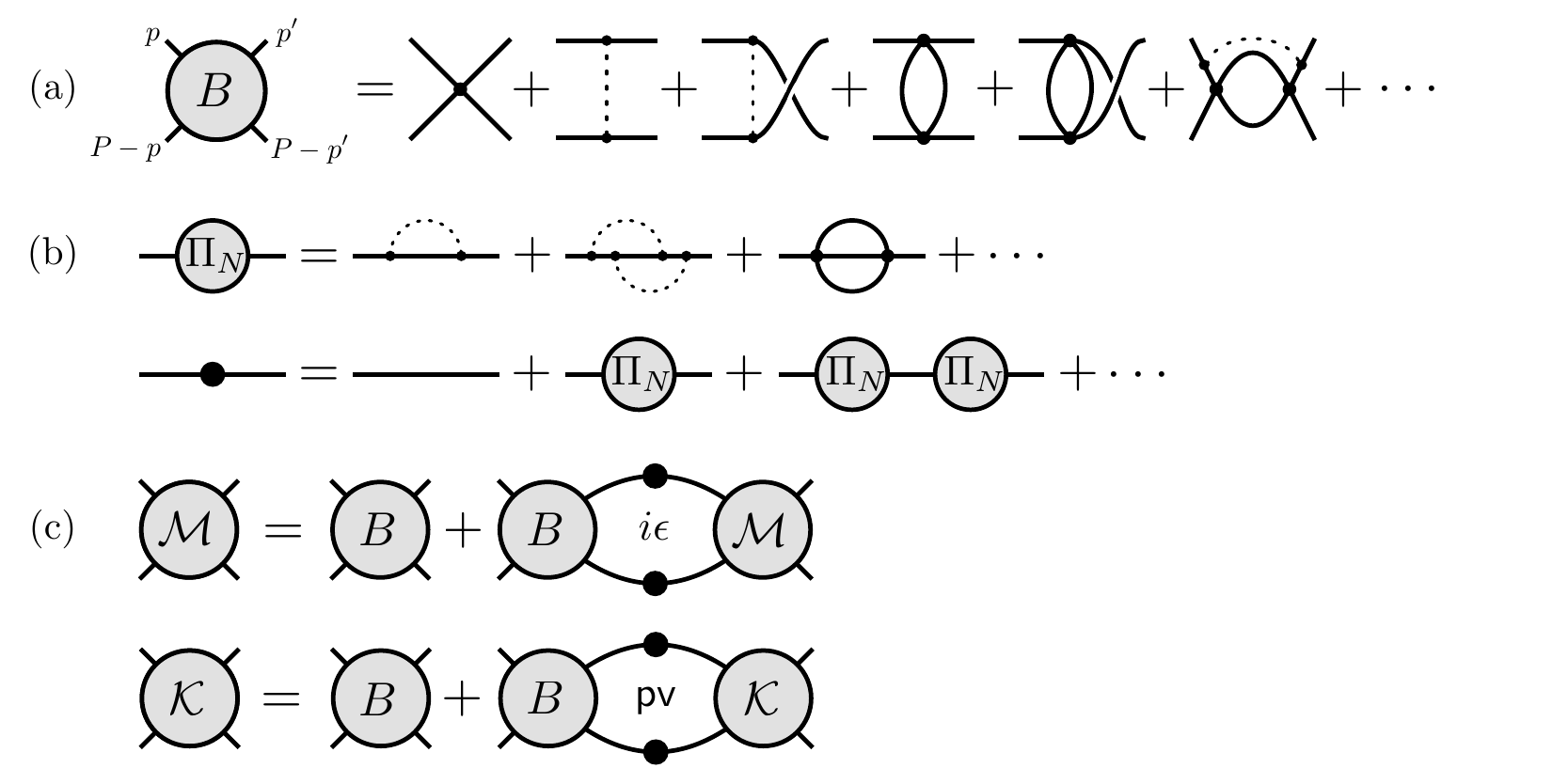}
\vspace{10pt}
\caption{
(a) Diagrammatic representation of the $NN$ Bethe-Salpeter kernel $B(P, p,p')$, built from all $NN \to NN$ diagrams that are two-nucleon irreducible in the $s$-channel. (b) Diagrammatic representation of the nucleon self-energy $\Pi_N (p^2)$ and of the dressed nucleon propagator. (c) Representation of the infinite-volume integral equations:~eqs.~\eqref{eq:m_integral_equation} and \eqref{eq:k_integral_equation} of the main text.
\label{fig:bskm_infvol}
}
\end{center}
\end{figure}

Next, introduce $B(P, p,p')$ as the Bethe-Salpeter kernel for $N N \to N N$ scattering, formally defined as the sum of all amputated two-to-two diagrams that are two-nucleon irreducible with respect to cuts intersecting the flow of total energy and momentum, see figure~\ref{fig:bskm_infvol}(a). As shown in the figure, the incoming nucleons carry four-momenta $p$ and $P-p$ and the outgoing nucleons carry $p'$ and $P-p'$. So $P$ is the total four-momentum in a given frame, related to the total energy and momentum via $P^\mu = (E, \boldsymbol P)$. We also define $s$ and $E^\star$ via
\begin{equation}
s = E^{\star 2} = P^2 = E^2 - \boldsymbol P^2 \,.
\end{equation}

In terms of the fully dressed nucleon propagator and the Bethe-Salpeter kernel, the scattering amplitude $\mathcal M$ and the K-matrix $\mathcal K$ for $ N N \to N N$ scattering can be expressed, respectively, via the integral equations
\begin{align}
\label{eq:m_integral_equation} i \mathcal M(P, p,p') & = i B(P, p,p') + \frac12 \int \! \frac{d^4k}{(2\pi)^4} \, i B(P, p, k) \Delta_{N, i \epsilon}(P-k) \Delta_{N, i \epsilon}(k) i \mathcal M(P, k, p') \,, \\[5pt]
\begin{split}
i \mathcal K(P, p,p') & = i B(P, p,p')
\\[3pt] & \hspace{0pt} + \frac12 \, {\sf pv} \! \int \! \! \frac{d^3 \boldsymbol k}{(2\pi)^3} \bigg [ \int \! \frac{d k^0}{2\pi} i B(P, p, k) \Delta_{N, i \epsilon}(P-k) \Delta_{N, i \epsilon}(k) i \mathcal K(P, k, p') \bigg ]_{\epsilon \to 0^+}
\end{split}\,, \label{eq:k_integral_equation}
\end{align}
where the label ${\sf pv}$ indicates that the principal-value pole prescription is used to define the K-matrix.
In eq.~\eqref{eq:k_integral_equation} the $i \epsilon$ prescription is only required to evaluate the $k^0$ integral, as indicated. See figure~\ref{fig:bskm_infvol}(b) for a representation of these equations.

Iteratively substituting $\mathcal M$ and $\mathcal K$ on the right-hand sides of these equations allows us to write a geometric series. Introducing a compact notation, this can be written as
\begin{align}
\label{eq:Mseries} i \mathcal M(s) & = i B(s) + i B(s) \circ_{i \epsilon} i \mathcal M(s) = \sum_{n=0}^\infty i B(s) [ \circ_{i \epsilon} \, i B(s) ]^n \,,
\\
\label{eq:Kseries} i \mathcal K(s) & = i B(s) + i B(s) \circ_{\sf pv} i \mathcal K(s) = \sum_{n=0}^\infty i B(s) [ \circ_{\sf pv} \, i B(s) ]^n \,,
\end{align}
where the $\circ$ indicates that the quantities are combined via an integral as
\begin{align}
\label{eq:B_ieps_compact} i B(s) \circ_{i \epsilon} i B(s) & = \frac12 \int \! \frac{d^4k}{(2\pi)^4} i B(P, p, k) \Delta_{N, i \epsilon}(P-k) \Delta_{N, i \epsilon}(k) i B(P, k, p') \,, \\[5pt]
i B(s) \circ_{\sf pv} i B(s) & = \frac12 \, {\sf pv} \! \int \! \! \frac{d^3 \boldsymbol k}{(2\pi)^3} \bigg [ \int \! \frac{d k^0}{2\pi} i B(P, p, k) \Delta_{N, i \epsilon}(P-k) \Delta_{N, i \epsilon}(k) i B(P, k, p') \bigg ]_{\epsilon \to 0^+} \,.
\label{eq:B_pv_compact}
\end{align}

To make use of these expressions, we now set all external four-momenta on shell so that they satisfy $p^2 = M_N^2$, etc. We also restrict attention to the regime of elastic scattering, defined via $(2 M_{N})^2 < s < (2 M_N + M_\pi)^2$. Then one can show
\begin{equation}
\label{eq:ieps_pv_diff} i B(s) \circ_{i \epsilon} i B(s) - i B(s) \circ_{{\sf pv}} i B(s) =
\frac12 \int \! \frac{d^3 \boldsymbol k}{(2 \pi)^3} i B(P, p, k) \frac{\pi \delta(E - \omega_{N}(\boldsymbol k)- \omega_{N}(\boldsymbol P - \boldsymbol k))}{2 \omega_{N}(\boldsymbol k) 2 \omega_{N}(\boldsymbol P - \boldsymbol k)} i B(P, k, p') \,.
\end{equation}
This holds because, when the $k^0$ integral is evaluated via contour integration, various contributions cancel between the principal-value and $i \epsilon$ integrals. In particular, contributions that are analytic in $\boldsymbol k$ and do not require a pole prescription will be identical between the two terms on the left-hand side. The exception is the contribution arising from the pole at $k^0 = \omega_{N}(\boldsymbol k)$. For this term, only the real part of the $i \epsilon$ contribution cancels, leaving the imaginary part to generate the term shown.

Next we project both sides of eq.~\eqref{eq:ieps_pv_diff} to definite orbital angular momentum. To do so define $\boldsymbol p^\star$ as the spatial part of the four-momentum resulting from boosting $p^\mu = (\omega_N(\boldsymbol p), \boldsymbol p)$ with boost velocity $\boldsymbol \beta = - \boldsymbol P/E$. [See also eq.~\eqref{eq:boost_explicit_def} below.] Then the expressions above can be expressed in terms of $\boldsymbol p'^\star$ and $\boldsymbol p^\star$ and one can define
\begin{equation}
\big [ i \widetilde B(s) \circ_{i \epsilon} i \widetilde B(s) \big ]_{\ell m, \ell' m'}
= \frac{1}{\vert \boldsymbol p^\star \vert^{\ell} \, \vert \boldsymbol p'^\star \vert^{\ell'}}
\int \! \frac{d \Omega_{\hat{\boldsymbol p}^\star}}{4 \pi} \! \int \! \frac{d \Omega_{\hat{\boldsymbol p}'^\star}}{4 \pi}
\, Y_{\ell m}(\hat {\boldsymbol p}^\star) \,
[ i B \circ_{i \epsilon} i B ](\boldsymbol p^\star, \boldsymbol p'^\star) \, Y^*_{\ell' m'}(\hat{\boldsymbol p}'^\star) \,,
\end{equation}
where the $Y_{\ell m}$ are standard spherical harmonics. Here we have also divided out a factor of $\vert \boldsymbol p'^\star \vert^{\ell'} \, \vert \boldsymbol p^\star \vert^{\ell}$. This is the leading-order behavior of this combination as $\vert \boldsymbol p'^\star \vert , \vert \boldsymbol p^\star \vert \to 0$ such that, by removing it, we have defined a quantity that approaches a constant in this limit. We use the tilde to indicate this rescaling. (See also, e.g.~ref.~\cite{\rescaleBarrier}.)

From here it follows that
\begin{align}
\label{eq:BrhoB} \big [ i \widetilde B(s) \circ_{i \epsilon} i \widetilde B(s) - i \widetilde B(s) \circ_{{\sf pv}} i \widetilde B(s) \big ]_{\ell m, \ell' m'} & = \big [ i \widetilde B(s) \, \widetilde \rho(s) \, i \widetilde B(s) \big ]_{\ell m, \ell' m'} \,,
\end{align}
where, on the right-hand side, each of the three-factors is a diagonal matrix, with each entry depending only on $s$. In particular
\begin{align}
\widetilde \rho_{\ell m, \ell' m'}(s) & = 2 \pi (k_{\sf os}^\star)^{\ell + \ell'}
\int \frac{d^3 \boldsymbol k}{(2 \pi)^3} Y_{\ell m}(\hat{\boldsymbol{k}}^\star) \frac{\pi \delta(E - \omega_{N}(\boldsymbol k)- \omega_{N}(\boldsymbol P - \boldsymbol k))}{2 \omega_{N}(\boldsymbol k) 2 \omega_{N}(\boldsymbol P - \boldsymbol k)} Y^*_{\ell' m'}(\hat{\boldsymbol{k}}^\star) \,,
\\
& = \delta_{\ell \ell'} \delta_{m m'} \, (k_{\sf os}^\star)^{2 \ell} \, \rho(s) \,,
\label{eq:rho_tilde_def}
\end{align}
where $\rho(s)$ is the volume of phase space
\begin{equation}
\rho(s) = \frac{k^\star_{\sf os}}{16 \pi \sqrt{s}}\,, \qquad \text{for} \qquad k^\star_{\sf os} = \sqrt{s/4 - M_{N}^2} \,, \label{eq:rhoN_pN_def}
\end{equation}
and ${\sf os}$ stands for on shell.

A non-trivial feature of eq.~\eqref{eq:BrhoB} is that the $B$ factors on the right-hand side no longer depend on the magnitude of integrated momentum. This follows from re-expressing quantities in terms of $\boldsymbol k^\star = \vert \boldsymbol k^\star \vert \hat {\boldsymbol k}^\star$. Then the Dirac delta function sets $\vert \boldsymbol k^\star \vert = k^\star_{\sf os}$ and the directional degrees of freedom are decomposed into the spherical harmonics.

Equation~\eqref{eq:BrhoB} holds for any factors on either side of $ \circ_{i \epsilon}$ and $ \circ_{{\sf pv}}$ and can be interpreted as an identity relating the pole prescriptions: $ \circ_{i \epsilon} =\circ_{{\sf pv}} + \rho$. Combining this with eqs.~\eqref{eq:Mseries} and \eqref{eq:Kseries} then allows one to relate the scattering amplitude and the K-matrix via
\begin{align}
\widetilde{\mathcal M}(s) & = \sum_{n=0}^\infty \widetilde{B}(s) \big [ \big ( \circ_{{\sf pv}} + \widetilde \rho(s) \big ) \, i \widetilde{B}(s) \big ]^n
\\
& = \sum_{n=0}^\infty \widetilde{\mathcal K}(s) \big [ i \widetilde \rho(s) \, \widetilde{\mathcal K}(s) \big ]^n \,, \qquad \qquad \qquad \text{for} \quad \widetilde{\mathcal K}(s) = \sum_{n'=0}^\infty \widetilde{B}(s) \big [ \circ_{{\sf pv}} \, i \widetilde{B}(s) \big ]^{n'} \,,\\
& = \frac{1}{\widetilde{\mathcal K}(s)^{-1} - i \widetilde \rho(s)} \,. \label{eq:final_M_to_K}
\end{align}
Taking the standard parametrization
\begin{equation}
\label{eq:Ktilde_delta_relation}
\widetilde{\mathcal K}_{\ell m, \ell' m'}(s) = \delta_{\ell \ell'} \delta_{m m'} (k^\star_{\sf os})^{-2 \ell} \, \mathcal K^{(\ell)}(s) = \delta_{\ell \ell'} \delta_{m m'} (k^\star_{\sf os})^{-2 \ell} \, 16 \pi \sqrt{s} \, \frac{\tan \delta_{\ell}( k^\star_{\sf os})}{k^\star_{\sf os}} \,,
\end{equation}
then gives
\begin{equation}
\mathcal M^{(\ell)}(s) = \frac{16 \pi \sqrt{s}}{k^\star_{\sf os} \cot \delta_{\ell}( k^\star_{\sf os}) - i k^\star_{\sf os}} \,,
\end{equation}
where $\widetilde {\mathcal M}_{\ell m, \ell' m'}(s) = \delta_{\ell \ell'} \delta_{m m'} (k^\star_{\sf os})^{-2 \ell} \mathcal M^{(\ell)}(s) $ defines the $\ell^{\text{th}}$ partial-wave component of the scattering amplitude.

So far we have assumed $(2 M_{N})^2 < s < (2 M_{N} + M_\pi)^2$. The case of $ s > (2 M_{N} + M_\pi)^2 $ requires explicit treatment of the $N N \pi$ states within the Bethe-Salpeter kernel. This leads to significantly more complicated unitarity constraints that have, however, received significant attention recently and are by now well understood \cite{\threeBodyIntegralEquations}.

In this work, we are instead interested in the other direction: $s < (2 M_N)^2$. In the next subsection, we describe the properties of the infinite-volume scattering amplitude and K-matrix continued into this region, with a particular focus on the left-hand cut. This will set the background and the notation for our finite-volume discussion of this region in section \ref{sec:FVformalism}.

\subsection{Analytic continuation and the left-hand cut}\label{sec:inf_vol_cut}

The K-matrix is generally expected to be a meromorphic function in a region of the complex plane containing $(k^\star_{\sf os})^2 = 0$, corresponding to $s = (2 M_N)^2$. As a result, one can expand $k \cot \delta_{\ell}(k)$ about this point as a series of polynomials and poles
\begin{equation}
k^\star_{\sf os} \cot \delta_{\ell}(k^\star_{\sf os}) = - \frac{1}{a_{\ell} (k^\star_{\sf os})^{2 \ell}} \Big [1 + \sum_{j=1}^{\infty} c_j (k^\star_{\sf os})^{2 j} \Big ] + \sum_{n=1}^P \sum_{j=0}^\infty \frac{A_{n,j} (k^\star_{\sf os})^j}{(k^\star_{\sf os})^2 - C_n} \,.
\end{equation}
This then defines a sub-threshold analytic continuation for the K-matrix.

In the case where interactions of the form $\pi(x)N^\dagger(x) N(x)$, $\pi(x)^2 N^\dagger(x) N(x)$ are included, as in eq.~\eqref{eq:lagrangian_density}, this expansion breaks down due to branch cuts, generically referred to as left-hand cuts, that run along the negative real axis of the complex $s$ plane. The nearest cut starts at $s = 4M_N^2 - M_\pi^2$ and arises due to single pion exchanges.

\begin{figure}
\begin{center}
\includegraphics[width=0.7\textwidth]{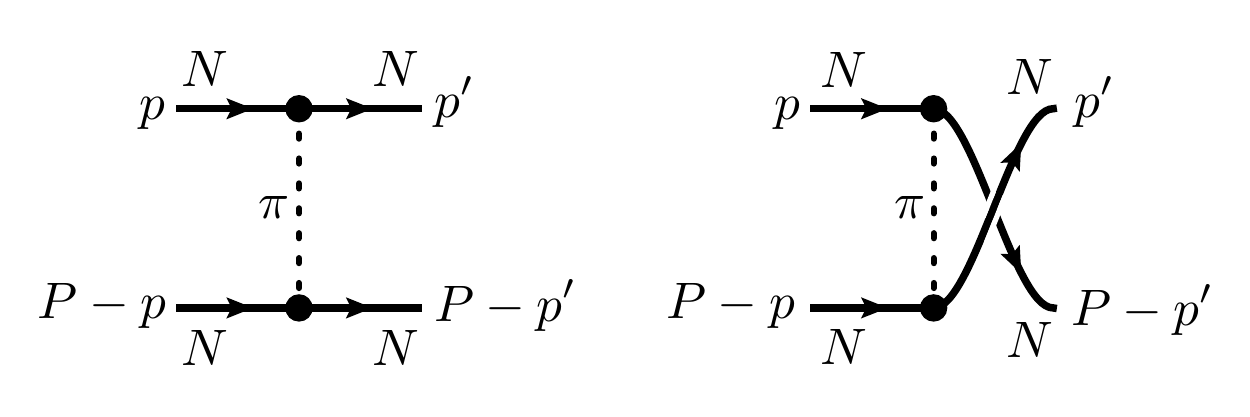}
\vspace{10pt}
\caption{Single-meson exchange contribution to the $NN \to NN$ scattering amplitude and K-matrix.\label{fig:t_channel_exchange}}
\end{center}
\end{figure}

To see this in more detail, we consider the tree-level $t$- and $u$-channel exchange diagrams shown in figure~\ref{fig:t_channel_exchange}. First assuming a fully dressed pion propagator and general vertex functions, these diagrams have the form
\begin{align}
D_{t}(P, p, p') & \equiv G(p,p') \ \Delta_{\pi, i \epsilon}(p'-p) \ G(P-p, P-p')\,, \label{eq:tchan_ker_infinite_volume_dressed} \\
D_{u}(P, p, p') & \equiv G(p,P-p') \ \Delta_{\pi, i \epsilon}(P - p' - p) \ G(p,P-p') \,, \label{eq:uchan_ker_infinite_volume_dressed}
\end{align}
where the middle propagator factor in each expression was introduced in eq.~\eqref{eq:Delta_full_def}. Here we have also introduced the form factor $G(p,p')$ defined as the amputated sum of all $N \to N \pi$ diagrams or, equivalently, via the matrix element
\begin{equation}
G(p,p') \equiv \Delta_{\pi, i \epsilon}(p'-p)^{-1} \langle N, \boldsymbol p', q' \vert \pi(0) \vert N, \boldsymbol p, q \rangle \,.
\label{eq:me_g_def}
\end{equation}

Because our concern is the singularities in the diagrams $D_{t}(P, p,p')$ and $D_{u}(P, p,p')$, it will be convenient to identify simpler expressions that contain these relevant features but differ from the full diagram by analytic terms. To define these first note that, for the case where all fields have spin zero, $G(p,p')$ is a Lorentz scalar and only depends on Lorentz scalar momentum combinations. Further, noting that $p^2 = p'^2 = M_N^2$ in eq.~\eqref{eq:me_g_def}, we see that the function in fact only depends on the momentum transfer. This allows us to define $\mathcal G$ as the same function of a single Lorentz invariant
\begin{equation}
\mathcal G\Big ((p'-p)^2 \Big ) = G(p,p') \,.
\end{equation}
We then define the coupling as
\begin{equation}
g \equiv \lim_{q^2 \to M_\pi^2}\mathcal G(q^2) \,.
\end{equation}

With this coupling defined we now observe that the differences
\begin{align}
\label{eq:delta_t_sep}
\delta D_{t}(P,p,p') & \equiv D_{t}(P,p,p') - i g^2 \mathcal T(P, p, p') \,, \\
\delta D_{u}(P,p,p') & \equiv D_{u}(P,p,p') - i g^2 \mathcal U(P, p, p') \,,
\end{align}
are analytic in the vicinity of the pole, where
\begin{align}
i g^2 \mathcal T(P, p, p') & \equiv -\frac{ig^2}{(p' - p)^2 - M_\pi^2 + i\epsilon} = -\frac{ig^2}{t - M_\pi^2 + i\epsilon}\ , \label{eq:tchan_ker_infinite_volume} \\
i g^2 \mathcal U(P, p, p') & \equiv -\frac{ig^2}{(P - p' - p)^2 - M_\pi^2 + i\epsilon} = -\frac{ig^2}{u - M_\pi^2 + i\epsilon}\ , \label{eq:uchan_ker_infinite_volume}
\end{align}
correspond precisely to the $t$- and $u$-channel exchange poles. Here we have also made use of the Mandelstam invariants $t$ and $u$. Recall that, for incoming particles with momenta $p$ and $P-p$ scattering to outgoing particles with $p'$ and $P-p'$ these are defined as
\begin{align}
s & = (p + P-p)^2 = P^2 \,,
\\
t & = (p'-p)^2 \,,
\\
u & = (P - p' - p)^2 \,,
\end{align}
and satisfy
\begin{equation}
s + t + u = p^2 + p'^2 + (P-p)^2 + (P-p')^2 \underset{{\sf on\ shell}}{=} 4 M_N^2 \,,
\end{equation}
where the second equality only holds when all momenta are on the mass shell.

For the following section, it will be convenient to define versions of the exchanges of eqs.~\eqref{eq:tchan_ker_infinite_volume} and \eqref{eq:uchan_ker_infinite_volume} in which the four-vectors $p$ and $p'$ are on shell and the corresponding three-momenta, as well as $P$, are expressed in the two-particle center-of-mass (CM) frame, e.g.~for the $t$-channel exchange
\begin{equation}
\mathcal T(s, \boldsymbol p^\star, \boldsymbol p'^\star) \equiv \mathcal T(P, p, p') \Big \vert_{p^0 = \omega_N(\boldsymbol p), \ p'^0 = \omega_N(\boldsymbol p')} \,. \label{eq:tchan_ker_onshell}
\end{equation}
Here we abuse notation by letting the nature of the arguments distinguish the two functions. This can then be decomposed in partial waves as
\begin{equation}
\label{eq:T_partial_wave_decom}
\mathcal T(s, \boldsymbol p^\star, \boldsymbol p'^\star)= \sum_{\ell=0}^\infty \mathcal T_{\ell} (s, \vert \boldsymbol p^\star \vert, \vert \boldsymbol p'^\star \vert) \, (2 \ell +1) \, P_{\ell}(\cos \theta^\star) \,,
\end{equation}
where $\cos \theta^\star \equiv (\boldsymbol p^\star \cdot \boldsymbol p'^\star) / (\vert \boldsymbol p^\star \vert \vert \boldsymbol p'^\star \vert)$ and $P_{\ell}(x)$ is a standard Legendre polynomial.

The final step in reducing $\mathcal T_{\ell} (s, \vert \boldsymbol p^\star \vert, \vert \boldsymbol p'^\star \vert)$ is to note that, when the particles are scattering with physical back-to-back momenta, the magnitudes are constrained to satisfy $\vert \boldsymbol p^\star \vert = \vert \boldsymbol p'^\star \vert = k^\star_{\sf os}$. For this reason, the partial-wave projected amplitude reduces to a single coordinate function that we denote by
\begin{equation}
\mathcal T^{\sf os}_\ell(s) \equiv \mathcal T_{\ell} \big (s,k^\star_{\sf os}, k^\star_{\sf os} \big ) \,.
\end{equation}
Equivalent expressions hold for the $u$-channel exchange.

\begin{figure}
\begin{center}
\includegraphics[width=\textwidth]{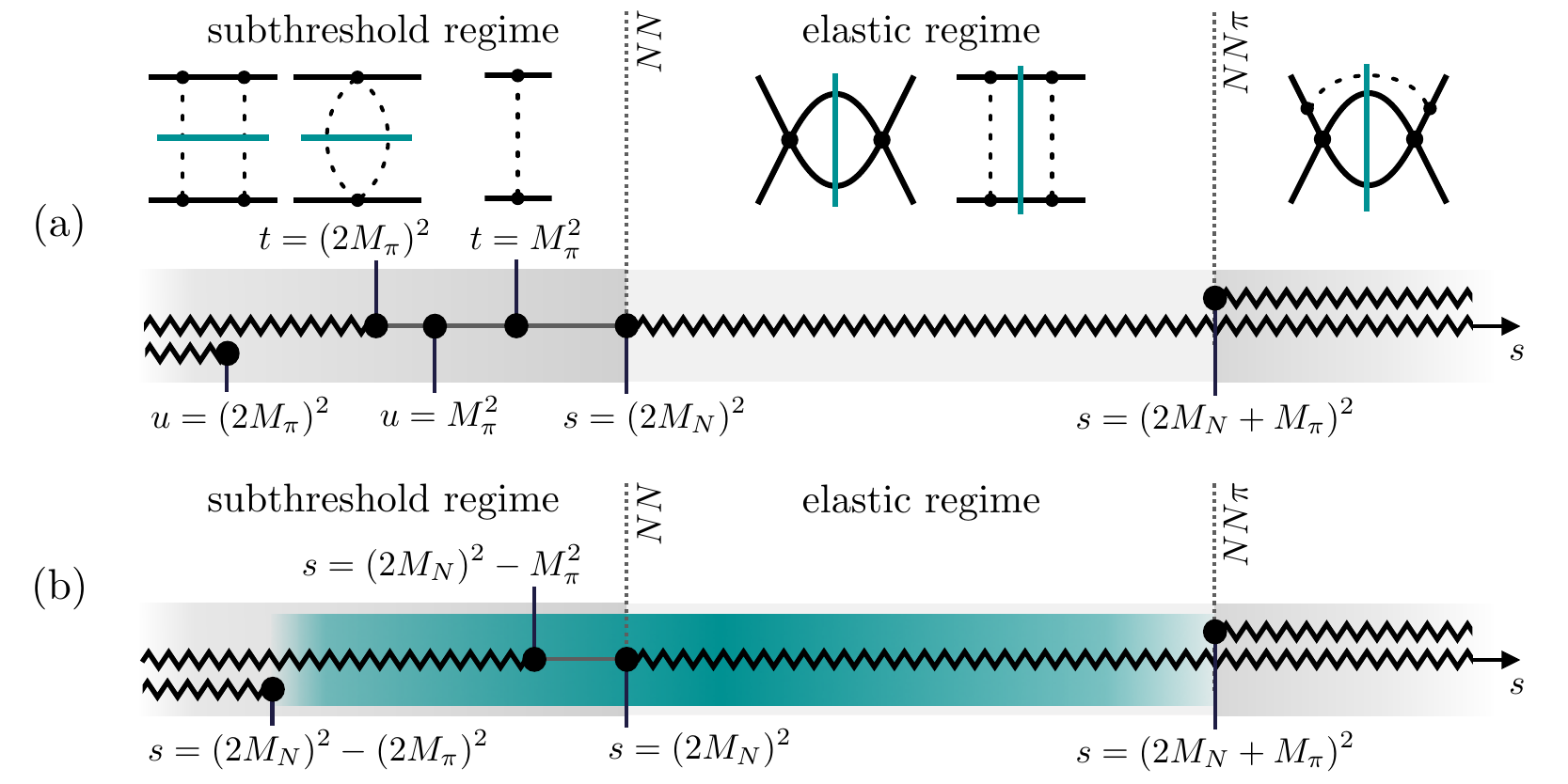}
\vspace{6pt}
\caption{Summary of the analytic structure of the $NN\to NN$ amplitude in the complex $s$ plane: (a) at fixed CM scattering angle $\theta^\star$ and (b) angular-momentum projected to a given partial wave. In both cases, we have the usual $s$-channel cuts starting at the $NN$ and $NN\pi$ thresholds (indicated as vertical dashed lines). Both poles and cuts, associated with the $t$-and $u$-channel exchanges, arise below threshold for the fixed $\theta^\star$ case of (a). These give rise to the left-hand cuts shown in (b). The formalism derived in this work holds for the region shown in cyan, i.e.~for $ (2 M_N)^2 - (2M_\pi)^2 \lesssim s \lesssim (2M_N + M_\pi)^2 $ where the limits are not sharp due to an enhancement of neglected exponentially suppressed terms near the branch points on either side. \label{fig:complex_plane}}
\end{center}
\end{figure}

Combining eqs.~\eqref{eq:tchan_ker_infinite_volume} and \eqref{eq:T_partial_wave_decom} then gives
\begin{align}
\mathcal T^{\sf os}_{\ell}(s) & = \frac{1}{2} \int_{-1}^1 d(\cos \theta^\star) \, \frac{P_{\ell}(\cos \theta^\star)}{2 (s/4 - M_N^2) (1 - \cos \theta^\star) + M_\pi^2 - i \epsilon} \,, \\
\mathcal U^{\sf os}_{\ell}(s) & = \frac{1}{2} \int_{-1}^1 d(\cos \theta^\star) \, \frac{P_{\ell}(\cos \theta^\star)}{2 (s/4 - M_N^2) (1+ \cos \theta^\star )+ M_\pi^2 - i \epsilon} \,,
\end{align}
where we have used that $t = - 2 (s/4 - M_N^2) (1 - \cos \theta^\star)$. This integral is straightforward to evaluate for any fixed $\ell$. For $\ell=0$, for example, one finds
\begin{equation}
\mathcal T^{\sf os}_0(s) = \frac{1}{s - 4 M_N^2} \log \bigg [ 1 + \frac{s - 4M_N^2}{M_\pi^2} - i \epsilon \bigg ] \,.
\label{eq:t_channel_cut_on_shell}
\end{equation}
The expression makes the logarithmic $t$-channel cut in the amplitude manifest. See figure~\ref{fig:complex_plane} for a summary of this analytic structure. Note that the expression is regular for $s = 4 M_N^2$ with an expansion about that point giving
\begin{equation}
\mathcal T^{\sf os}_0(s) = \frac{1}{M_\pi^2} - \frac{s - 4 M_N^2}{2 M_\pi^4} + {\mathcal O} \! \left(\left(s - 4 M_N^2\right)^2\right) \,.
\end{equation}
$\mathcal U^{\sf os}_0(s)$ satisfies the same expansion.

We stress that the partial wave expansion is expected to exhibit arbitrarily poor convergence near $t = M_\pi^2$ or $u = M_\pi^2$. This is intuitive, because the expansion must reproduce an angular dependence that diverges at the pion exchange poles, but is otherwise finite. In addition, the $S$-wave partial-wave-projected amplitudes diverge at the branch point $s = 4 M_N^2 - M_\pi^2$, as can be seen from eq.~\eqref{eq:t_channel_cut_on_shell}. This generically does not match a divergence of the unprojected amplitude. In the following sections, we will circumvent these issues by introducing an intermediate quantity in which both the pole and the branch cut are removed. This is then related to the scattering amplitude via integral equations that recover the physical singularities as required.

\subsection{Incorporating spin}
\label{sec:inf_vol_spin}

The arguments given above go through with few modifications when intrinsic spin is included for the nucleon field.

For spin-half nucleons, the first adjustment is that eq.~\eqref{eq:N_scalaR_Normalization} is modified to
\begin{align}
\langle \overline N, \boldsymbol p', s' \vert N, \boldsymbol p, s \rangle &= 0 \,,
\\
\langle N, \boldsymbol p', s' \vert N, \boldsymbol p, s \rangle &= \delta_{s's} \, 2 \omega_N(\boldsymbol p) \, (2 \pi)^3 \delta^3(\boldsymbol p' - \boldsymbol p) \,,
\end{align}
where the distinction between $N$ and $\overline N$ has replaced the $q$ label, and we have introduced the two-component spin label $s \in \{-1/2, \, 1/2 \}$.

The second modification is that the scalar nucleon propagator is replaced with a fully dressed Dirac propagator
\begin{equation}
\Delta^{\alpha \beta}_{N, i \epsilon}(k) = \bigg [ \frac{i}{A_N(k^2) \slashed k - B_N(k^2) + i \epsilon} \bigg ]^{\alpha \beta} \,,
\end{equation}
where $\alpha$ and $\beta$ are Dirac indices. Instead of the single self-energy function of the scalar case, here we have two self-energy functions $A_N(k^2)$ and $B_N(k^2)$. These are constrained by renormalization conditions to ensure that $M_N$ is the pole mass and that in an expansion about the pole position one finds
\begin{equation}
\Delta^{\alpha \beta}_{N, i \epsilon}(k) = \bigg [ \frac{i (\slashed k + M_N)}{k^2 - M_N^2 + i \epsilon} \bigg ]^{\alpha \beta} + {\mathcal O} \left ( \left(k^2 - M_N^2 \right)^0 \right )\,.
\end{equation}

Equation \eqref{eq:lagrangian_density} is similarly modified to
\begin{equation}
{\mathcal L}_{\overline N \! N \pi}(x) = \overline N(x) \mathcal F \big ( \pi(x) \big ) N(x) \,,
\end{equation}
where $\mathcal F$ may now carry Dirac structures as allowed by the quantum numbers.

Otherwise, with the exception of eq.~\eqref{eq:ieps_pv_diff} for which an explanation is given below, eqs.~\eqref{eq:m_integral_equation}--\eqref{eq:final_M_to_K} hold as written with the caveat that $B$, $\mathcal K$ and $\mathcal M$ all carry implicit indices that are contracted between adjacent factors. As an example, eq.~\eqref{eq:B_ieps_compact} can be written more explicitly as
\begin{equation}
[i B \circ_{i \epsilon} i B]^{\sigma \delta, \sigma' \delta'} = \frac12 \int \! \frac{d^4 k}{(2\pi)^4} i B^{\sigma \delta, \alpha \beta}(P, p, k) \Delta^{\alpha \alpha'}_{N, i \epsilon}(k) \Delta^{\beta \beta'}_{N, i \epsilon}(P-k) i B^{\alpha' \beta', \sigma' \delta'}(P, k, p') \,.
\end{equation}
The key relation of eq.~\eqref{eq:ieps_pv_diff} does require some modification. The result with intrinsic spin takes the form
\begin{multline}
\label{eq:BrhoB_spin}
\big [ i B \circ_{i \epsilon} i B - i B \circ_{{\sf pv}} i B \big ]_{tv, t' v'} =
\\
\frac12 \int \frac{d^3 \boldsymbol k}{(2 \pi)^3} i B_{tv,rs} (P, p, k) \frac{\pi \delta(E - \omega_{N}(\boldsymbol k)- \omega_{N}(\boldsymbol P - \boldsymbol k)) \, \delta_{rr'} \delta_{ss'}}{2 \omega_{N}(\boldsymbol k) 2 \omega_{N}(\boldsymbol P - \boldsymbol k)} i B_{r's',t'v'}(P, k, p') \,,
\end{multline}
where
\begin{align}
{B}_{tv,rs} (P,p,k) & = \overline u_t^{\sigma}(\boldsymbol p) \overline u_v^{\delta}(\boldsymbol P - \boldsymbol p) B^{\sigma \delta , \alpha \beta} (P,p,k) u_r^{\alpha}(\boldsymbol k) u_s^{\beta}(\boldsymbol P - \boldsymbol k) \,,
\end{align}
are spin-projected versions of the Bethe-Salpeter kernels. To define these we have introduced the four-component spinors $\bar u_r^{\alpha}(\boldsymbol k)$ and $ u_r^{\alpha}(\boldsymbol k)$, which satisfy the relation
\begin{equation}
[\slashed k + M_N]^{\alpha \beta} = \sum_{r = \pm} u_r^{\alpha}(\boldsymbol k) \bar u_r^{\beta}(\boldsymbol k) \,,
\end{equation}
provided that on-shell momentum $k = (\omega_N(\boldsymbol k), \boldsymbol k)$ is used on the left-hand side. The sum runs over the two possible spin states, labeled here by the sign of the spin magnetic quantum number $r = \pm \frac 12$.

The key point here is that the spin-state indices appear through the Kronecker deltas in eq.~\eqref{eq:BrhoB_spin}. This translates to a corresponding modification to the phase-space factor, which can be written as
\begin{equation}
\rho_{r s, r' s'}(s) = \frac{k^\star_{\sf os}}{16 \pi \sqrt{s}} \, \delta_{rr'} \delta_{ss'} \,. \label{eq:rho_N_spin}
\end{equation}
Completing the derivation as above yields
\begin{align}
\widetilde{\mathcal M}(s) & = \frac{1}{\widetilde{\mathcal K}(s)^{-1} - i \widetilde \rho (s) } \,,
\end{align}
exactly matching eq.~\eqref{eq:final_M_to_K}, but where all objects are now matrices in the combined orbital angular momentum and spin index space. The elements of these matrices are
\begin{align}
\widetilde{\rho}_{\ell m r s, \ell'm' r' s'} &= \delta_{\ell \ell'}\delta_{mm'} (k^\star_{\sf os})^{2\ell} \rho_{rs, r's'} \,, \\
\widetilde{\mathcal K}_{\ell m r s, \ell'm' r' s'} &= (k^\star_{\sf os})^{\ell+\ell'} \, \mathcal K_{\ell m rs, \ell' m'r's'} \,, \\
\widetilde{\mathcal M}_{\ell m r s, \ell'm' r' s'} &= (k^\star_{\sf os})^{\ell+\ell'} \, \mathcal M_{\ell m rs, \ell' m'r's'} \,.
\end{align}
Unlike the spinless case, we now need two sets of indices in the K-matrix and scattering amplitude because orbital angular momentum and spin orientation are not individually conserved and can thus change between incoming and outgoing states.

The elements $\mathcal M_{\ell m r s, \ell' m' r' s'}$ are projections of the amplitude to definite incoming and outgoing orbital angular momentum and single-particle spin states. We can write this projection as
\begin{align}
\mathcal M_{\ell m r s, \ell' m' r' s'} = \big( \langle \ell' m'\vert \otimes \langle s_1 r', s_2 s' \vert \, \big) \, \mathcal M \, \big( \, \vert \ell m\rangle \otimes \vert s_1 r, s_2 s \rangle \big) \,,
\end{align}
where $\vert \ell m \rangle$ is a state of definite orbital angular momentum and $\vert s_1 r, s_2 s \rangle$ (with $s_1 = s_2 = \frac12$ for nucleons) encodes the two-particle spin states. We can arrive at more useful set of indices by subsequent changes of basis, for example to definite total spin $\ell m S m_S, \ell' m' S' m_S'$ via
\begin{align}
\vert \ell m, S m_S \rangle &\equiv \vert \ell m \rangle \otimes \vert S m_S, s_1 s_2 \rangle \\
&= \sum_{r,s} \big( \, \vert \ell m\rangle \otimes \vert s_1 r, s_2 s \rangle \big) \langle s_1 r, s_2 s \vert S m_S, s_1s_2 \rangle \,,
\end{align}
where $\langle s_1 r, s_2 s \vert s_1s_2, S m_S \rangle $ are the Clebsch-Gordon coefficients for the spin addition. One can further combine these indices to label the amplitude in terms of total angular momentum, using
\begin{equation}
\vert J m_J, \ell S \rangle = \sum_{m,m_S} \vert \ell m, S m_S \rangle \langle \ell m , S m_S \vert J m_J, \ell S \rangle \,.
\end{equation}
The advantage is that total angular momentum is conserved, and the amplitude is diagonal in the indices $J, m_J$. The standard expressions for the K-matrix and amplitude in terms of the phase shift can be used with little alteration:
\begin{equation}
\label{eq:Ktilde_delta_relation2}
\mathcal K_{Jm_J\ell S, J' m_{J}' \ell'S'}
= \delta_{JJ'} \delta_{m_J m_{J}'} \mathcal K^{(J)}_{\ell S, \ell'S'}
= \delta_{JJ'} \delta_{m_J m_{J}'} \, 16 \pi \sqrt{s} \, \frac{\tan \delta^{(J)}_{\ell S,\ell'S'} (k^\star_{\sf os})}{k^\star_{\sf os}} \,,
\end{equation}
and
\begin{equation}
\mathcal M_{Jm_J\ell S, J' m_{J}' \ell'S'}
= \delta_{JJ'} \delta_{m_J m_{J}'}
\mathcal M^{(J)}_{\ell S,\ell'S'}
= \delta_{JJ'} \delta_{m_J m_{J}'} \frac{16 \pi \sqrt{s}}{k^\star_{\sf os} \cot \delta^{(J)}_{\ell S,\ell'S'} ( k^\star_{\sf os}) - i k^\star_{\sf os}} \,,
\end{equation}
where the phase shifts $\delta^{(J)}_{\ell S,\ell' S'}$ are labelled by the total angular momentum and total incoming and outgoing orbital angular momentum and spin. For a more detailed discussion on the inclusion of spin, see for example refs.~\cite{\SpinBases}.

Having completed our discussion of $NN \to NN$ in the infinite-volume context, we now turn to the effects of a finite, periodic, spatial volume.

\section{Finite-volume formalism} \label{sec:FVformalism}

In this section we review the analysis of finite-volume correlation functions, originally worked out in refs.~\cite{Luscher:1986pf,Kim:2005gf}, which leads to the quantization condition relating two-particle finite-volume energies to elastic scattering amplitudes. We then illustrate the problem with the derivation that arises in the case of a subthreshold left-hand cut, and derive a new formalism that can be applied for finite-volume energies extracted either near or on the cut. In sections~\ref{sec:skel_exp}~--~\ref{sec:os_subs}, we focus on energies in the above-threshold elastic regime, while commenting occasionally on what changes below threshold. We detail the issues that arise on the left-hand cut in sections \ref{sec:subthr} and \ref{sec:bs_kernel}. The derivation of the new quantization condition is presented in section \ref{sec:derivation} and the result is given in \ref{sec:result}. Finally, the incorporation of intrinsic spin is explained in section~\ref{sec:spin}.

\subsection{Skeleton expansion for the finite-volume correlator} \label{sec:skel_exp}

Consider the system introduced in the previous section, now constrained to a finite, periodic volume with periodicity $L$ in each of the three spatial directions. Define a time-ordered Minkowski correlator
\begin{equation}
C_L(P) \equiv \int dx^0 \int_L d^3 \boldsymbol x\, e^{iEx^0} e^{-i \boldsymbol P\cdot \boldsymbol x} \< 0 \vert \text{T} \mathcal{A}(x) \mathcal{A}^\dagger (0) \vert 0 \>_L \ , \label{eq:corr_def}
\end{equation}
where we use the notation $x \equiv (x^0, \boldsymbol x)$ for the spatial coordinate and $P \equiv (E,\boldsymbol P)$ for the total four-momentum, as in the previous section. The operators $\mathcal{A}(x)$ and $\mathcal{A}^\dagger(x)$ are annihilation and creation operators, respectively, carrying the quantum numbers of the $NN$ states of interest. The $\text{T}$ symbol denotes standard time ordering and the subscript $L$ on the integral indicates that integration is performed over a single cell of the periodic volume, while the same subscript on the matrix element emphasizes that this is evaluated in the finite-volume theory.

In a finite volume the theory has a discrete energy spectrum. Looking at the spectral representation of the correlator $C_L(P)$, it can be shown that it has poles at the finite-volume energy levels of the system (see e.g.~refs.~\cite{\polesInCL}). In a suitably defined infinite-volume ($L\to\infty$) limit, these poles will accumulate to form the $s$-channel branch cuts, matching the analytic structure of the amplitude shown in figure \ref{fig:complex_plane}(a). However, the structure of the correlator below the (infinite-volume) elastic threshold, in both finite and infinite-volume versions, differs significantly from that of the amplitude: $C_L(P)$ contains no $t$- or $u$-channel poles or cuts. Particularly relevant for us later in this section is the fact that the single-meson-exchange cut described in section \ref{sec:inf_vol_cut} is not present in the correlator, and only becomes an issue because of its presence in the two-to-two scattering amplitude.

\begin{figure}
\begin{center}
\includegraphics[width=\textwidth]{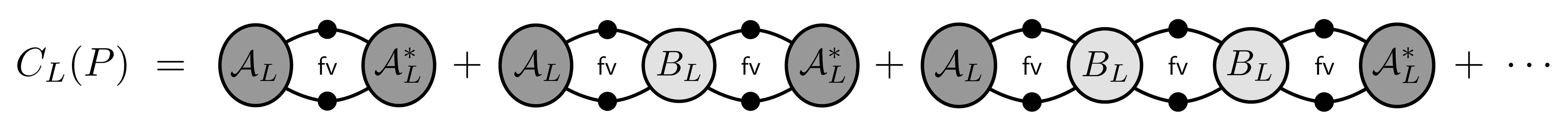}
\vspace{0pt}
\caption{Skeleton expansion for the finite-volume correlator $C_L(P)$, built from Bethe-Salpeter kernels $B_L(P, p,p')$, endcap functions $\mathcal A_L (P, p)$ and $\mathcal A_L^* (P, p) $ and fully-dressed nucleon propagators. Here we include the $L$-subscript on all quantities to indicate that these are volume-dependent, albeit with exponentially suppressed scaling in the elastic regime $(2 M_N)^2 < s < (2 M_N + M_\pi)^2$. The finite-volume two-particle loops are denoted $\sf fv$. Depending on the details of the operators used in $C_L(P)$, one may also have a contribution from a term with no two-particle loops (i.e.~no intermediate $NN$ states), which we omit here.\label{fig:skel_exp}}
\end{center}
\end{figure}

As with the scattering amplitude and the K-matrix in the infinite-volume case, $C_L(P)$ admits a diagrammatic representation that we organize into a skeleton expansion, shown in figure \ref{fig:skel_exp}. The expansion employs the building blocks introduced in the infinite-volume context of section~\ref{sec:scattering_K_phase}, namely the fully dressed propagator and the Bethe-Salpeter kernel. We denote these by $\Delta_{N,L}(p)$ and $B_{L}(P,p,p')$, respectively, in the finite-volume theory. As is well known (e.g.~from refs.~\cite{Luscher:1986pf,Kim:2005gf}), and will be recalled in some detail below, this decomposition is useful because it exposes the power-like $L$ dependence of $C_L(P)$ in the elastic regime: $(2 M_N)^2 < s < (2 M_N + M_\pi)^2$.

A consequence of the $L$ periodicity is that all spatial momenta are discretized to values of type $\boldsymbol k = {2\pi \boldsymbol n}/{L}$, with $\boldsymbol n \in \mathbb{Z}^3$ a three-vector of integers. Thus, a diagrammatic definition of $\Delta_{N, L}(p)$, $B_{L}(P,p,p')$, and $C_L(P)$ itself, is given by taking a generic low-energy effective theory, formally writing all diagrams to all orders in perturbation theory, and replacing the integrals of spatial loop momenta with sums over the corresponding discretized set:
\begin{equation}
\int \! \frac{d^4 k}{(2\pi)^4} ~ \longrightarrow ~ \int \! \frac{dk^0}{2\pi} \frac{1}{L^3} \! \sum_{\boldsymbol k \in 2 \pi \mathbb{Z}^3/L} \, , \label{eq:FV_presc}
\end{equation}
where we use the notation $k = (k^0, \boldsymbol k)$.

Beyond the $L$-dependence, the only new ingredients relative to the previous section are the endcap ``blobs'' shown in figure~\ref{fig:skel_exp}. These represent functions in momentum-space resulting from the creation and annihilation operators. We write these as $\Tilde{\mathcal A}_L^*(P, p)$ and $\Tilde{\mathcal A}_L(P, p)$, respectively. As we review in section~\eqref{sec:fv_effects} based on the work of refs.~\cite{Luscher:1986pf,Kim:2005gf}, these quantities have only exponentially suppressed volume dependence of all energies (including those on the left-hand cut). We will therefore drop the $L$ subscripts in section~\ref{sec:derivation}, but keep them here to emphasize that the functions are defined in the finite-volume theory.

The $n$-loop contributions to the skeleton expansion of the correlator can thus be written as
\begin{align}
C^{(1)}_L(P) & = \frac{1}{2} \int \frac{dk_1^0}{(2\pi)} \frac{1}{L^3} \sum_{\boldsymbol k_1} \Tilde{\mathcal A}_L(P, k_1) \, \Delta_{N,L}(k_1) \, \Delta_{N,L}(P - k_1) \Tilde{\mathcal A}_L^*(P, k_1) \qquad \text{for $n=1$} \,, \\ \nonumber
C^{(n)}_L(P) & = \frac{1}{2^n} \int \frac{dk_1^0\, dk_2^0\,...\, dk_n^0}{(2\pi)^n} \frac{1}{(L^3)^n} \sum_{\boldsymbol k_1, \boldsymbol k_2, ..., \boldsymbol k_n} \Tilde{\mathcal A}_L(P, k_1) \, \Delta_{N,L}(k_1) \, \Delta_{N,L}(P - k_1) \\
& \times \prod_{j = 2}^{n} \left[ iB_{L} (P, k_{j-1}, k_j) \, \Delta_{N,L}(k_j) \, \Delta_{N,L}(P - k_j) \right] \Tilde{\mathcal A}_L^*(P, k_n) \qquad \qquad \, \text{for $n\geq 2$} \,.
\label{eq:FV_corR_Nloop}
\end{align}
This can be expressed compactly by introducing $\circ_{\sf fv}$ as a shorthand for the propagators and associated loop integration and summation operations, in analogy to $\circ_{i\epsilon}$ and $\circ_{\sf pv}$ as used in eqs.~\eqref{eq:B_ieps_compact} and \eqref{eq:B_pv_compact} above. The explicit definition for a single loop between two Bethe-Salpeter kernels is
\begin{equation}
B_{L} \circ_{\sf fv} B_{L} = \frac12 \int \! \frac{dk^0}{2\pi} \frac{1}{L^3} \! \sum_{\boldsymbol k} B_{L}(P, p, k) \, \Delta_{N,L}(P - k) \, \Delta_{N,L}(k) \, B_{L}(P, k, p') \, .
\end{equation}
Equation~\eqref{eq:FV_corR_Nloop} can then be written as
\begin{equation}
C_L^{(n)}(P) = \Tilde{\mathcal A}_L \, \circ_{\sf fv} \left[ iB_{L} \, \circ_{\sf fv} \right]^{n-1} \Tilde{\mathcal A}_L^\dagger \qquad \qquad \text{for $n\geq 1$}\,, \label{eq:FV_corR_Nloop2}
\end{equation}
and the full correlation function can be expressed compactly by summing over the loops
\begin{equation}
\label{eq:corr_skel}
C_L(P) = C^{(0)}_L(P) + \sum_{n=0}^\infty \Tilde{\mathcal A}_L \, \circ_{\sf fv} \left[ iB_{L} \, \circ_{\sf fv} \right]^n \Tilde{\mathcal A}_L^\dagger \,,
\end{equation}
where $C^{(0)}_L(P)$ is a potential contribution with no $NN$ intermediate states. In the following subsections we illustrate the utility of this expression for identifying power-like volume dependence in the correlator. To do so, we first review the basic strategy for distinguishing power-like and exponentially suppressed volume effects.

\subsection{Classifying finite-volume effects} \label{sec:fv_effects}

The prescription of summing rather than integrating the spatial loop momenta is the only distinction between the finite- and infinite-volume Feynman rules. Thus, the relation between Feynman diagrams in the two contexts is conveniently understood by studying sum-integral differences.

A first key observation in this direction is as follows: If, for any direction of $\boldsymbol k$, a generic function, $f(\boldsymbol k)$, has a strip of analyticity in the complex $\vert \boldsymbol k \vert$ plane
then one can use the Poisson summation formula to show that
\begin{equation}
\label{eq:sum_int_f}
\left[ \frac{1}{L^3} \sum_{\boldsymbol k} - \int \frac{d^3 \boldsymbol k}{(2\pi)^3} \right] f(\boldsymbol k) = {\mathcal O}\left( e^{- \mu L} \right) \,,
\end{equation}
for a characteristic scale $\mu$, governed by the width of the strip.
In words, the function has exponentially suppressed volume effects.%
\footnote{An alternative condition is that if $f(\boldsymbol k)$ is smooth, i.e.~infinitely differentiable along the real axis, then the sum-integral difference falls faster than any power of $1/L$.}
As is common to a large body of work quantifying finite-volume effects for scattering states, we will neglect such exponentially suppressed scaling throughout this work whenever $\mu \sim M_\pi$.

If, instead, one considers a function, $g(\boldsymbol k)$, with a singularity on the real axis, then the difference has power-like scaling in $L$:
\begin{equation}
\label{eq:sum_int_g}
\left[ \frac{1}{L^3} \sum_{\boldsymbol k} - \, {\sf presc.} \int \frac{d^3 \boldsymbol k}{(2\pi)^3} \right] g(\boldsymbol k) = {\mathcal O}(L^{-n})\ ,
\end{equation}
for some integer $n$. This expression assumes that a suitable prescription has been used to make the integral of $g(\boldsymbol k)$ well-defined, represented by the abbreviated ${\sf presc.}$ acting on the integral. In the case of a single pole, for example, both a principal-value and an $i \epsilon$ pole prescription may be used. In our analysis of Feynman diagrams, we will mainly use a principal value prescription, denoted by $\sf pv$ below. We stress that $n$ can also be negative, in which case the sum-integral difference has a divergent infinite-volume limit.

Returning to the skeleton expansion of eq.~\eqref{eq:corr_skel}, our task is to identify when summands are analytic and when these are singular with respect to the summed momentum coordinates $\boldsymbol k_i$. As has been shown in refs.~\cite{\onShellSingularity}, the singularity condition is related to the question of intermediate states going on shell. Whenever a set of internal propagators can go on shell (i.e.~can carry the total external energy and momentum while satisfying $p^2 = M^2$ for each propagator, with $M$ denoting the appropriate propagator mass), then singularities arise in the loop summands. By contrast, if a given propagator never appears in an on-shell set, then the summand has a strip of analyticity in the spatial loop momenta, and the sum can be replaced with an integral up to neglected $e^{- \mu L}$ effects.

\begin{figure}
\begin{center}
\includegraphics[width=0.72\textwidth]{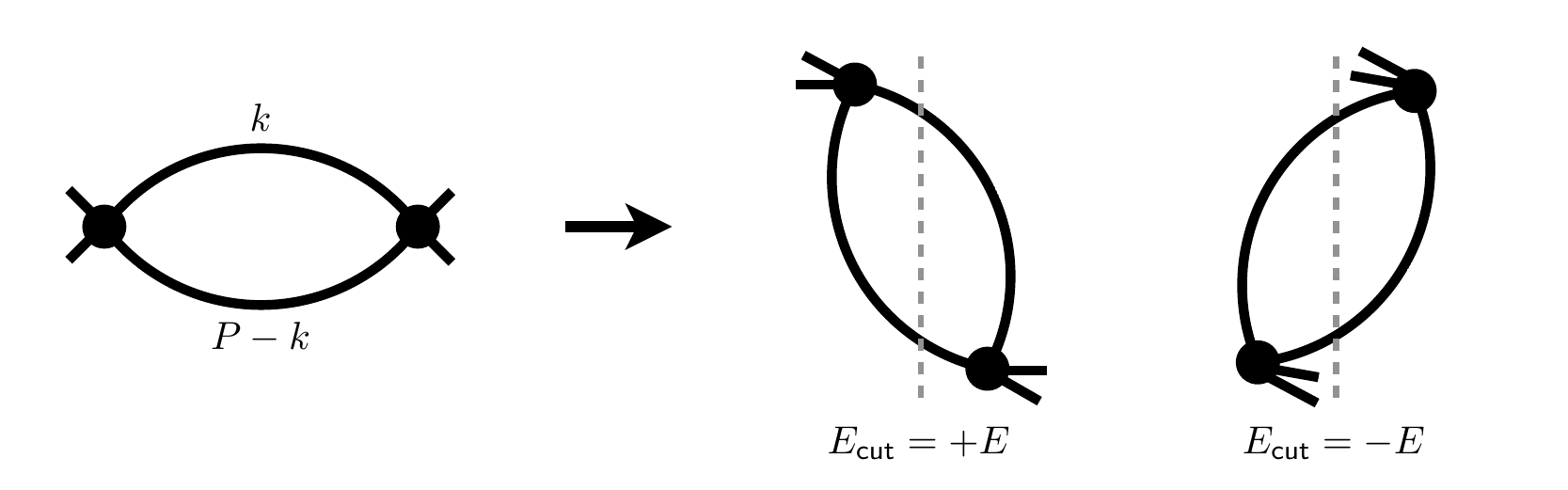}
\vspace{0pt}
\caption{Example application of time-ordered perturbation theory (TOPT) to a simple diagram. This can represent a contribution to $C_L(P)$ (in which the vertices are generated by the operators $\mathcal A(x)$ and $\mathcal A^\dagger(x)$) or a contribution to the finite-volume amplitude introduced in section~\ref{sec:int_eqs} below. The diagram on the left has two possible time orderings of the vertices, shown on the right. The cuts between consecutive vertices are shown in blue. Defining the total energy flowing into the left diagram in a given frame as $E = P^0$, we note that the energy flowing across the cuts is $E_{\sf cut} = + E$ and $E_{\sf cut} = - E$ for the first and second diagram on the right, respectively. TOPT gives a framework to identify the singularities after $k^0$ integration. In the indicated case these occur at $E = \omega_N(\boldsymbol k) + \omega_N(\boldsymbol P - \boldsymbol k)$ and $ - E = \omega_N(\boldsymbol k) + \omega_N(\boldsymbol P - \boldsymbol k)$. \label{fig:TOPT}}
\end{center}
\end{figure}

As is discussed in refs.~\cite{\TOPT}, both based on the more general discussion in ref.~\cite{sterman_1993}, the statement of the on-shell condition can be made more precise via time-ordered perturbation theory (TOPT). TOPT is a method for generating expressions from Feynman diagrams with all $k_i^0$ integrals already performed. Figure~\ref{fig:TOPT} shows an example, with each ordering of the vertices corresponding to a distinct term. For a given time ordering, one identifies all cuts between consecutive vertices, and then the TOPT rules generate a factor of
\begin{equation}
\frac{1}{E_{\sf cut} - \sum_i \omega_{f(i)}(\boldsymbol p_i)} \,,
\end{equation}
associated to each cut. Here $E_{\sf cut}$ is the total energy flow across the cut, defined as the sum of inflowing $p^0$ components, from all external propagators, into all vertices to the left of (or ``earlier than'') the cut location. (See, for example, eq.~(9.71) of ref.~\cite{sterman_1993}.) In the subtracted term, $i$ labels the $i^{\rm th}$ propagator intersected by the cut; $\boldsymbol p_i$ is its corresponding momentum, and $f(i)$ is a flavor label. Note that $\sum_i \boldsymbol p_i = \boldsymbol P$ by total momentum conservation.

Provided $\mathcal A(x)$ and $\mathcal A^\dagger(x)$ are local in time, then any diagram contributing to $C_L(P)$ only has two vertices with external energy and momentum flow: $(E, \boldsymbol P)$ flows in via $\mathcal A^\dagger(0)$ and then back out via the Fourier transformed $\mathcal A(x)$. Thus, $E_{\sf cut} = \pm E$ or $0$ for any cut in the diagram. In this way, the condition of an intermediate propagator set going on shell is simply reformulated as the condition that the TOPT pole is hit for some real-valued $\boldsymbol p_i$. If this is the case, then the summand is of the type of $g(\boldsymbol k)$ in eq.~\eqref{eq:sum_int_g}, leading to power-like $L$ scaling. If not, then the summand is like $f(\boldsymbol{k})$ in eq.~\eqref{eq:sum_int_f} and the $L$ dependence is neglected.

Restricting attention to the elastic scattering regime, $(2M_N)^2 < s < (2M_N + M_\pi)^2$, guarantees that only intermediate $NN$ states can go on shell, implying that all sums arising in self-energy diagrams and Bethe-Salpeter kernels can be replaced by integrals: $B_L \to B$ and $\Delta_{N, L} \to \Delta_{N,i \epsilon}$, where the latter are understood as infinite-volume objects. Analogous replacements can also be made for the endcap factors. The power-like $L$ dependence from momentum sums over $NN$ states is captured in the decomposition of eq.~\eqref{eq:corr_skel}. We emphasize the parallel between $\circ_{\sf fv}$ insertions leading to volume dependence in $C_L(P)$ and $\circ_{i\epsilon}$ insertions leading to imaginary contributions in the scattering amplitude. Both effects originate from poles in the spatial momentum dependence.

\subsection{Reduction of the finite-volume two-particle loop}
\label{sec:loop_cont}

To continue the analysis of power-like $L$ dependence in $C_L(P)$ for $(2 M_N)^2 < s < (2 M_N + M_\pi)^2$, we now analyze a single $NN$ loop, arising at any location of any skeleton expansion term, as shown in figure~\ref{fig:skel_exp}. We introduce generic functions to the left and right of the loop, denoted $\mathcal L(P, k)$ and $\mathcal R^* (P, k)$ respectively.%
\footnote{Here the ${}^*$ on $\mathcal R^*(P, k)$ indicates complex conjugation. This convention is useful when the functions represent the endcaps of a correlator constructed from an operator and its hermitian conjugate, as in eq.~\eqref{eq:corr_def}. In this case, $\mathcal L (P, k) = \mathcal R (P, k)$.}
These correspond to either endcap functions or Bethe-Salpeter kernels, and we omit other possible momentum arguments not relevant for the loop considered. The loop contribution can then be written as
\begin{equation}
\mathcal L \, \circ_{\sf fv} \, \mathcal R^\dagger = \frac{1}{2} \int \frac{d k^0}{2 \pi} \frac{1}{L^3} \sum_{\boldsymbol k} \mathcal L(P, k) \, \Delta_{N,i \epsilon}(k) \, \Delta_{N,i \epsilon}(P - k) \, \mathcal R^*(P, k) \,.
\end{equation}

The first step in simplifying the expression is to separate the propagator into the simple pole and a remainder
\begin{align}
\Delta_{N,i \epsilon}(p) & = \frac{i}{p^2 - M_N^2 + i\epsilon} + R_N(p) \, ,
\label{eq:prop_red} \\
R_N(p) & \equiv \frac{i}{p^2 - M_N^2 - \Pi_N(p^2) + i\epsilon} - \frac{i}{p^2 - M_N^2 + i\epsilon} \, . \label{eq:remainder}
\end{align}
The poles cancel in the second equation so that the remainder $R_N(p)$ is analytic in a neighborhood around $p^2 = M_N^2$.%
\footnote{This relies on the on-shell renormalization scheme established after eq.~\eqref{eq:Delta_full_def} which sets $M_N$ to be the pole mass and the residue to be unity.}

Substituting this decomposition of each propagator, and denoting the sum of all terms containing at least one factor of $R_N$ by $\mathcal I^{[1]}(P)$, we write
\begin{equation}
\mathcal L \, \circ_{\sf fv} \, \mathcal R^\dagger = \mathcal I^{[1]}(P) + \frac{1}{2} \int \frac{d k^0}{2 \pi} \frac{1}{L^3} \sum_{\boldsymbol k} \frac{\mathcal L(P, k) \, i^2 \, \mathcal R^*(P, k)}{[k^2 - M_N^2 + i\epsilon][(P-k)^2 - M_N^2 + i\epsilon]} \,.
\label{eq:loop_cont2}
\end{equation}
The term $\mathcal I^{[1]}(P)$ is defined via summands that are analytic with respect to $\vert \boldsymbol k \vert$ in a strip including the real axis.
Similar replacements will be made below and the superscript on $\mathcal I^{[i]}(P)$ will be incremented to indicate the change. In all cases, exponentially suppressed volume effects are being dropped.

Performing the integration over $k^0$ (closing the contour in the lower half of the complex plane) gives
\begin{equation}
\mathcal L \, \circ_{\sf fv} \, \mathcal R^\dagger = \mathcal I^{[2]}(P) + \sum_{\boldsymbol k} \frac{\mathcal L (P,k) \, i \, \mathcal R^* (P,k)}{2L^3 \cdot 2 \omega_N(\boldsymbol k) \big[ (E - \omega_N(\boldsymbol k) )^2 - \omega_N (\boldsymbol P - \boldsymbol k)^2 \big]} \, \Bigg \vert_{k^0 = \omega_N(\boldsymbol k)} \,. \label{eq:Cloop_intm}
\end{equation}
Note that, at this stage, $k$ has been set on shell by $k^0 = \omega_N(\boldsymbol k)$, but $P-k$ is itself not on shell, i.e.~does not satisfy $(P-k)^2 = M_N^2$.

Additional contributions from the $k^0$ integral can be picked up from poles or branch cuts within $\mathcal L (P,k)$ and $\mathcal R^* (P,k)$, but these lead to analytic functions of $\boldsymbol k$ that can be absorbed into the remainder term via $ \mathcal I^{[1]}(P) \to \mathcal I^{[2]}(P)$.
This holds for all objects that can appear in place of $\mathcal L(P,k)$ and $\mathcal R^*(P,k)$, up to a caveat discussed in appendix~\ref{app:BS_kernel_sing}. The issue has to do with the distinction between $t$- and $u$-channel loops within kernels, defined according to the choice of momentum routing in the adjacent two-particle loops. The upshot is that it is always possible to find a routing for which $\mathcal I^{[2]}(P)$, $\mathcal L(P,k)$, and $\mathcal R^*(P,k)$ have the desired analyticity and exponentially suppressed volume dependence.
This also holds for the $t$- and $u$-channel single-pion exchanges (the problem with these first appears at a later step) and for objects containing lower left-hand cuts. The complete justification is given in section~\ref{sec:bs_kernel} below, together with appendix~\ref{app:BS_kernel_sing}.

Returning to eq.~\eqref{eq:Cloop_intm}, we next decompose the functions $\mathcal L (P,k)$ and $\mathcal R^* (P,k)$, with on-shell $k = (\omega_N(\boldsymbol k), \boldsymbol k)$, into spherical harmonics with respect to the direction of the three-momentum in the CM frame:
\begin{align}
\label{eq:harm_proj_L}
\mathcal L (P,k) \big \vert_{k^0 = \omega_N (\boldsymbol k)} & \equiv \sqrt{4\pi} \, Y_{\ell m} (\hat{\boldsymbol k}^\star) \vert \boldsymbol k^\star \vert^\ell \, \widetilde{\mathcal L}_{\ell m} (P, \vert \boldsymbol k^\star \vert ) \ ,
\\
\mathcal R^* (P,k) \big \vert_{k^0 = \omega_N (\boldsymbol k)} & \equiv \sqrt{4\pi} \, Y^*_{\ell m} (\hat{\boldsymbol k}^\star) \vert \boldsymbol k^\star \vert^\ell \, \widetilde{\mathcal R}^*_{\ell m} (P, \vert \boldsymbol k^\star \vert ) \ ,
\label{eq:harm_proj_R}
\end{align}
with sums over the indices $\ell, m$ left implicit. As introduced in section \ref{sec:scattering_K_phase}, the superscript $\star$ indicates the boost to the CM frame. In particular $k^\star = (\omega_N(\boldsymbol k^\star), \boldsymbol k^\star)$ is the four-momentum reached by boosting $k = (\omega_N(\boldsymbol k), \boldsymbol k)$ to with boost velocity $\boldsymbol \beta = - \boldsymbol P/E$:
\begin{equation}
\label{eq:boost_explicit_def}
\begin{pmatrix} \omega_N(\boldsymbol k^\star)
\\
\boldsymbol k^\star \end{pmatrix}
= {\Lambda} (\boldsymbol \beta)
\begin{pmatrix} \omega_N(\boldsymbol k)
\\
\boldsymbol k \end{pmatrix} \,,
\end{equation}
where $\Lambda(\boldsymbol \beta)$ is a standard Lorentz boost matrix.

We include the factors of $\vert \boldsymbol k^\star \vert^\ell$ because $Y_{\ell m} (\hat{\boldsymbol k}^\star)$ without this factor is a singular function of $\boldsymbol k^\star$.%
\footnote{For example
\begin{equation}
Y_{10}( \hat{\boldsymbol k}^\star) = \sqrt{\frac{3}{4\pi}} \, \frac{k_z^\star}{\sqrt{(k_x^\star)^2 + (k_y^\star)^2 + (k_z^\star)^2}} \,,
\end{equation}
vanishes for $k_z^\star \to 0$ with fixed $(k_x^\star,k_y^\star)$ but can take on non-zero values if the origin is approached from a different direction.}
As the functions on the left-hand sides of eqs.~\eqref{eq:harm_proj_L} and \eqref{eq:harm_proj_R} are non-singular, the coefficient of the spherical harmonic must vanish with the corresponding power of $\vert \boldsymbol k^\star \vert^\ell$. For the subsequent steps taken below, it is more convenient to separate out this scaling in our definition of $\widetilde{\mathcal L}_{\ell m} (P, \vert \boldsymbol k^\star \vert )$ and $\widetilde{\mathcal R}^*_{\ell m} (P, \vert \boldsymbol k^\star \vert )$ as we have done. The tilde notation denotes this rescaling, matching also the notation used in section \ref{sec:scattering_K_phase}. A consequence of this definition is that $\widetilde{\mathcal L}_{\ell m} (P, \vert \boldsymbol k^\star \vert )$ and $\widetilde{\mathcal R}^*_{\ell m} (P, \vert \boldsymbol k^\star \vert )$ approach a constant as $\vert \boldsymbol k^\star \vert \to 0$, rather than vanishing as $\vert \boldsymbol k^\star \vert^\ell$.

It is also useful to introduce an index notation for the momenta. We find it more convenient to index with $\boldsymbol k^\star$ rather than $\boldsymbol k$. It is then understood that $\boldsymbol k^\star$ takes on a discrete set of allowed values inherited from $\boldsymbol k \in (2 \pi/L) \mathbb Z^3$, via the boost of eq.~\eqref{eq:boost_explicit_def}. We write
\begin{equation}
\mathcal L \, \circ_{\sf fv} \, \mathcal R^\dagger = \mathcal I^{[3]}(P) + \widetilde{\mathcal L}_{\boldsymbol k^\star \ell m}(P) \, iS_{\boldsymbol k^\star \ell m , \boldsymbol k'^\star \ell' m'}(P,L) \, \widetilde{\mathcal R}^*_{\boldsymbol k'^\star \ell' m'}(P) \ , \label{eq:Cloop_compact}
\end{equation}
where sums over all repeated indices (including momentum indices) are implied, and we have introduced the matrix
\begin{equation}
\label{eq:S_def}
S_{\boldsymbol k^\star \ell m , \boldsymbol k'^\star \ell' m'} (P, L) = \frac{1}{2L^3} \, \frac{4\pi \, Y_{\ell m}(\hat{\boldsymbol k}^\star) \, Y^*_{\ell' m'}(\hat{\boldsymbol k}^\star) \, \delta_{\boldsymbol k^\star \boldsymbol k'^\star} \, \vert \boldsymbol k^\star \vert^{\ell + \ell'} \, H(\boldsymbol k^\star)}{4 \omega_N(\boldsymbol k) \, \big [ (k_{\sf os}^\star)^2 - (\boldsymbol k^\star)^2\big ]} \,.
\end{equation}
A few manipulations are required to reach the form shown here from the preceding relations. These are detailed in appendix~\ref{sec:manipulating_S}.

We have also introduced a regulator function, $H(\boldsymbol k^\star)$, which must equal $1$ whenever
\begin{equation}
(k_{\sf os}^\star)^2 - (\boldsymbol k^\star)^2 = 0 \,,
\end{equation}
and should decay exponentially, with increasing $(\boldsymbol k^\star)^2$, to ensure that our final results are numerically tractable. The value of $1 - H(\boldsymbol k^\star)$ away from the pole (as well as a term resulting from the manipulations leading to eq.~\eqref{eq:S_def}) leads to a smooth summand that is absorbed by replacing $\mathcal I^{[2]}(P) \to \mathcal I^{[3]}(P)$, as shown.

One possible choice is the regulator used in ref.~\cite{Kim:2005gf}
\begin{equation}
\label{eq:h_example}
H(\boldsymbol k^\star) = \exp \! \Big ( \!\! - \! \alpha \big [(\boldsymbol k^\star)^2 - (k_{\sf os}^\star)^2 \big ] \Big ) \,,
\end{equation}
where $\alpha > 0$ should be chosen sufficiently small to not enhance the neglected volume dependence. Alternatively, one can make use of a piecewise definition of $H(\boldsymbol{k}^\star)$ that is identically zero above some cutoff, similar to that used in the relativistic field-theoretic three-particle quantization condition \cite{\ThreeBodyRFT}. This type of function necessarily has no strip of analyticity in the complex-$\vert \boldsymbol k^\star \vert$ plane and thus generically leads to neglected volume effects that are not exponentially suppressed, though still decaying faster than any power of $1/L$. In contrast to the three-particle formalism, in this application there is no reason to require the function to vanish identically, and we expect a function as in eq.~\eqref{eq:h_example} will be more useful. It may seem surprising here that $H(\boldsymbol k^\star) = 1$ is only required at $(\boldsymbol k^\star)^2 = (k_{\sf os}^\star)^2$ and not, for example, in the subthreshold region where one encounters the $t$- and $u$-channel pion exchange poles. In a nutshell, this is valid because it leads to a correct representation of the singularities within $C_L(P)$ and the scattering amplitude.

Suppressing the indices, eq.~\eqref{eq:Cloop_compact} can be written in a compact matrix form
\begin{equation}
\mathcal L \, \circ_{\sf fv} \, \mathcal R^\dagger = \mathcal I^{[3]}(P) + \widetilde{\mathcal L}(P) \, iS(P, L) \, \widetilde{\mathcal R}(P)^\dagger\ , \label{eq:Cloop_compact2}
\end{equation}
with $S(P, L)$ denoting the matrix with elements $S_{\boldsymbol k^\star \ell m; \boldsymbol k'^\star \ell' m'} (P,L)$, and $\widetilde{\mathcal L}(P)$ and $\widetilde{\mathcal R}(P)^\dagger$ denoting a row and column vector with components $\widetilde{\mathcal L}_{\boldsymbol k^\star \ell m}(P)$ and $\widetilde{\mathcal R}^*_{\boldsymbol k'^\star \ell' m'}(P)$, respectively. We also define the notation
\begin{equation}
\mathcal L \, \circ_{\sf rm} \, \mathcal R^\dagger \equiv \mathcal I^{[3]}(P) \,,
\end{equation}
for the remainder term, such that we can write the loop contribution as
\begin{equation}
\mathcal L \, \circ_{\sf fv} \, \mathcal R^\dagger = \mathcal L \left[ \circ_{\sf rm} + iS(P, L) \right] \mathcal R^\dagger \, ,
\label{eq:Cloop_compact3}
\end{equation}
with $S(P, L)$ encoding the singular part of the summand and $\circ_{\sf rm}$ the smooth part of the summand in the loop momentum sum. In a slight abuse of notation, we have dropped the arguments on $\widetilde{\mathcal L}(P)$ and $\mathcal R^\dagger(P)$ multiplying $S(P,L)$ so that we can group the remainder term, $\circ_{\sf rm} $, with the singular factor as shown.

Our result for the contribution of a generic two-particle finite-volume loop can be applied to all loops in the skeleton expansion of eq.~\eqref{eq:Cloop_compact3}, provided that the momentum routing issue of appendix~\ref{app:BS_kernel_sing} is handled correctly. Note, however, that the factors multiplying $S(P,L)$ in eq.~\eqref{eq:Cloop_compact3} still depend on off-shell values of $P-k$. In the next section, we discuss how to eliminate this issue. We highlight this intermediate result because partial off-shellness will be of great importance for treating the left-hand cuts, as detailed in sections~\ref{sec:bs_kernel} and \ref{sec:derivation}.

\subsection{On-shell intermediate states}
\label{sec:os_subs}

In the elastic regime, $(2 M_N)^2 < s < (2 M_N + M_\pi)^2$, the singularity in the summand of the loop contribution, expressed compactly in eqs.~\eqref{eq:S_def} and \eqref{eq:Cloop_compact2}, corresponds to the on-shell condition for the intermediate $NN$ state. In terms of quantities in the finite-volume frame, this condition is $E = \omega_N(\boldsymbol k) + \omega_N(\boldsymbol P - \boldsymbol k)$, and in the CM frame, this reduces to $\vert \boldsymbol k^\star \vert = k^\star_{\sf os}$.
Due to the singularity, the $L$-dependence of $\widetilde{\mathcal L}(P) \, S(P,L) \, \widetilde{\mathcal R}(P)^\dagger$ is dominated by the left and right functions evaluated at these on-shell kinematics.

To express the on-shell-value dominance more explicitly, define
\begin{align}
\widetilde{\mathcal L}^{\sf os}_{\ell m}(P) & \equiv \widetilde{\mathcal L}_{\ell m}(P, k^\star_{\sf os}) \,,
\\
\widetilde{\mathcal R}^{\sf os *}_{\ell m}(P) & \equiv \widetilde{\mathcal R}^*_{\ell m}(P, k^\star_{\sf os}) \,,
\end{align}
and make the straightforward separation
\begin{multline}
\widetilde{\mathcal L}(P) \, S(P,L) \, \widetilde{\mathcal R}(P)^\dagger = \widetilde{\mathcal L}^{\sf os} (P) \, \xi \, S (P,L) \, \xi^\dagger \, \widetilde{\mathcal R}^{\sf os}(P)^\dagger \\[5pt] + \Big[ \widetilde{\mathcal L}(P) \, S(P,L) \, \widetilde{\mathcal R}(P)^\dagger - \widetilde{\mathcal L}^{\sf os} (P) \, \xi \, S(P,L) \, \xi^\dagger \, \widetilde{\mathcal R}^{\sf os}(P)^\dagger \Big ] \,. \label{eq:os_sep2}
\end{multline}
On the right-hand side we have introduced the trivial row vector $\xi$ in the momentum index space, with elements $\xi_{\boldsymbol k^\star} = 1$, and the vectors ${\mathcal L}^{\sf os}(P)$ and ${\mathcal R}^{\sf os}(P)^\dagger$ in angular momentum index space, with components $\widetilde{\mathcal L}^{\sf os}_{\ell m}(P)$ and $\mathcal R^{\sf os*}_{\ell m}(P)$. The vectors $\xi$ and $\xi^\dagger$ contract only with the momentum indices of $S(P,L)$, while $\widetilde{\mathcal L}^{\sf os} (P)$ and $\widetilde{\mathcal R}^{\sf os} (P)^\dagger$ contract with the angular momentum indices
\begin{align}
\widetilde{\mathcal L}^{\sf os} (P) \, \xi \, S (P,L) \, \xi^\dagger \widetilde{\mathcal R}^{\sf os}(P)^\dagger & = \widetilde{\mathcal L}^{\sf os}_{\ell m}(P) \, \xi_{\boldsymbol k^\star} \, S_{\boldsymbol k^\star \ell m; \boldsymbol k'^\star \ell' m'}(P,L) \, \xi_{\boldsymbol k'^\star} \widetilde{\mathcal R}^{\sf os *}_{\ell' m'}(P) \,, \\[5pt] & = \widetilde{\mathcal L}^{\sf os}_{\ell m}(P) \bigg ( \sum_{\boldsymbol k^\star, \, \boldsymbol k'^\star} S_{\boldsymbol k^\star \ell m; \boldsymbol k'^\star \ell' m'}(P,L) \bigg ) \widetilde{\mathcal R}^{\sf os *}_{\ell' m'}(P) \,.
\label{eq:sum_over_s}
\end{align}

We note that the square-bracketed term in eq.~\eqref{eq:os_sep2} represents the sum over an analytic summand, with the singularities within $S(P,L)$ canceling in the difference between terms.
As a result, the sum can again be replaced with an integral up to exponentially suppressed terms, and the latter can be absorbed by replacing $\mathcal I^{[3]}(P) \to \mathcal I^{[4]}(P)$. The loop contribution can thus be written as
\begin{equation}
\mathcal L \, \circ_{\sf fv} \, \mathcal R^\dagger = \mathcal I^{[4]}(P) + \widetilde{\mathcal L}^{\sf os} (P) \, \xi \, iS(P,L) \, \xi^\dagger \, \widetilde{\mathcal R}^{\sf os} (P)^\dagger \,. \label{eq:os_loop_S}
\end{equation}
Provided we are in the elastic regime, this expression is always valid up to exponentially suppressed $L$-dependence absorbed in $\mathcal I^{[4]}(P)$. However, in the case of a nearby left-hand cut present in the on-shell projected left and right functions, the scale $\mu$ in the $e^{-\mu L}$ scaling is set by the cut and can be significantly smaller than $M_\pi$ such that the neglected volume effects can be numerically large. The situation is more dramatic in the subthreshold regime where the on-shell projection can lead to neglected power-like effects, thus leading to a quantization condition that is no longer valid. In both cases the generalized quantization condition presented in section \ref{sec:result} is required.

In summary, on-shell dominance allows one to derive a simpler form for the loop with the $L$-dependence contained completely in $\xi \, S(P, L) \, \xi^\dagger$, provided the left and right functions can be safely put on-shell. We now consider the effect of analytically continuing below threshold, first on the $S(P,L)$ factor (in the next subsection) and then on the Bethe-Salpeter kernel (in section~\ref{sec:bs_kernel}).

\subsection{Subthreshold regime} \label{sec:subthr}

The skeleton expansion for the correlator is applicable in the subthreshold region $s < (2 M_N)^2$, and employing the cutting rules of TOPT allows us to conclude that no intermediate states can go on-shell. Consequently, the summands of spatial loop momenta have the required strips of analyticity for all finite-volume effects to be exponentially suppressed:
\begin{equation}
\label{eq:subthresh_exp_supp}
C_L(P) - C_\infty(P) = {\mathcal O}(e^{- \mu L}) \,, \qquad \text{for $s<(2 M_N)^2$} \,,
\end{equation}
for some, potentially $s$-dependent scale, $\mu$.
Here $C_\infty(P)$ is the correlator evaluated in infinite volume, defined through the same diagrammatic series as $C_L(P)$, but with integrated loops. Because the relation is given below threshold, no pole prescription is required. Alternatively the definition is equivalent to the analytic continuation of the $i \epsilon$ prescription below threshold, on the physical Riemann sheet.

To make eq.~\eqref{eq:subthresh_exp_supp} more concrete, we return to the partially off-shell loop decomposition of eq.~\eqref{eq:Cloop_compact2}. For $(2M_N)^2 < s < (2M_N + M_\pi)^2$, we have argued that the terms generating $\mathcal I^{[3]}(P)$ have negligible exponentially suppressed scaling, while the second term, $\widetilde{\mathcal L}(P) \, S(P, L) \, \widetilde{\mathcal R}(P)^\dagger$, has power-like $L$ dependence. Below threshold, the latter term is given by
\begin{multline}
\widetilde{\mathcal L}(P) \, S(P, L) \, \widetilde{\mathcal R}(P)^\dagger =
\\
- \frac{1}{2} \frac{1}{L^3} \sum_{\boldsymbol k} \widetilde{\mathcal L}_{\ell m} (P, \vert \boldsymbol k^\star \vert ) \frac{4\pi \, Y_{\ell m}(\hat{\boldsymbol k}^\star) Y^*_{\ell' m'}(\hat{\boldsymbol k}^\star) \, \vert \boldsymbol k^\star \vert^{\ell + \ell'} H(\boldsymbol k^\star)}{4 \omega_N(\boldsymbol k) \, \big [ \kappa(s)^2 + (\boldsymbol k^\star)^2\big ]} \widetilde{\mathcal R}^*_{\ell m} (P, \vert \boldsymbol k^\star \vert ) \,,
\label{eq:subthr_loop}
\end{multline}
where we have introduced the subthreshold binding momentum
\begin{equation}
\kappa(s) = \sqrt{M_N^2 - s/4} \,,
\end{equation}
and the functions $\widetilde{\mathcal L}_{\ell m} (P, \vert \boldsymbol k^\star \vert )$ and $\widetilde{\mathcal R}^*_{\ell m} (P, \vert \boldsymbol k^\star \vert )$ are defined as analytic continuations in the energy of the above-threshold objects. In words, the above-threshold pole at $\vert \boldsymbol k^\star \vert = k^\star_{\sf os}$ moves to the complex plane at $\pm i \kappa(s)$ and the function is analytic for $- \kappa(s) < \text{Im}(\vert \boldsymbol k^\star \vert) < \kappa(s)$. This leads to volume effects satisfying
\begin{equation}
\left [1 - \lim_{L \to \infty} \, \right ] \widetilde{\mathcal L}(P) \, S(P, L) \, \widetilde{\mathcal R}(P)^\dagger = \mathcal {\mathcal O}(e^{- \kappa(s) L}) \,,
\hspace{50pt} \text{for $s < (2 M_N)^2$} \,,
\end{equation}
where we have used that the $L \to \infty$ limit is well-defined without a pole prescription for $s < (2 M_N)^2$.

A key point for the derivation of the sub-threshold quantization condition below is that it is only well motivated if a hierarchy exists between the size of volume effects we include (within $\widetilde{\mathcal L}(P) \, S(P, L) \, \widetilde{\mathcal R}(P)^\dagger$) and those we neglect (within $\mathcal I^{[3]}(P)$). Roughly, this holds provided the binding momentum is below the pion mass: $\kappa(s) < M_\pi \ \Longrightarrow \ e^{- M_\pi L} < e^{- \kappa(s) L}$.

The partially off-shell form of eq.~\eqref{eq:subthr_loop} is valid, even for energies overlapping the left-hand cut. However, the subsequent step of $\widetilde{\mathcal L}_{\ell m}(P, k^\star_{\sf os}) \to \widetilde{\mathcal L}^{\sf os}_{\ell m}(P) $ and $ \widetilde{\mathcal R}^*_{\ell m}(P, k^\star_{\sf os}) \to \widetilde{\mathcal R}^{\sf os *}_{\ell m}(P)$, described in section~\ref{sec:os_subs}, fails in this region, as we now describe.

\subsection{Analytic structure of the Bethe-Salpeter kernel} \label{sec:bs_kernel}

The infinite-volume Bethe-Salpeter kernel, $B(P,p,p')$, was defined in section~\ref{sec:scattering_K_phase} (see figure~\ref{fig:bskm_infvol}) as the sum of all amputated $NN \to NN$ diagrams that are two-particle irreducible in the $s$-channel. In our analysis of the correlator $C_L(P)$ in sections \ref{sec:loop_cont}--\ref{sec:subthr}, this kernel can appear as the left or right function of the finite-volume loops, in different versions with various on- and off-shell kinematics. This is of great importance as the nature and location of the $t$- and $u$-channel singularities depends on these.

We identify five different configurations of relevance:
\begin{enumerate}
\item {\bf Most general case}:~all external four-momenta $p,\, P-p,\, p',\, P-p'$ are off shell;
\item {\bf Partially off-shell}:~two of the external four-momenta, $p \equiv (p^0, \boldsymbol p)$ and $p' \equiv (p'^0, \boldsymbol p')$, are placed on-shell, with $p^0 = \omega_N(\boldsymbol p)$ and $p'^0 = \omega_N(\boldsymbol p')$ respectively, and the two remaining four-momenta are off-shell, leaving $P$, $\boldsymbol p$ and $\boldsymbol p'$ as the only free arguments;
\item {\bf Fully on-shell}:~as in Case 2, but with the remaining external four-momenta, $(P - p)$ and $(P - p')$, also on-shell, constraining the magnitude of CM momenta $\vert \boldsymbol p^\star \vert = \vert \boldsymbol p'^\star \vert = k_{\sf os}^\star$ and leaving $P$, $\hat{\boldsymbol p}^\star$ and $\hat{\boldsymbol p}'^\star$ as free arguments;
\item {\bf Partially off-shell, angular momentum projected}:~Case 2 above, with $\hat {\boldsymbol p}^\star$ and $\hat {\boldsymbol p}'^\star$ integrated with spherical harmonics to derive projected components (labelled by $\ell m, \ell' m'$) which are functions of $P$, $\vert \boldsymbol p^\star \vert$ and $\vert \boldsymbol p'^\star \vert$;
\item {\bf Fully on-shell, angular momentum projected}:~Case 3, projected in the same way as Case~4. In this case the components (again labelled by $\ell m, \ell' m'$) are functions only of Mandelstam $s$.
\end{enumerate}

The most general case (Case 1) appears in our expression for the correlator in eq.~\eqref{eq:corr_skel}. The partially off-shell version (Case 2) then arises from performing the $k^0$ contour integration leading to eq.~\eqref{eq:Cloop_intm} while the angular momentum projection (Case 4) arises in eqs.~\eqref{eq:harm_proj_L} and \eqref{eq:harm_proj_R}. Finally, the fully on-shell and angular-momentum projected kernel (Case 5) arises from the separation discussed in subsection \ref{sec:os_subs}. Case 3 (the on-shell but not-projected kernel) has not appeared in our analysis, but is useful as an intermediate step in understanding the analytic structure of its angular-momentum-projected counterpart.

The analytic continuation of the on-shell kernel below threshold has common features with that of the full scattering amplitude, discussed in section~\ref{sec:inf_vol_cut} and shown in figure~\ref{fig:complex_plane}(a). In particular, the single $t$- and $u$-channel pion exchanges are included in the kernel and give poles at $t = M_\pi^2$ and $u = M_\pi^2$. Similarly, multiple pion exchanges, also included in the kernel, lead to branch cuts.

It is instructive to study the on-shell kernel as a function of Mandelstam $s$ at fixed CM scattering angle $\theta^\star$. The single exchange poles lie at
\begin{align}
s & = 4M_N^2 - \frac{M_\pi^2}{\sin^2(\theta^\star / 2)} \qquad \qquad \qquad \qquad \qquad (t\text{-channel\ pole}) \, , \\
s & = 4M_N^2 - \frac{M_\pi^2}{\cos^2(\theta^\star / 2)} \qquad \qquad \qquad \qquad \qquad (u\text{-channel\ pole}) \, .
\end{align}
For both channels, the pole position varies between $s = -\infty$ and $s = 4M_N^2 - M_\pi^2$ as $\theta^\star$ is scanned from $0$ to $\pi$. The integration required to project to definite angular momentum therefore generates a branch cut along $s \in (- \infty, \ 4M_N^2 - M_\pi^2]$. The multiple meson exchanges lead to cuts in both the fixed $\theta^\star$ and the angular-momentum-projected Bethe-Salpeter kernel. For the latter, the corresponding branch points sit at $s = 4M_N^2 - (n M_\pi)^2$ for each $n \geq 2$.

The remainder of this subsection is dedicated to two important, but somewhat technical points. First we argue that partially off-shell Bethe-Salpeter kernels (i.e.~those with the kinematics of Cases 2 and 4) have no $t$-channel singularities. That is, provided that the spatial momenta $\boldsymbol p$ and $\boldsymbol p'$ are real, the partial off-shellness removes the $t$-channel branch cut. Second, we explain that the story is more subtle for the $u$-channel singularities, but can be addressed with a trick based on the invariance of Feynman diagrams under rerouting of internal loop momenta.

To understand the first point, consider the single-pion $t$-channel diagram in isolation. This is given in eq.~\eqref{eq:tchan_ker_infinite_volume_dressed} for the infinite-volume case, and the finite-volume analog can be written as
\begin{align}
D_{t,L}(P, p, p') \equiv G_L(p,p') \ \Delta_{\pi, L}(p'-p) \ G_L(P-p, P-p')\,,
\end{align}
where we have also introduced finite-volume versions of the $NN\pi$ form factors and the pion propagator. As discussed above, the $L$-depencence within $G_L(p,p')$ and $\Delta_{\pi, L}(p'-p)$ is exponentially suppressed as long as the energy and momentum flowing through the diagrams is below the lowest multi-particle threshold. We can then additionally separate out the singularity from a remainder, as in eqs.~\eqref{eq:delta_t_sep}~and~\eqref{eq:tchan_ker_infinite_volume}, to write
\begin{align}
i g^2 \mathcal T(P,p,p') = - \frac{i g^2}{\big ( \omega_N(\boldsymbol p) - \omega_N(\boldsymbol p') \big )^2 - (\boldsymbol p - \boldsymbol p')^2 - M_\pi^2 + i \epsilon} \,,
\label{eq:tchan_ker_finite_volume}
\end{align}
where we have already applied the partially off-shell kinematics of Case 2.
As in the infinite-volume section, $M_\pi$ is the physical pion mass and $g$ is the residue of poles in the full scattering amplitude. This is the same residue as in the Bethe-Salpeter kernel as follows from the fact that the pole arises from the single-Bethe-Salpeter term of the skeleton expansion.

The first key observation for eq.~\eqref{eq:tchan_ker_finite_volume} is that it has no $s$ dependence, by virtue of the fact that it is evaluated at $p = (\omega_N(\boldsymbol p), \boldsymbol p)$ and $p' = (\omega_N(\boldsymbol p'), \boldsymbol p')$, with the magnitude of $\boldsymbol p$ and $\boldsymbol p'$ left independent of $E$ (by allowing $(E - \omega_N(p), \boldsymbol P - \boldsymbol p)$ and its primed counterpart to be off shell). In this sense it cannot generate any singularities in the complex $s$ plane. In addition to this, the momenta entering the $t$-channel exchange here guarantee that $t < 0$ and thus that the singularity at $t = M_\pi^2$ is not encountered. This, in turn, implies that no power-like finite-volume effects arise in $C_L(P)$ due to this contribution. We will argue that this statement generalizes to any diagram that can be expressed as some number of $t$-channel exchanges.

\begin{figure}
\begin{center}
\includegraphics[width=\textwidth]{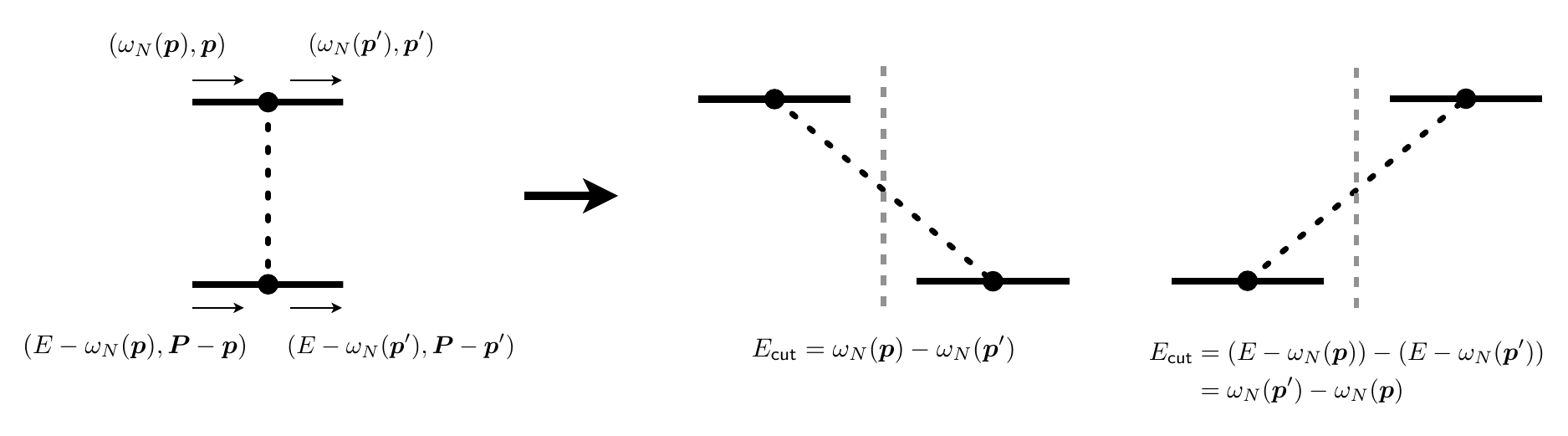}
\vspace{-20pt}
\caption{Illustration of time-ordered perturbation theory (TOPT), applied to the single-meson $t$-channel exchange diagram. In contrast to figure~\ref{fig:TOPT}, here the energy flowing across the cut takes on values other than $\pm E$ and $0$. As discussed in the text, the analogous $u$-channel construction (not shown) leads to spurious poles in the partially off-shell case that are removed once all legs are set on-shell.
\label{fig:TOPT_single_exchange}
}
\end{center}
\end{figure}

We can also analyze the single $t$-channel exchange using TOPT. Following the rules described in ref.~\cite{sterman_1993}, we note that the diagram has two vertices and one internal propagator (see the left diagram of figure~\ref{fig:TOPT_single_exchange}). For Case 2 kinematics, momentum $p - p' = \big (\omega_N(\boldsymbol p) - \omega_N(\boldsymbol p'), \boldsymbol p - \boldsymbol p' \big )$ flows into one vertex and out of the other (right two diagrams of figure~\ref{fig:TOPT_single_exchange}). Thus, the energy flowing across the cut is given by $E_{\sf cut} = \omega_N(\boldsymbol p) - \omega_N(\boldsymbol p')$ for one of the two time orderings, and by $E_{\sf cut} = - \omega_N(\boldsymbol p) + \omega_N(\boldsymbol p')$ for the other. Combining with the single pion propagator intersected by each cut, we expect the denominators $ \omega_N(\boldsymbol p) - \omega_N(\boldsymbol p') - \omega_\pi(\boldsymbol p - \boldsymbol p')$ and $ - \omega_N(\boldsymbol p) + \omega_N(\boldsymbol p') - \omega_\pi(\boldsymbol p - \boldsymbol p')$ to appear. These combine with other factors in the TOPT construction to reproduce eq.~\eqref{eq:tchan_ker_finite_volume}, which can be rewritten as
\begin{multline}
\label{eq:tchan_ker_finite_volume2}
i g^2 \mathcal T(P,p,p') = - \frac{i g^2}{2 \omega_\pi(\boldsymbol p - \boldsymbol p')} \bigg [ \frac{1}{\omega_N(\boldsymbol p) - \omega_N(\boldsymbol p') - \omega_\pi(\boldsymbol p - \boldsymbol p') + i \epsilon} \\ + \frac{1}{- \omega_N(\boldsymbol p) + \omega_N(\boldsymbol p') - \omega_\pi(\boldsymbol p - \boldsymbol p') + i \epsilon} \bigg ] \,.
\end{multline}

More generally, for any diagram that contributes to the Bethe-Salpeter kernel, we can enumerate all time orderings and then identify the (possibly empty) set of cuts for which $E_{\sf cut} = \pm \big (\omega_N(\boldsymbol p) - \omega_N(\boldsymbol p') \big )$. Each will lead to a denominator of the form
\begin{equation}
E_{\sf cut}- \sum_{i} \omega_{f(i)}(\boldsymbol p_i) = \pm \big (\omega_N(\boldsymbol p) - \omega_N(\boldsymbol p') \big ) - \sum_{i} \omega_{f(i)}(\boldsymbol p_i) \,,
\end{equation}
where the sum runs over all internal cut propagators, $f(i)$ is either $N$ or $\pi$, and $\sum_i \boldsymbol p_i = \pm (\boldsymbol p - \boldsymbol p')$, with the sign matching that of $E_{\sf cut}$. For these cases, no singularity is encountered for real three-momenta, since it would correspond to the decay of a nucleon into a multi-particle state, which is always forbidden either by kinematics or by the conserved quantum numbers.

Note the contrast of the discussion here to that in section~\ref{sec:fv_effects} and refs.~\cite{\TOPT}. If we are only interested in the singularities of the full correlator, $C_L(P)$, then the cuts always satisfy $E_{\sf cut} = \pm E,0$. However, if we are also concerned with singularities of the Bethe-Salpeter kernel itself, for various choices of external four-momenta, then it is necessary to revisit TOPT at the level of diagrams with four-external legs (as in figure~\ref{fig:t_channel_exchange}). Then new energies can flow across the cuts, such as the case with $E_{\sf cut} = \pm \big (\omega_N(\boldsymbol p) - \omega_N(\boldsymbol p') \big )$ considered above. If this were the only possibility, then the Bethe-Salpeter kernel would have no singularities on the real $s$-axis for $p = (\omega_N(\boldsymbol p), \boldsymbol p)$ and $p'=(\omega_N(\boldsymbol p'), \boldsymbol p')$.

The second and final message of this section is that the $u$-channel exchanges, such as the right diagram in figure~\ref{fig:t_channel_exchange}, are more subtle. Such diagrams have singularities when evaluated with Case 2 kinematics, i.e.~with $k = (\omega_N(\boldsymbol k), \boldsymbol k)$, but with $P-k$ off shell. These manifest in TOPT from orderings with the vertices that have inflowing $p$ and outflowing $P-p'$ both appearing either to the left or to the right of a particular cut. Then one has $E_{\sf cut} = \pm \big ( \omega_N(\boldsymbol p) - E + \omega_N(\boldsymbol p') \big )$ and, as a result, the denominator has the form
\begin{equation}
E_{\sf cut} - \sum_{i} \omega_{f(i)}(\boldsymbol p_i) = \pm \big ( \omega_N(\boldsymbol p) - E + \omega_N(\boldsymbol p') \big ) - \sum_{i} \omega_{f(i)}(\boldsymbol p_i) \,.
\end{equation}
As we discuss in more detail in appendix~\ref{app:BS_kernel_sing}, this can lead to singularities in the partly off-shell Bethe-Salpeter kernel that do not correspond to singularities of $C_L(P)$. This is an artifact of the partial off-shellness and can be treated most easily by defining an auxiliary Bethe-Salpeter kernel, denoted by $B^{\mathfrak{T}}(P,p,p')$, in which all $u$-channel exchanges are replaced by $t$-channel exchanges.

The explicit construction of this kernel is given in appendix~\ref{app:BS_kernel_sing}, so we do not provide the details here, but only emphasize three key properties: First, $B^{\mathfrak T}(P,p,p')$, like $ \mathcal T(P,p,p')$ in eq.~\eqref{eq:tchan_ker_finite_volume}, has no singularities on the real $s$-axis when kinematics are assigned in the partially off-shell manner of Case 2 and Case 4. Second, $B^{\mathfrak T}(P,p,p')$ does have singularities for fully on-shell kinematics, in the subthreshold regime, and for $(2 M_N)^2 - (2M_\pi)^2 < s < (2 M_N)^2 - M_\pi^2$, these are completely captured by $\mathcal T(P,p,p')$.
Third, the auxiliary kernel is equal to the original Bethe-Salpeter kernel after symmetrization:
\begin{align}
B(P,p,p') = \frac12 \Big [B^{\mathfrak T}(P,p,p') + B^{\mathfrak T}(P,p,P - p') \Big ]\,. \label{eq:bs_sym}
\end{align}

We now return to the loop-reduction equations, eqs.~\eqref{eq:Cloop_compact2} and \eqref{eq:os_loop_S}, recalling that while the first depends on the partially off-shell quantity: $\widetilde{\mathcal L}(P) S(P,L) \widetilde{\mathcal R}(P)^\dagger$, the second is expressed with its fully on-shell counterpart: $\widetilde{\mathcal L}^{\sf os} (P) \xi S(P,L) \xi^\dagger \widetilde{\mathcal R}^{\sf os} (P)^\dagger$. If the left or right function is taken to be a $B^{\mathfrak T}$ insertion then we see the subthreshold cut is absent from the first version but present in the second, and therefore also present in the remainder $\mathcal{I}^{[4]}(P)$. Neglecting this singularity in the remainder invalidates the subsequent steps in the standard derivation of a quantization condition.

To address the issue caused by $\vert \boldsymbol k^\star \vert \to k^\star_{\sf os}$ on the cut, we introduce
\begin{equation}
\overline{B}^{\mathfrak T}(P, p, p') \equiv B^{\mathfrak T}(P, p, p') - 2 g^2 \mathcal T (P, p, p') \,.
\end{equation}
By construction, this does not contain single-exchange poles, and it is safe to perform the angular momentum projections and set the arguments on shell.
Each partial-wave amplitude of this kernel is then a function of $s$ with an analytic strip along the real axis for $s > (2M_N)^2 - (2M_\pi)^2$.
We cannot do the same for $\mathcal T(P,p,p')$ and thus reach a new quantization condition by keeping this party off-shell, as we explain in the next subsection.

Before turning to this, it is useful to define a subtracted version of the full Bethe-Salpeter kernel
\begin{equation}
\overline{B}(P, p, p') \equiv B(P, p, p') - g^2 \mathcal E (P, p, p') \, , \label{eq:ker_sep}
\end{equation}
where $\mathcal E (P, p, p')$ is the sum of the $t$- and $u$-channel single-pion exchanges
\begin{equation}
\label{eq:cal_E_definition}
\mathcal E (P, p, p') \equiv \mathcal T (P, p, p') + \mathcal U (P, p, p') \,,
\end{equation}
and $\mathcal U(P, p, p')$ is defined in eq.~\eqref{eq:uchan_ker_infinite_volume}.
A symmetrization relation holds between the subtracted kernels, directly inherited from the unsubtracted cases
\begin{equation}
\overline{B}(P, p, p') = \frac12 \Big [ \overline{B}^{\mathfrak T}(P, p, p') + \overline{B}^{\mathfrak T}(P, p, P - p') \Big ] \,.
\end{equation}

\subsection{Full decomposition of the finite-volume correlator}
\label{sec:derivation}

We are finally in position to put together the results of the previous sections into the final derivation.
Recalling the expression for the finite-volume correlator, eq.~\eqref{eq:corr_skel}, and substituting the separation of the kernel introduced in eq.~\eqref{eq:ker_sep} above, we reach
\begin{align}
\label{eq:CL_decomposed_minusone}
C_L(P) & = C^{(0)}(P) + \sum_{n=0}^\infty \Tilde{\mathcal A} \, \circ_{\sf fv} \, \left[ \big( i\overline{B} + ig^2\mathcal E \big) \, \circ_{\sf fv} \, \right]^n \, \Tilde{\mathcal A}^\dagger \,,
\\ & = C^{(0)}(P) + \sum_{n=0}^\infty \Tilde{\mathcal A} \, \circ_{\sf fv} \, \left[ \big( i\overline{B}^{\mathfrak T} + 2 ig^2\mathcal T \big) \, \circ_{\sf fv} \, \right]^n \, \Tilde{\mathcal A}^\dagger \, .
\label{eq:CL_decomposed_one}
\end{align}
Here and in the following we drop the $L$ subscripts of objects (such as $\Tilde{\mathcal A}_L$) that are known to have exponentially suppressed volume dependence. Since the loops and the endcaps are exchange-symmetric, we have noted that both separations can be substituted for the Bethe-Salpeter kernel.

To further reduce this expression, consider a generic $\mathcal T$ factor with a finite-volume loop on both sides of it. In each loop, the contour integration leading to eq.~\eqref{eq:Cloop_intm} places one of the two momenta on shell. The resulting object is (cf.~Case 2):
\begin{equation}
\mathcal T (P, p, p') \Big\vert_{\begin{subarray}{l} p^0 = \omega_N(\boldsymbol p),\\
p'^0 = \omega_N(\boldsymbol p') \end{subarray}}
= \frac{1}{2\omega_N(\boldsymbol p^\star) \omega_N(\boldsymbol p'^\star) - 2 \vert \boldsymbol p^\star \vert \vert \boldsymbol p'^\star \vert \cos \theta^\star - 2M_N^2 + M_\pi^2 - i\epsilon} \, .
\label{eq:tchan_ker2}
\end{equation}
This quantity has a strip of analyticity about the real-axis in the $\vert \boldsymbol p^\star \vert$ and $\vert \boldsymbol p'^\star \vert$ complex planes.%
\footnote{We note that the contour integration also encircles the poles in $\mathcal T (P, p, p')$ itself, but this is not an issue as it simply leads to an $NN\pi$ intermediate state pole, which is above the elastic region and thus gives a smooth summand contribution that can be absorbed in the definition of $\mathcal I^{[2]}$ in eq.~\eqref{eq:Cloop_intm}.}
As a result, the loop separation identity of eq.~\eqref{eq:Cloop_compact3}, $\circ_{\sf fv} = \circ_{\sf rm} + iS(P, L)$, can be applied to two-particle loops involving $\mathcal T$ exchanges, and thus to all finite-volume loops appearing in \eqref{eq:CL_decomposed_one}.
Making the substitution, we reach
\begin{equation}
C_L(P) = C^{(0)}(P) + \sum_{n=0}^\infty \Tilde{\mathcal A} \, \big[ \! \circ_{\sf rm} + \, iS(P, L) \big] \, \Big [ \big( i\overline{B}^{\mathfrak T} + 2 ig^2\mathcal T \big) \, \big [ \! \circ_{\sf rm} +\, iS(P, L) \big ] \, \Big ]^n \, \Tilde{\mathcal A}^\dagger \, .
\label{eq:CL_decomposed_two}
\end{equation}

Next, we rearrange the series by $S(P,L)$ insertions to find
\begin{equation}
C_L(P) = \mathcal I^{[1]}_C(P) + \sum_{n=0}^\infty A^{[1]}(P) \, iS(P, L) \, \Big [ \big( i\overline{\mathcal K}^{[1]}(P) + 2 ig^2\mathcal T(P) \big) \, iS(P, L) \Big ]^n \, {A}^{[1]\dagger}(P) \,,
\label{eq:CL_decomposed_three}
\end{equation}
where all $\circ_{\sf rm}$ operations are absorbed into the four new $L$-independent quantities:
\begin{align}
\mathcal I^{[1]}_C(P) & = C^{(0)}(P) + \sum_{n=0}^\infty \Tilde{\mathcal A} \, \circ_{\sf rm} \, \Big [ \big( i\overline{B}^{\mathfrak T} + 2 ig^2\mathcal T \big) \, \circ_{\sf rm} \Big ]^n \, \Tilde{\mathcal A}^\dagger \,, \\
A^{[1]}(P) & = \sum_{n=0}^\infty \Tilde{\mathcal A} \, \Big [ \circ_{\sf rm} \big( i\overline{B}^{\mathfrak T} + 2 ig^2\mathcal T \big) \Big ]^n \,, \\
A^{[1]\dagger}(P) & = \sum_{n=0}^\infty \, \Big [ \big( i\overline{B}^{\mathfrak T} + 2 ig^2\mathcal T \big) \circ_{\sf rm} \Big ]^n \, \Tilde{\mathcal A}^\dagger \,, \\
i\overline{\mathcal K}^{[1]}(P) & = \sum_{n=0}^\infty \Big [ \big( i\overline{B}^{\mathfrak T} + 2 ig^2\mathcal T \big) \, \circ_{\sf rm} \Big ]^n \, \big( i\overline{B}^{\mathfrak T} + 2 ig^2\mathcal T \big) - 2 i g^2 \mathcal T \,.
\label{eq:CL_decomposed_Kbar}
\end{align}
Here the ${}^{[1]}$ superscripts are used, as in section~\ref{sec:loop_cont}, to indicate intermediate quantities.

In contrast to eq.~\eqref{eq:CL_decomposed_two}, in eq.~\eqref{eq:CL_decomposed_three} the integral operator does not appear explicitly. All factors are vectors or matrices on the $\boldsymbol k^\star \ell m$ space, multiplied with indices contracted in the usual way. In particular, $\mathcal T(P)$ is defined from the projection of $\mathcal T (P, p, p')$, as defined in eq.~\eqref{eq:tchan_ker2}, to definite angular momentum [see also eq.~\eqref{eq:T_partial_wave_decom}]:
\begin{align}
\mathcal T (P, p, p') \Big\vert_{\begin{subarray}{l} p^0 = \omega_N(\boldsymbol p),\\
p'^0 = \omega_N(\boldsymbol p') \end{subarray}} & \equiv 4\pi \, \vert \boldsymbol p^\star \vert^{\ell} \, Y^*_{\ell m} (\hat{\boldsymbol p}^\star) \, \mathcal T_{\ell m , \ell' m'} (P, \vert \boldsymbol p^\star \vert, \vert \boldsymbol p'^\star \vert) \, \vert \boldsymbol p'^\star \vert^{\ell'} \, Y_{\ell' m'}(\hat{\boldsymbol p}'^\star) \,.
\label{eq:tchan_ker4}
\end{align}
This leads us to define
\begin{align}
\mathcal T_{\boldsymbol k^\star \ell m,\boldsymbol k'^\star \ell' m'}(P) & = \mathcal T_{\ell m , \ell' m'} (P, \vert \boldsymbol k^\star \vert, \vert \boldsymbol k'^\star \vert) \,. \label{eq:t_matrix_def}
\end{align}
We emphasize again that the object we are projecting here is not the on-shell exchange containing the sub-threshold pole leading to the left-hand cut. As remarked at the end of section~\ref{sec:inf_vol_cut}, such an expansion is expected to have poor convergence on the cut. Instead, what we project is the partially off-shell exchange of eq.~\eqref{eq:tchan_ker2} (cf.~the Case 2 kinematics described in the previous section). This object does not contain a pole for real $\vert \boldsymbol p^\star \vert$, $\vert \boldsymbol p'^\star\vert$, and can therefore be expanded in partial waves without convergence problems.

It is straightforward to work out expressions for particular angular-momentum components. The $S$-wave result, for example, is given by
\begin{equation}
\mathcal T_{\boldsymbol k^\star 0 0,\boldsymbol k'^\star 0 0}(P) = \frac{1}{4\vert \boldsymbol k^\star \vert \vert \boldsymbol k'^\star \vert} \log \left( \frac{2\omega_N(\boldsymbol k^\star) \omega_N(\boldsymbol k'^\star) + 2 \vert \boldsymbol k^\star \vert \vert \boldsymbol k'^\star \vert - 2M_N^2 + M_\pi^2 - i\epsilon}{2\omega_N(\boldsymbol k^\star) \omega_N(\boldsymbol k'^\star) - 2 \vert \boldsymbol k^\star \vert \vert \boldsymbol k'^\star \vert - 2M_N^2 + M_\pi^2 - i\epsilon} \right) \,,
\label{eq:tchan_swave}
\end{equation}
which matches eq.~\eqref{eq:t_channel_cut_on_shell} when we set $\vert \boldsymbol k^\star \vert$ and $\vert \boldsymbol k'^\star \vert$ to their complex, subthreshold on-shell values.

The final step is to note that all insertions of $A^{[1]}(P)$, $\overline{\mathcal K}^{[1]}(P)$, and $A^{[1]\dagger}(P)$ can be set on shell, as described in sections~\ref{sec:os_subs} and \ref{sec:subthr}. This requires a complicated set of redefinitions for the $L$-independent quantities, as detailed in appendix~\ref{app:derivation_details}. The resulting expression is
\begin{align}
\hspace{-22pt} C_L(P) & = \mathcal I_C(P) + \sum_{n = 0}^\infty A^{\sf os}(P) \, \xi \, iS(P,L) \left[ \left( \xi^\dagger\, i\overline{\mathcal K}^{\sf os}(P) \, \xi + 2 i g^2 \mathcal T(P) \right) iS(P,L) \right]^n \xi^\dagger \, A^{\sf os}(P)^{\dagger} \,, \label{eq:final_geometric_series}
\end{align}
where $\mathcal I_C(P)$, $A^{\sf os}(P)$, $A^{\sf os}(P)$, and $\overline{\mathcal K}^{\sf os}(P)$, yet another set of infinite-volume quantities, are the final objects entering our decomposition of $C_L(P)$. Their detailed definitions are given in appendix~\ref{app:derivation_details} where we also prove eq.~\eqref{eq:final_geometric_series}. Note that the definitions are not particularly relevant to the main derivation, since all quantities besides $\overline{\mathcal K}^{\sf os}(P)$ do not appear in the final result. The latter is more usefully defined through its relation to the scattering amplitude, derived in section~\ref{sec:int_eqs}. One crucial point, also addressed in appendix~\ref{app:derivation_details}, is that the asymmetry from $\overline B^{\mathfrak T}$ is removed at this stage, so that $\overline{\mathcal K}^{\sf os}(P)$ is defined with the same exchange symmetry as the scattering amplitude.

Summing the geometric series in eq.~\eqref{eq:final_geometric_series}, we finally obtain
\begin{align}
C_L(P) & = \mathcal I_C(P) + A^{\sf os}(P) \, \xi \frac{i}{S(P,L)^{-1} + \xi^\dagger \, \overline{\mathcal K}^{\sf os}(P) \, \xi + 2 g^2 \mathcal T(P)} \xi^\dagger \, A^{\sf os}(P)^\dagger \,. \label{eq:corr_final_form}
\end{align}
This is the main result of this subsection. Up to neglected exponentially suppressed contributions, the volume dependence of $C_L(P)$ is contained entirely in the second term, both through the explicit $L$ dependence of $S(P,L)$ and the implicit dependence in
the matrix multiplications, which involve sums over discretized spatial momenta.

\subsection{Result}
\label{sec:result}

For a given total spatial momentum $\boldsymbol P$ and volume $L$, let us denote the finite volume energy levels by $E_j(\boldsymbol P,L)$. Our main motivation for examining $C_L(P)$ was that it has poles at these energy levels and from eq.~\eqref{eq:corr_final_form}, we see that these poles arise at the values of the energy for which the matrix in the denominator has a vanishing eigenvalue. This gives us the following quantization condition:
\begin{equation}
\det_{\boldsymbol k^\star \ell m} \left[ S(P_j, L)^{-1} + \xi^\dagger \, \overline{\mathcal K}^{\sf os}(P_j) \, \xi + 2 g^2 \mathcal T(P_j) \right] = 0\,, \label{eq:qcnew}
\end{equation}
where $P_j = (E_j(\boldsymbol P,L), \boldsymbol P)$ and the determinant is taken over the whole $\boldsymbol{k}^\star \ell m, \boldsymbol{k}'^\star \ell' m'$ index space, as indicated. This condition, together with the integral equations of section \ref{sec:int_eqs}, is the main result of this work. It can be used to constrain $\overline{\mathcal K}^{\sf os}(P)$, as well as the trilinear coupling $g$, via lattice-QCD-determined energies.%
\footnote{We note that, in certain cases, it may be more feasible or practical to constrain $g$ from fits to form factors, e.g.~the nucleon axial form factor in the case of the physical $NN$ system. In all cases, one must be certain that the extracted coupling satisfies the defining relation of being the residue of the physical scattering amplitude at the $t$- and $u$-channel poles.}

To make this a stand-alone section, we collect all definitions required to numerically evaluate the quantization condition.

The matrix $\mathcal T(P)$, which encodes the angular-momentum projected $t$-channel exchange, has matrix elements given by
\begin{multline}
\mathcal T_{\boldsymbol k^\star \ell m,\boldsymbol k'^\star \ell' m'}(P) = \frac{1}{4 \pi \vert \boldsymbol k^\star \vert^{\ell} \vert \boldsymbol k'^\star \vert^{\ell'}} \int d \Omega_{\hat{\boldsymbol k}^\star} d \Omega_{\hat{\boldsymbol k}'^\star} Y_{\ell m}(\hat{\boldsymbol k}^\star) \, Y^*_{\ell' m'} (\hat{\boldsymbol k}'^\star) \\
\times \bigg [ \frac{1}{2\omega_N(\boldsymbol k^\star) \omega_N(\boldsymbol k'^\star) - 2 \vert \boldsymbol k^\star \vert \vert \boldsymbol k'^\star \vert \cos \theta^\star - 2M_N^2 + M_\pi^2 - i\epsilon} \bigg ] \,,
\end{multline}
where $\cos \theta^\star = \hat{\boldsymbol k}^\star \cdot \hat{\boldsymbol k}'^\star$.
This expression is reached by combining eqs.~\eqref{eq:tchan_ker2}, \eqref{eq:tchan_ker4} and \eqref{eq:t_matrix_def} and using the orthogonality of spherical harmonics.

The matrix $S(P,L)$, which encodes the intermediate two-nucleon on-shell pole, was already defined in eq.~\eqref{eq:S_def}. We repeat the expression here
\begin{equation}
S_{\boldsymbol k^\star \ell m , \boldsymbol k'^\star \ell' m'} (P, L) = \frac{1}{2L^3} \, \frac{4\pi \, Y_{\ell m}(\hat{\boldsymbol k}^\star) \, Y^*_{\ell' m'}(\hat{\boldsymbol k}^\star) \, \delta_{\boldsymbol k^\star \boldsymbol k'^\star} \, \vert \boldsymbol k^\star \vert^{\ell + \ell'} \, e^{- \alpha [ (\boldsymbol k^\star)^2 - (k_{\sf os}^\star)^2]}}{4 \omega_N(\boldsymbol k) \, \big [ (k_{\sf os}^\star)^2 - (\boldsymbol k^\star)^2\big ]} \,,
\label{eq:s_with_h_example}
\end{equation}
where we have substituted the possible form for $H(\boldsymbol k^\star)$ from eq.~\eqref{eq:h_example}. Recall that $\boldsymbol k^\star$ takes on all values reached by boosting $(\omega_N(\boldsymbol k), \boldsymbol k)$ to the CM frame, where $\boldsymbol k \in (2 \pi / L) \mathbb Z^3$ is a finite-volume momentum. Recall also the definition $k_{\sf os}^\star = \sqrt{s/4 - M_N^2}$, given in eq.~\eqref{eq:rhoN_pN_def}.

It remains to describe the middle term of eq.~\eqref{eq:qcnew}. First, $\overline{\mathcal K}^{\sf os}(P)$ (the quantity one aims to determine from the spectrum) is a diagonal matrix in angular-momentum space so that we can write
\begin{equation}
\overline{\mathcal K}_{\ell m, \ell' m'}^{\sf os}(P) = \delta_{\ell \ell'} \, \delta_{m m'} \, \overline{\mathcal K}^{{\sf os}(\ell)}(s) \,,
\end{equation}
where we have also used that each component is a Lorentz invariant, depending only on the CM energy, encoded here via the Mandelstam variable $s = P^2$. Unlike the standard K-matrix, this quantity is analytic in the extended range $(2M_N)^2 - (2M_\pi)^2 < s < (2 M_N + M_\pi)^2$ and the usual parametrizations involving polynomials and poles can be used in this region. Finally, as introduced in section~\ref{sec:os_subs}, $\xi$ is a trivial vector with $\boldsymbol k^\star$ indices, defined as
\begin{equation}
\xi_{\boldsymbol k^\star} = 1 \,,
\end{equation}
used to promote $\mathcal K^{\sf os}(P)$ from the $\ell m$ space to the $\boldsymbol k^\star \ell m$ space.

This completes our discussion of the generalized quantization condition for spin-zero nucleons. In section~\ref{sec:int_eqs}, we discuss how $g$ and $\overline{\mathcal K}^{\sf os}(P)$ can then be used to then obtain the $NN\to NN$ scattering amplitude. Before turning to this, we first discuss the extension to spinning particles.

\subsection{Incorporating spin}
\label{sec:spin}

Thus far, our finite-volume analysis has been restricted to particles without intrinsic spin. In this subsection, we describe how the derivation of \eqref{eq:qcnew} can be readily modified to accommodate spinning particles. We first present the extension to identical spin-half particles, relevant for physical two-nucleon systems, and then briefly comment on the generalization to arbitrary spin. These generalizations follow readily from the extension of the usual scattering formalism to include spin, as described in refs.~\cite{\spinPapers}.

Consider a realistic, generic low-energy theory of QCD with Dirac spinors $N$ representing nucleons and anti-nucleons coupled to pseudoscalar fields $\pi$ representing the spin-zero pions. The interactions in the Lagrangian are modified appropriately to contain all vacuum-quantum-number combinations of $N$, $\overline N$, and $\pi$.

The dressed propagator of the nucleon in a finite volume can be written as
\begin{align}
\Delta_{N,L}(k)
& = \frac{i(\slashed k + M_N)}{k^2 - M_N^2 + i\epsilon} + R_{N,L}(k) \ ,
\end{align}
where $R_{N,L}(k)$ encodes the self-energy contributions. As above, we require $M_N$ to be the physical pole mass and assume renormalization such that the residue at the pole of $\Delta_{N,L}(k)$ is the same as that of the first term on the right-hand side. This has the consequence that $R_{N,L}(k)$ is analytic in a neighborhood of the pole, up to the exponentially suppressed volume effects on the pole mass, which we neglect.

As in eq.~\eqref{eq:corr_def}, we define a correlator $C_L(P)$ of operators carrying the quantum numbers of an $NN$ state with spin and consider the corresponding skeleton expansion. Following the steps of section~\ref{sec:loop_cont}, the contribution from a generic two-nucleon loop after $k^0$ integration is given by
\begin{equation}
\mathcal L \, \circ_{\sf fv} \, \mathcal R^\dagger = \mathcal I^{[1]}(P) + \sum_{\boldsymbol k} \frac{i}{2L^3} \frac{\mathcal L^{\alpha\beta}(P,k) (\slashed k + M_N)^{\alpha\alpha'} (\slashed P - \slashed k + M_N)^{\beta\beta'} \mathcal R^{\alpha'\beta'}(P,k)^*}{2 \omega_N(\boldsymbol k) \left[ (E - \omega_N(\boldsymbol k))^2 - \omega_N (\boldsymbol P - \boldsymbol k)^2 \right]} \bigg\vert_{k^0 = \omega_N(\boldsymbol k)}
\label{eq:Cloop_intm_ferm}
\end{equation}
where we have absorbed the $R_{N,L}(k)$ dependent terms into $\mathcal I^{[1]}(P)$, since these only generate exponentially suppressed contributions.
Here $\alpha,\beta,\alpha',\beta'$ are Dirac spinor indices and sums over repeated indices are implied.

Noting that $k$ is on shell, we identify the spin sum relation
\begin{equation}
(\slashed k + M_N)^{\alpha\alpha'} \Big\vert_{k^0 = \omega_N(\boldsymbol k)} = \sum_{r = \pm} u^\alpha_r(\boldsymbol k) \bar u^{\alpha'}_r(\boldsymbol k) \ ,
\end{equation}
where $u^\alpha_r(\boldsymbol k)$ and $\bar u^{\alpha'}_r(\boldsymbol k)$ are the Dirac spinors describing a free $N$ state with definite spin. The spin-up and down states are labelled by $r = \pm$.
By contrast,
the four-vector $(P - k) \vert_{k^0 = \omega_N(\boldsymbol k)} = (E - \omega_N(\boldsymbol k), \boldsymbol P - \boldsymbol k)$ is not on-shell. To address this we consider an on-shell vector $(P - k)_{\sf os} \equiv (\omega_N(\boldsymbol P - \boldsymbol k), \boldsymbol P - \boldsymbol k)$ such that
\begin{align}
(\slashed P - \slashed k + M_N)^{\beta\beta'} & = \left( (\slashed P - \slashed k)_{\sf os} + M_N \right)^{\beta\beta'} + \left[ (\slashed P - \slashed k) - (\slashed P - \slashed k)_{\sf os}\right]^{\beta\beta'} \ ,
\\
& = \sum_{s = \pm} u^\beta_s(\boldsymbol P - \boldsymbol k) \bar u^{\beta'}_s(\boldsymbol P - \boldsymbol k) + \left[ E - \omega_N(\boldsymbol k) - \omega_N (\boldsymbol P - \boldsymbol k) \right] (\gamma^0)^{\beta\beta'} \ , \label{eq:Pkos_exp}
\end{align}
where we have used the spin sum relation on $(P - k)_{\sf os}$. This intermediate step differs from the approach of refs.~\cite{\spinPapers} but is more convenient here, given our focus on sub-threshold kinematics.

Substituting these expressions for $(\slashed k + M_N)^{\alpha\alpha'}$ and $(\slashed P - \slashed k + M_N)^{\beta\beta'}$ into \eqref{eq:Cloop_intm_ferm}, we see that the term $\left[ E - \omega_N(\boldsymbol k) - \omega_N (\boldsymbol P - \boldsymbol k) \right] (\gamma^0)^{\beta\beta'}$ in \eqref{eq:Pkos_exp} leads to a cancellation of the pole, so we can include this contribution in the remainder term and obtain
\begin{align}
\begin{split}
\mathcal L \, \circ_{\sf fv} \, \mathcal R^\dagger & =
\label{eq:ferm_loop}
\mathcal I^{[2]}(P) + \sum_{\boldsymbol k} \frac{i}{2L^3} \frac{{\mathcal L}_{rs}(P, \boldsymbol k^\star) \, \delta_{rr'} \, \delta_{ss'} \, {\mathcal R}^*_{r's'}(P, \boldsymbol k^\star )}{2 \omega_N(\boldsymbol k) \left[
(E - \omega_N(\boldsymbol k))^2 - \omega_N (\boldsymbol P - \boldsymbol k)^2
\right]} \, ,
\end{split}
\end{align}
where we have defined the objects
\begin{align}
\begin{split} & {\mathcal L}_{rs}(P, \boldsymbol k^\star) \equiv \mathcal L^{\alpha\beta}(P,k) \big\vert_{k^0 = \omega_N(\boldsymbol k)} \, u^\alpha_r(\boldsymbol k) u^\beta_s(\boldsymbol P - \boldsymbol k) \ ,
\\
& {\mathcal R}^*_{r's'}(P, \boldsymbol k^\star) \equiv \bar u^{\alpha'}_{r'}(\boldsymbol k) \bar u^{\beta'}_{s'} (\boldsymbol P - \boldsymbol k) \, \mathcal R^{* \alpha' \beta'}(P,k)\big\vert_{k^0 = \omega_N(\boldsymbol k)} \ , \end{split}
\end{align}
which we choose to write as functions of the CM spatial loop momentum $\boldsymbol k^\star$.

We next decompose these functions into spherical harmonics
\begin{align}
\begin{split} & {\mathcal L}_{rs} (P, \boldsymbol k^\star) \equiv \sqrt{4\pi} \, Y_{\ell m}(\hat{\boldsymbol k}^\star) \, \vert \boldsymbol k^\star \vert^\ell \, \widetilde{\mathcal L}_{rs,\ell m} (P, \vert \boldsymbol k^\star \vert ) \ ,
\\
& {\mathcal R}^*_{r's'} (P, \boldsymbol k^\star) \equiv \sqrt{4\pi} \, Y^*_{\ell' m'}(\hat{\boldsymbol k}^\star) \, \vert \boldsymbol k^\star \vert^{\ell'} \, \widetilde{\mathcal R}^*_{r's',\ell' m'} (P, \vert \boldsymbol k^\star \vert ) \,, \end{split} \label{eq:harm_proj}
\end{align}
and rearrange the pole term in eq.~\eqref{eq:ferm_loop} to reach
\begin{equation}
\mathcal L \, \circ_{\sf fv} \, \mathcal R^\dagger = \mathcal L \circ_{\sf rm} \mathcal R^\dagger + \widetilde{\mathcal L}(P) \, iS(P,L) \, \widetilde{\mathcal R}(P)^\dagger \,,\label{eq:Cloop_ferm}
\end{equation}
where, as in the spin zero case, we have absorbed extra terms in the in a redefinition of $\mathcal I^{[2]}(P) \to \mathcal I^{[3]}(P)$, and also introduced $\mathcal I^{[3]}(P) \equiv \mathcal L \circ_{\sf rm} \mathcal R^\dagger$. This result exactly matches eq.~\eqref{eq:Cloop_compact3} above.

The objects $\widetilde{\mathcal L}(P)$ and $\widetilde{\mathcal R}(P)^\dagger$ are vectors in the combined space of spin, spatial loop momentum, and angular momentum. The vector elements can be written explicitly as $\widetilde{\mathcal L}_{rs, \boldsymbol k^\star \ell m} (P)$ and $\widetilde{\mathcal R}^*_{r's', \boldsymbol k'^\star \ell' m'} (P)$, respectively. The matrix $S(P,L)$, in turn, has elements
\begin{equation}
\label{eq:S_def_spin}
S_{rs, \boldsymbol k^\star \ell m ; \, r's', \boldsymbol k'^\star \ell' m'} (P,L) = \frac{1}{2L^3} \, \frac{4\pi \, Y_{\ell m}(\hat{\boldsymbol k}^\star) \, Y^*_{\ell' m'}(\hat{\boldsymbol k}^\star) \, \delta_{rr'} \, \delta_{ss'} \, \delta_{\boldsymbol k^\star \boldsymbol k'^\star} \, \vert \boldsymbol k^\star \vert^{\ell + \ell'} \, H(\boldsymbol k^\star)}{4 \omega_N(\boldsymbol k) \, \big [ (k_{\sf os}^\star)^2 - (\boldsymbol k^\star)^2\big ]} \,.
\end{equation}
This matches eq.~\eqref{eq:S_def} up to the additional Kronecker deltas in the spin components.

To give the effect of spin on $\mathcal T$, the off-shell Bethe-Salpeter kernel is split into the contribution from the exchanges and a remainder, as before. The off-shell spin projection described above appears in this separation and leads to the following quantity:
\begin{equation}
\begin{split}
\mathcal T_{rs,r's'} (P,\boldsymbol p^\star, \boldsymbol p'^\star) & \equiv - \bar u^\alpha_r (\boldsymbol p) \bar u^\beta_s (\boldsymbol P - \boldsymbol p) \, \gamma_5^{\alpha\alpha'} \, \gamma_5^{\beta\beta'} \, u^{\alpha'}_{r'}(\boldsymbol p') u^{\beta'}_{s'}(\boldsymbol P - \boldsymbol p')
\\
& \hspace{130pt} \times \ \mathcal T(P, p, p') \big\vert_{p^0 = \omega_N(\boldsymbol p), \, p'^0 = \omega_N(\boldsymbol p')} \ .
\end{split}
\end{equation}
Here we have also included the $\gamma_5$ factors that appear in the $N N \pi$ vertex,
since the exchanged pion is a pseudoscalar.
Projecting this to definite angular momentum, we reach a modified matrix of off-shell log functions, denoted by $\mathcal T$ as in the spin-zero case. The elements $\mathcal T_{rs, \boldsymbol k^\star \ell m ; \, r's', \boldsymbol k'^\star \ell' m'} (P)$ of the matrix correspond to the projected functions.

In addition to $S$ and $\mathcal T$, the matrix $\overline{\mathcal K}^{\sf os}$ is also modified and the elements with spin included are denoted by $\overline{\mathcal K}^{\sf os}_{rs, \ell m ; \, r's', \ell' m'} $. The symmetry properties of this quantity match those of a physical two-to-two amplitude with an incoming spin state of orbital angular momentum $\ell', m'$ and spin components $r's'$ scattering to a state with $\ell, m$ and $r,s$. As with the usual two-to-two scattering formalism with spin, once a truncation in $\ell, \ell'$ is set one can identify the non-zero components and their relations. Here it may be useful to work in the basis of total spin and total angular momentum $J$.

Having extended our results to spin-half particles, we now comment on the generalization to identical particles of arbitrary spin. If we take $N$ to have spin $S$, the dressed propagator can be expanded about the on-shell point $k^2 = M_N^2$ as
\begin{equation}
\Delta_{N, L} (k) = \frac{i \mathcal P_N (\boldsymbol k)}{k^2 - M_N^2 + i\epsilon} + R_{N,L}(k)\ ,
\end{equation}
where $\mathcal P_N (\boldsymbol k)$ is a volume-independent matrix carrying two sets of field indices $A,A'$. In the spin-half case, $\mathcal P_N (\boldsymbol k)$ corresponds to $(\slashed k + M_N) \vert_{k^0 = \omega_N(\boldsymbol k)}$. As in the Dirac spinor case, this can be decomposed into projectors that relate the $A,A'$ indices to the spin state indices $r,r'$,
\begin{equation}
\mathcal P_N (\boldsymbol k)^{AA'} = U_r^A (\boldsymbol k) \, \delta_{rr'} \, \overline U_{r'}^{A'} (\boldsymbol k) \,. \label{eq:gen_sumrule}
\end{equation}
where $r$ and $r'$ are summed over the allowed values
$r,r' = -S,-S+1,...,S-1,S$, and the bar on $\overline U$ denotes the appropriate conjugation.
Using eq.~\eqref{eq:gen_sumrule}, we are left with the same form for the loop contribution as the spin-half case in eq.~\eqref{eq:ferm_loop} above.

The details of the $\mathcal T$ factor will depend on how the light exchanged particle couples to the spin-$S$ nucleon. For example if envision the case of a scalar (rather than a pseudoscalar) exchange particle coupled to two real vectors, we find
\begin{equation}
\begin{split}
\mathcal T_{rs,r's'} (P,\boldsymbol p^\star, \boldsymbol p'^\star) & \equiv \bar \epsilon^\alpha_r (\boldsymbol p) \bar \epsilon^\beta_s (\boldsymbol P - \boldsymbol p) \, \delta^{\alpha\alpha'} \, \delta^{\beta\beta'} \, \epsilon^{\alpha'}_{r'}(\boldsymbol p') \epsilon^{\beta'}_{s'}(\boldsymbol P - \boldsymbol p')
\\
& \hspace{140pt} \times \ \mathcal T(P, p, p') \big\vert_{p^0 = \omega_N(\boldsymbol p), \, p'^0 = \omega_N(\boldsymbol p')} \ .
\end{split}
\end{equation}

\section{Relating \texorpdfstring{$\overline{\mathcal K}^{\sf os}$}{the K-matrix} to the scattering amplitude}
\label{sec:int_eqs}

In section~\ref{sec:result}, we presented a generalized quantization condition, eq.~\eqref{eq:qcnew}, which allows one to constrain the volume-independent object $\overline{\mathcal K}^{\sf os}(P)$ using the finite-volume spectrum. As emphasized in previous sections, this quantity differs from the standard K-matrix, for example due to its dependence on the cutoff function $H(\boldsymbol k^\star)$, appearing in the definition of $S(P,L)$. In this section, we show how the scheme-dependent $\overline{\mathcal K}^{\sf os}(P)$ is related to the infinite-volume $NN \to NN$ scattering amplitude, denoted $\mathcal M (P, p, p')$ and thus also to more standard K-matrix definitions. The basic method given here has been used extensively, for example in refs.~\cite{Hansen:2015zga,Briceno:2017tce,Briceno:2018mlh,Briceno:2018aml}.

\subsection{Finite-volume auxiliary amplitude}

We begin by defining a finite-volume quantity $\mathcal M_L (P, p, p')$, that we loosely refer to as a ``finite-volume amplitude''. Diagrammatically it is related to the sum of all $NN \to NN$ four-leg amputated diagrams, but with all diagrams evaluated in the finite volume, and with external arguments not necessarily set on shell. This quantity is useful because we can formally recover the physical amplitude from it by introducing an $i\epsilon$ prescription on the propagators and taking an ordered double limit: first $L \to \infty$ at fixed $\epsilon$, causing the spatial loop momentum sums to converge to the usual $i\epsilon$-prescription Feynman integrals, followed by $\epsilon \to 0$ and the on-shell limit, to reach the standard scattering amplitude.

Using the notation introduced in the previous section, we write
\begin{equation}
i \mathcal M_L
= \sum_{n = 0}^\infty \left[ iB \, \circ_{\sf fv} \right]^n iB + i \Delta \mathcal M_L \, ,
\label{eq:FV_amp}
\end{equation}
where, for simplicity, we omit the dependence on total four-momentum $P$ and external momentum arguments $p$ and $p'$.
Here we have introduced $\Delta \mathcal M_L$ as an additional, volume-dependent term that vanishes when the external momenta are set on shell. This represents an additional freedom we have in the definition of $\mathcal M_L$, since any quantity that vanishes in the on-shell limit will be irrelevant for the value of the extracted amplitude.
Below we will make a particular choice of $\Delta \mathcal M_L$ that simplifies the relation between $\mathcal M_L$ and $\overline{\mathcal K}^{\sf os}$.

As in the previous section, one can express $\mathcal M_L$ in terms of the $t$-exchange modified kernel $B^{\mathfrak T}$ which can be separated into the subtracted part $\overline B^{\mathfrak T}$ and the exchange term $2 g^2 \mathcal T$. The only issue here is that, unlike with $C_L(P)$, the lack of exchange symmetry within $B^{\mathfrak T}$ will have an effect on the amplitude constructed with it. To handle this, we define a finite-volume auxiliary amplitude $\mathcal M^{\sf aux}_L (P, p, p')$, as follows:
\begin{align}
i \mathcal M^{\sf aux}_L & = \sum_{n = 0}^\infty \left[ iB^{\mathfrak T} \, \circ_{\sf fv} \right]^n iB^{\mathfrak T} + i \Delta \mathcal M^{\sf aux}_L \,,
\\
& = \sum_{n = 0}^\infty \left[ \left (i \overline B^{\mathfrak T} + 2 i g^2 \mathcal T \right )\, \circ_{\sf fv} \right]^n \left ( i \overline B^{\mathfrak T} + 2 i g^2 \mathcal T \right ) + i \Delta \mathcal M^{\sf aux}_L \,.
\label{eq:aux_amp}
\end{align}

We additionally define the infinite-volume counterpart of $\mathcal M^{\sf aux}_L$ via
\begin{equation}
i \mathcal M^{\sf aux} = \sum_{n = 0}^\infty \left[ iB^{\mathfrak T} \, \circ_{i \epsilon} \right]^n iB^{\mathfrak T}
+ i \Delta \mathcal M^{\sf aux}\,,
\label{eq:aux_amp_fv}
\end{equation}
and observe this satisfies the ordered double limit described in the first paragraph of this subsection
\begin{equation}
i \mathcal M^{\sf aux} = \lim_{\epsilon \to 0} \lim_{L \to \infty} i \mathcal M^{\sf aux}_L \Big \vert_{E \to E + i \epsilon} \, .
\label{eq:aux_amp_inf_lim}
\end{equation}
To recover the standard amplitude from the auxiliary, one simply symmetrizes under the exchange $p' \leftrightarrow (P - p')$:
\begin{equation}
\mathcal M (P, p, p') = \frac12 \left[ \mathcal M^{\sf aux} (P, p, p') + \mathcal M^{\sf aux} (P, p, P - p') \right] \ , \qquad \qquad \text{(bosons)} \, ,
\label{eq:aux_to_amp}
\end{equation}
with the corresponding result for fermions as required by the total spin state.

In the remainder of this section, we make extensive use of the angular momentum-projected versions of these amplitudes. The relations between $\mathcal{M}$ and $\mathcal{M}^{\sf aux}$ and their spherical harmonic projections are given in the usual way:
\begin{align}
\label{eq:M_proj}
\mathcal M (P, p, p') & = 4\pi \, \vert \boldsymbol p^\star \vert^\ell \, \vert \boldsymbol p'^\star \vert^{\ell'} \, Y^*_{\ell m}(\hat{\boldsymbol p}^\star) \, Y_{\ell' m'}(\hat{\boldsymbol p}'^\star) \, \widetilde{\mathcal M}_{\ell m, \ell' m'} (P, \vert \boldsymbol p^\star \vert, \vert \boldsymbol p'^\star \vert) \, , \\
\mathcal M^{\sf aux} (P, p, p') & = 4\pi \, \vert \boldsymbol p^\star \vert^\ell \, \vert \boldsymbol p'^\star \vert^{\ell'} \, Y^*_{\ell m}(\hat{\boldsymbol p}^\star) \, Y_{\ell' m'}(\hat{\boldsymbol p}'^\star) \, \widetilde{\mathcal M}^{\sf aux}_{\ell m, \ell' m'} (P, \vert \boldsymbol p^\star \vert, \vert \boldsymbol p'^\star \vert) \, .
\label{eq:aux_proj}
\end{align}
These projections have a straightforward relation to the partial wave expansion coefficients denoted by a single $(\ell)$ superscript:
\begin{align}
\widetilde{\mathcal M}_{\ell m, \ell' m'} (P, \vert \boldsymbol p^\star \vert, \vert \boldsymbol p'^\star \vert) & = \delta_{\ell \ell'}\delta_{m m'} \frac{1}{\vert \boldsymbol p^\star \vert^{\ell} \vert \boldsymbol p'^\star \vert^{\ell}} \, {\mathcal M}^{(\ell)} (P, \vert \boldsymbol p^\star \vert, \vert \boldsymbol p'^\star \vert) \, , \\
\widetilde{\mathcal M}^{\sf aux}_{\ell m, \ell' m'} (P, \vert \boldsymbol p^\star \vert, \vert \boldsymbol p'^\star \vert) & = \delta_{\ell \ell'}\delta_{m m'} \frac{1}{\vert \boldsymbol p^\star \vert^{\ell} \vert \boldsymbol p'^\star \vert^{\ell}} \, {\mathcal M}^{{\sf aux}(\ell)} (P, \vert \boldsymbol p^\star \vert, \vert \boldsymbol p'^\star \vert) \, .
\label{eq:M_proj_partial_wave_coeff}
\end{align}
The symmetry properties of the amplitude under momentum exchanges, mentioned above, allows us to derive the following relations for identical bosons:
\begin{align}
{\mathcal M}^{(\ell)} (P, \vert \boldsymbol p^\star \vert, \vert \boldsymbol p'^\star \vert) & = {\mathcal M}^{{\sf aux}(\ell)} (P, \vert \boldsymbol p^\star \vert, \vert \boldsymbol p'^\star \vert) \, , \qquad & \text{for even $\ell$} \, ,
\label{eq:M_plwaves1} \\
{\mathcal M}^{(\ell)} (P, \vert \boldsymbol p^\star \vert, \vert \boldsymbol p'^\star \vert) & = 0 \, , & \text{for odd $\ell$} \, .
\label{eq:M_plwaves2}
\end{align}
with the corresponding results for fermions as required by the total spin state.

\subsection{Integral equations}

We now derive an integral equation relating $\overline{\mathcal K}^{\sf os}(P)$ to the auxiliary amplitude $\mathcal M^{\sf aux}(P, p, p')$, making use of $\mathcal M^{\sf aux}_L$ to do so. The series given in eq.~\eqref{eq:aux_amp} can be manipulated in much the same way as the correlator series of eq.~\eqref{eq:corr_skel}. All steps detailed in the previous section can be applied in a straightforward manner and one obtains
\begin{equation}
i \mathcal M^{\sf aux}_L (P) \equiv
\sum_{n = 0}^\infty \Big[ \left( \xi^\dagger\, i\overline{\mathcal K}^{\sf os}(P) \, \xi + 2 i g^2 \mathcal T(P) \right) iS(P, L) \, \Big]^n \left( \xi^\dagger\, i\overline{\mathcal K}^{\sf os}(P) \, \xi + 2i g^2 \mathcal T(P) \right) \,.
\label{eq:aux_amp2}
\end{equation}
As with other, similarly notated quantities, $\mathcal M^{\sf aux}_L (P)$ is the matrix with elements $\mathcal M^{\sf aux}_{L,\boldsymbol p^\star \ell m; \boldsymbol p'^\star \ell'm'} (P) $. We note that the quantity only depends on the magnitudes of $\boldsymbol p^\star$ and $\boldsymbol p'^\star$ as the directional dependence has gone into the angular-momentum projection.%
\footnote{A subtlety arises from the fact that the momenta in $\mathcal M^{\sf aux}_L (P)$ are in the finite-volume set. Thus, one does not have the continuous rotational symmetry for the angular-momentum projection. This is resolved by noting that the dependence in fact only appears in infinite-volume quantities after the reduction to eq.~\eqref{eq:aux_amp2}, so the extension of $\mathcal M^{\sf aux}_L (P)$ to the continuous infinite-volume set of momenta is straightforward. See also the discussion above eq.~(35) of ref.~\cite{Hansen:2015zga}.\label{footnote:aux_amp}}

An important point, explained in more detail in appendix~\ref{app:derivation_details}, is that the right-hand side of eq.~\eqref{eq:aux_amp2} and the first term in eq.~\eqref{eq:aux_amp} differ by terms that vanish when all external momenta are set on shell. These are the terms collected in $\Delta \mathcal M^{\sf aux}_L$ and, in this sense, it is actually eq.~\eqref{eq:aux_amp2} that provides the precise definition of $\mathcal M^{\sf aux}_L (P)$. Since the final step in our approach requires setting external legs on shell anyway, this distinction is irrelevant for the scattering amplitude predicted. $\Delta \mathcal M^{\sf aux}_L$ is defined by taking the difference between eqs.~\eqref{eq:aux_amp} and \eqref{eq:aux_amp2} and $\Delta \mathcal M_L$ is defined from there by symmetrization, i.e.~by applying eq.~\eqref{eq:aux_to_amp}.

To see why this is necessary, consider the $n=0$ terms of eqs.~\eqref{eq:aux_amp} and \eqref{eq:aux_amp2} in turn. For eq.~\eqref{eq:aux_amp} this includes the off-shell parts of $\overline B^{\mathfrak T}$ while in \eqref{eq:aux_amp2} this same object appears inside $\overline {\mathcal K}^{\sf os}$ and is therefore defined with all momenta on the mass shell. The difference of the two terms is an example of the contributions collected in $\Delta \mathcal M_L$.

We can remove the infinite sum by substituting eq.~\eqref{eq:aux_amp2} into itself, yielding
\begin{equation}
i \mathcal M^{\sf aux}_L (P) = \left( \xi^\dagger\, i\overline{\mathcal K}^{\sf os}(P) \, \xi + 2i g^2 \mathcal T(P) \right) + i \mathcal M^{\sf aux}_L (P) \, iS(P, L) \left( \xi^\dagger\, i\overline{\mathcal K}^{\sf os}(P) \, \xi + 2i g^2 \mathcal T(P) \right) \, .
\label{eq:aux_amp3}
\end{equation}
Starting from this equation, we replace the energy as $E \to E + i\epsilon$ in the matrix $S(P, L)$, which introduces the $i\epsilon$ prescription to the relevant two-nucleon intermediate state, and take the infinite-volume limit $L \to \infty$. These steps lead to the following integral equation:
\begin{multline}
\widetilde {\mathcal M}^{\sf aux}_{\ell m, \ell' m'} (P, \vert \boldsymbol p^\star \vert, \vert \boldsymbol p'^\star \vert) = \mathcal K^{\mathcal T}_{\ell m, \ell' m'}(P, \vert \boldsymbol p^\star \vert, \vert \boldsymbol p'^\star \vert)
\\
- \frac12 \int_0^\infty \! \frac{d \vert \boldsymbol k^\star \vert}{(2\pi)^3} \widetilde {\mathcal M}^{\sf aux}_{\ell m, \ell'' m''} (P, \vert \boldsymbol p^\star \vert, \vert \boldsymbol k^\star \vert) \frac{4\pi \vert \boldsymbol k^\star \vert^{2\ell'' + 2} H(\boldsymbol k^\star)}{4\omega_N(\boldsymbol k^\star) \big[ (k_{\sf os}^\star)^2 - (\boldsymbol k^\star)^2 + i\epsilon \big]} \mathcal K^{\mathcal T}_{\ell'' m'', \ell' m'} (P, \vert \boldsymbol k^\star \vert, \vert \boldsymbol p'^\star \vert) \,,
\label{eq:int_eq1}
\end{multline}
where we have introduced
\begin{equation}
\mathcal K^{\mathcal T}_{\ell m, \ell' m'}(P, \vert \boldsymbol p^\star \vert, \vert \boldsymbol p'^\star \vert) = \overline{\mathcal K}^{\sf os}_{\ell m, \ell' m'} (P) + 2 g^2 \mathcal T_{\ell m, \ell' m'} (P, \vert \boldsymbol p^\star \vert, \vert \boldsymbol p'^\star \vert) \,.
\end{equation}
Note that the $L \to \infty$ limit is performed with fixed external momentum, requiring an extension of these to be outside the finite-volume set of a given $L$. See footnote \ref{footnote:aux_amp} and the indicated reference for further discussion. Observe also that no $L$-depence appears in eq.~\eqref{eq:int_eq1} and that we have replaced the indices indicating discrete CM spatial momenta with continuous arguments.
To reach this expression, we have also performed the angular integral resulting from the infinite-volume limit applied to the sums over $S(P, L)$.

As indicated at various points above, both $\overline{\mathcal K}^{\sf os}$ and $\mathcal M^{\sf aux}$ can be represented either as matrices with angular-momentum indices or as single functions of the scattering momenta. The two forms are connected in the usual way, already seen for $\mathcal M^{\sf aux}$ in eq.~\eqref{eq:aux_proj}. For $\overline{\mathcal K}^{\sf os}$, this is simply
\begin{equation}
\overline{\mathcal K}^{\sf os}(P, p, p') = 4\pi \, \vert \boldsymbol p^\star \vert^\ell \, \vert \boldsymbol p'^\star \vert^{\ell'} \, Y^*_{\ell m}(\hat{\boldsymbol p}^\star) \, Y_{\ell' m'}(\hat{\boldsymbol p}'^\star) \, \overline{\mathcal K}^{\sf os}_{\ell m, \ell' m'} (P) \, ,
\label{eq:Kos_proj}
\end{equation}
Combining these relations with eq.~\eqref{eq:int_eq1}, we obtain a second type of integral equation, directly for the unprojected auxiliary amplitude:
\begin{equation}
\mathcal M^{\sf aux} (P, p, p') = \mathcal K^{\mathcal T}(P, p, p')
- \frac12 \int \! \frac{d^3 \boldsymbol k^\star}{(2\pi)^3} \frac{\mathcal M^{\sf aux} (P, p, k) \, H(\boldsymbol k^\star) \, {\mathcal K}^{\mathcal T}(P, k, p')}{4\omega_N(\boldsymbol k^\star) \big[ (k_{\sf os}^\star)^2 - (\boldsymbol k^\star)^2 + i\epsilon \big]} \, ,
\label{eq:int_eq2}
\end{equation}
where
\begin{equation}
\mathcal K^{\mathcal T}(P, p, p') = \overline{\mathcal K}^{\sf os}(P, p, p') + 2 g^2 \mathcal T(P, p, p') \,,
\end{equation}
and
where the four-momentum arguments $p,p',k$ are all taken to be on shell.

Though the relations given here completely solve the task of relating $\overline{\mathcal K}^{\sf os}$ to $\mathcal M^{\sf aux}$ in principle, the singularities within $\mathcal M^{\sf aux} $ and $\mathcal T$ could well make numerical evaluation very difficult. For this reason, in the next section we consider an alternative, based on a divergence-free intermediate quantity.

\subsection{Divergence-free amplitude}

To find the physical scattering amplitude, $\mathcal M (P, p, p')$, from the intermediate quantity, $\overline{\mathcal K}^{\sf os} (P, p, p')$, one needs to set all external four-momenta, $(p,\, P-p,\, p',\, P-p')$, to their on-shell values.
This, in turn, requires setting the momenta within the auxiliary amplitude, $\mathcal M^{\sf aux} (P, p, p')$, on shell. However, this may lead to an unstable numerical evaluation, associated with the fact that the angular-momentum components of the auxiliary amplitude contain a branch point at $s = 4 M_N^2 - M_\pi^2$, and the unprojected amplitude contains a pole at $t = M_\pi^2$. As both of these singularities only arise for on-shell momenta, extrapolating or interpolating to physical kinematics could be challenging.

Moreover, the partial-wave expansion of $\mathcal M^{\sf aux} (P, p, p')$ is expected to be slowly-converging in the vicinity of the $t$-channel pole and, for this reason, an order-by-order determination of the partial-wave components through eq.~\eqref{eq:int_eq1} will not be useful near $t = M_\pi^2$. This is partly addressed with the unprojected integral equation, eq.~\eqref{eq:int_eq2}.
An alternative solution, again taking inspiration form refs.~\cite{\dfInspiration}, is to instead introduced a divergence-free scattering amplitude, defined as follows:
\begin{equation}
\mathcal M^{\sf df} (P, p, p') \equiv \mathcal M^{\sf aux} (P, p, p') - 2g^2 \mathcal T(P, p, p') \, .
\label{eq:df_amp}
\end{equation}
Projecting this quantity to spherical harmonics as in eq.~\eqref{eq:aux_proj} gives a faster converging series, down to the neighborhood of the second left-hand cut, i.e.~for $s \gtrsim (2M_N)^2 - (2M_\pi)^2$. We can substitute the divergence-free amplitude into eq.~\eqref{eq:int_eq2} to find
\begin{equation}
\mathcal M^{\sf df} (P, p, p') = \overline{\mathcal K}^{\sf os}(P, p, p') - \frac12 \int \! \frac{d^3 \boldsymbol k}{(2\pi)^3} \frac{H(\boldsymbol k^\star) \big [ \mathcal M^{\sf df} (P, p, k) - 2g^2 \mathcal T (P, p, k) \big ] {\mathcal K}^{\mathcal T}(P, k, p')}{4\omega_N(\boldsymbol k) \big[ (k_{\sf os}^\star)^2 - (\boldsymbol k^\star)^2 + i\epsilon \big]} \,.
\label{eq:int_eq3}
\end{equation}
The two terms in the numerator on the right-hand side (in square brackets) should be treated in practice as two separate integrals. The $\mathcal T$-dependent integral can be calculated numerically given knowledge of $\overline{\mathcal K}^{\sf os}(P, p, p')$ and $g$. We can then for $\mathcal M^{\sf df} (P, p, p')$ using standard integral-equation techniques, with the $\mathcal T$-dependent integral and $\overline{\mathcal K}^{\sf os}$ forming the driving term.

For completeness, we also provide the angular-momentum-projected version of eq.~\eqref{eq:int_eq3}:
\begin{align}
\begin{split}
\widetilde {\mathcal M}^{\sf df}_{\ell m, \ell' m'} (P, \vert \boldsymbol p^\star \vert, \vert \boldsymbol p'^\star \vert ) & = \overline{\mathcal K}^{\sf os}_{\ell m, \ell' m'}(P) \\[5pt]
& \hspace{-60pt} + \frac12 \int_0^\infty \! \frac{d \vert \boldsymbol k^\star \vert}{(2\pi)^3} \frac{4\pi \vert \boldsymbol k^\star \vert^{2\ell'' + 2} H(\boldsymbol k^\star) \, g^2 \mathcal T_{\ell m, \ell'' m''} (P, \vert \boldsymbol p^\star \vert, \vert \boldsymbol k^\star \vert) \, {\mathcal K}^{\mathcal T}_{\ell'' m'', \ell' m'} (P, \vert \boldsymbol k^\star \vert, \vert \boldsymbol p'^\star \vert)}{4\omega_N(\boldsymbol k^\star) \big[ (k_{\sf os}^\star)^2 - (\boldsymbol k^\star)^2 + i\epsilon \big]}
\, \\[5pt]
& \hspace{-60pt} - \frac12 \int_0^\infty \! \frac{d \vert \boldsymbol k^\star \vert}{(2\pi)^3} \frac{4\pi \vert \boldsymbol k^\star \vert^{2\ell'' + 2} H(\boldsymbol k^\star) \widetilde {\mathcal M}^{\sf df}_{\ell m, \ell'' m''} (P, \vert \boldsymbol p^\star \vert, \vert \boldsymbol k^\star \vert) \, {\mathcal K}^{\mathcal T}_{\ell'' m'', \ell' m'} (P, \vert \boldsymbol k^\star \vert, \vert \boldsymbol p'^\star \vert)}{4\omega_N(\boldsymbol k^\star) \big[ (k_{\sf os}^\star)^2 - (\boldsymbol k^\star)^2 + i\epsilon \big]}
\, .
\label{eq:int_eq4}
\end{split}
\end{align}
In this version we have separated out the driving term (first two lines) more explicitly.

Having obtained $\mathcal M^{\sf df} (P, p, p')$, one can recover the amplitude $\mathcal M (P, p, p')$ by combining eqs.~\eqref{eq:aux_to_amp} and \eqref{eq:df_amp}
\begin{equation}
\mathcal M (P, p, p') = \frac12 \left[ \mathcal M^{\sf df} (P, p, p') + \mathcal M^{\sf df} (P, p, P - p') \right] + g^2 \mathcal E(P, p, p') \, .
\end{equation}
We can see that, when putting external arguments on shell, the $\mathcal E(P, p, p')$ term (as defined in eq.~\eqref{eq:cal_E_definition}) is responsible for generating the $t$- and $u$-channel $\pi$ exchange poles present in the amplitude, and the divergence-free auxiliary amplitudes does not carry any singular behavior until one reaches the two-pion-exchange cut.

\subsection{Analytic continuation}

Evaluating the on-shell scattering amplitude below threshold requires analytically continuing the external momenta to be complex, such that $\vert \boldsymbol p^\star \vert = \vert \boldsymbol p'^\star \vert = i \sqrt{M_N^2 - s/4 \,}$. One method to achieve this is to first solve the integral equation \eqref{eq:int_eq3} for real external momenta, hence obtaining the amplitude, and then compute the sub-threshold on-shell values using the relation
\begin{align}
\begin{split}
\mathcal M (P, p, p') & = \overline{\mathcal K}^{\sf os}(P, p, p') + ig^2 \mathcal E(P, p, p') \\
& + \frac12 \int \! \frac{d^3 \boldsymbol k}{(2\pi)^3} \frac{H(\boldsymbol k^\star) \, {\mathcal K}^{\mathcal T}(P, p, k) \big [ \overline{\mathcal K}^{\sf os}(P, k, p') + ig^2 \mathcal E(P, k, p') \big ]}{4\omega_N(\boldsymbol k) \big[ (k_{\sf os}^\star)^2 - (\boldsymbol k^\star)^2 + i\epsilon \big]} \\
& \hspace{-30pt} - \frac14 \int\! \frac{d^3 \boldsymbol k}{(2\pi)^3} \int\! \frac{d^3 \boldsymbol k'}{(2\pi)^3} \frac{H(\boldsymbol k^\star) \, {\mathcal K}^{\mathcal T}(P, p, k) \, \mathcal M (P, k, k')}{4\omega_N(\boldsymbol k) \big[ (k_{\sf os}^\star)^2 - (\boldsymbol k^\star)^2 + i\epsilon \big]}
\frac{H(\boldsymbol k'^\star) {\mathcal K}^{\mathcal T}(P, k', p')}{4\omega_N(\boldsymbol k') \big[ ({k'}_{\sf os}^\star)^2 - ({\boldsymbol k'}^\star)^2 + i\epsilon \big]} \, .
\end{split}
\end{align}
The momenta $p$, $p'$, $P-p$ and $P-p'$ can be put on shell and taken below threshold here. The key point is that the amplitude on the right-hand side is nested between two loops and thus its arguments are always real. Knowledge of $\mathcal M$ for real momenta can thus be exploited to obtain its subthreshold on-shell value. These methods can similarly be applied to analytically continue the amplitude into the complex plane. Whether using the method above or directly trying to solve the integral equations for complex momenta, it is important to account for the singularities contained in the amplitude and $\mathcal K^{\mathcal T}$. Depending on the specific external momenta considered, these can cross the path of integration on the real line, requiring contour deformation.

\section{Exploring the new formalism}
\label{sec:explore}

In this section we study the new quantization condition and integral equations that we have derived. In section~\ref{sec:alternative_form}, we present an alternative form of the quantization condition that is superficially more similar to the standard formalism. Then, in section~\ref{sec:compare_to_luscher}, we demonstrate that the standard L\"uscher formalism is exactly recovered, also for $s < 4 M_N^2 - M_\pi^2$, in the limit where the $NN\pi$ coupling is set to zero. Finally, in section~\ref{sec:s_wave_dominance}, we give expressions in the case where all partial-wave components of $\overline{\mathcal K}^{\sf os}$, besides the $S$-wave, are negligible.

\subsection{Alternative form}
\label{sec:alternative_form}

An instructive rewriting of the quantization condition is reached by first noting that eq.~\eqref{eq:qcnew} is satisfied whenever the following matrix has a divergent eigenvalue:
\begin{equation}
\Xi_L(P,L) = \frac{1}{1 + \xi^\dagger \, \overline{\mathcal K}^{\sf os}(P) \, \xi \big [ S(P, L)^{-1} + 2 g^2 \mathcal T(P) \big ]^{-1}} \xi^\dagger \, \overline{\mathcal K}^{\sf os}(P) \, \xi \,.
\end{equation}
Expanding order by order and rearranging the factors of $\xi$, one finds that $\Xi_L(P,L) $ can be exactly rewritten as
\begin{equation}
\Xi_L(P,L) = \xi^\dagger \frac{1}{\overline{\mathcal K}^{\sf os}(P)^{-1} + \xi \big [ S(P, L)^{-1} + 2 g^2 \mathcal T(P) \big ]^{-1} \xi^\dagger} \xi \,.
\label{eq:qcnew_alt}
\end{equation}
We thus reach the following alternative to eq.~\eqref{eq:qcnew}:
\begin{equation}
\label{eq:new_ellm_qc}
\det_{\ell m} \left[ \overline{\mathcal K}^{\sf os}(P_j)^{-1} + F^{\mathcal T}(P_j, L) \right] = 0 \,,
\end{equation}
where we have introduced
\begin{align}
\label{eq:FT_def}
F^{\mathcal T}(P, L) = \xi S(P, L) \frac{1}{1 + 2 g^2 \mathcal T(P) S(P, L)} \xi^\dagger \,.
\end{align}

Crucially, this quantization condition is now defined as a determinant only on the orbital angular momentum, exactly as in refs.~\cite{Luscher:1986pf,Kim:2005gf}. We have included the $\ell m$ subscript on the determinant to emphasize this point. One nice feature of this representation is that it allows one to use the standard technology for projection to irreps.
See, for example, refs.~\cite{\groupTheory}.

This rewriting also holds for the case of particles with intrinsic spin. The only adjustment is that the spin indices remain in the final determinant condition, directly inherited from the quantities introduced in section~\ref{sec:spin}. We summarize this as
\begin{equation}
\label{eq:new_ellm_qc_spin}
\det_{\ell m r s} \left[ \overline{\mathcal K}^{\sf os}(P_j)^{-1} + F^{\mathcal T}(P_j, L) \right] = \det_{J m_J \ell S} \left[ \overline{\mathcal K}^{\sf os}(P_j)^{-1} + F^{\mathcal T}(P_j, L) \right] = 0 \,.
\end{equation}
Here we have indicated two of the possible bases for particles with spin. In the first expression, the labels $r$ and $s$ refer to the components of individual particle spin, as in eq.~\eqref{eq:S_def_spin}. For example, for two spin-half particles, the allowed values are given by $r,s = \pm 1/2$. The middle expression is labelled in terms of total angular momentum $J$ as well as orbital angular momentum $\ell$ and total spin $S$. To reach this form one uses Clebsch-Gordon coefficients to first combine the individual spin states to definite total spin, labelled as $S, m_S$. Then, in a second step, $\ell, m$ and $S, m_S$ are combined to $J, m_J, \ell, S$. This just amounts to changes of basis on the matrices, as described, for example, in refs.~\cite{Briceno:2014oea,Briceno:2015csa}. As in the spin-zero cases, the projections to definite irreps is inherited from the standard formalism.

Also common to the standard approach is the mixing of angular momenta due to the reduced symmetry of the finite-volume system. This means that, even after projection to a given irrep, the matrices entering the determinant are formally infinite dimensional. As a result, numerical evaluation is only possible with truncation: one sets $\overline{\mathcal K}^{\sf os}_{\ell m \ell' m'} = 0$ for $\ell , \ell' > \ell^{\sf max}$. Having truncated $\overline{\mathcal K}^{\sf os}$ in this way, one can do the same for $F^{\mathcal T}(P,L)$ without any additional approximations. It is, however, an additional approximation to truncate $S(P,L)$ and $\mathcal T(P)$ within $F^{\mathcal T}(P,L)$.

Further work is needed to investigate the numerical convergence in the evaluation of components of $F^{\mathcal T}(P,L)$ as a function of the truncations of its building blocks. At the same time, we view the separation of the quantization condition into $\overline{\mathcal K}^{\sf os}(P)$, $S(P,L)$, and $\mathcal T(P)$ as optimal, given the expectations of underlying physical system. $\overline{\mathcal K}^{\sf os}(P)$ is a generic analytic function and requires parametrization to extract scattering predictions from finite-volume energies. In that sense, this is the most challenging input and the one for which partial-wave convergence is most important. For this reason it is important that the single-exchange left-hand cut is removed from $\overline{\mathcal K}^{\sf os}(P)$, such that this object does not suffer the partial wave convergence problems of the amplitude itself.

By contrast, while the convergence for $S(P,L)$ and $\mathcal T(P)$ is less obvious, no new free parameters are introduced if one varies the size of these quantities. One can thus envision cases where only the $S$-wave component of $\overline{\mathcal K}^{\sf os}(P)$ is kept but many more components of the other factors. This will still lead to an $(n+1)$-parameter description of the finite-volume energies, with $n$ being the number of free parameters used in the $S$ component of $\overline{\mathcal K}^{\sf os}(P)$.

\subsection{Recovering the standard formalism}
\label{sec:compare_to_luscher}

We now show that, in the case of vanishing trilinear coupling: $g = 0$, our modified quantization condition, eq.~\eqref{eq:qcnew}, is equivalent to the well-known L\"uscher scattering formalism \cite{Luscher:1986pf} and its extensions to nonzero spatial momentum in the finite-volume frame \cite{Rummukainen:1995vs,Kim:2005gf}. To this end it is convenient to use the alternative form given in eq.~\eqref{eq:qcnew_alt}.
In the $g \to 0$ limit, the result becomes
\begin{equation}
\det_{\ell m} \left[ \overline{\mathcal K}^{\sf os}(P_j)^{-1} + \xi S(P_j, L) \xi^\dagger \right] = 0 \,,
\label{eq:qcnew_gzero_alt}
\end{equation}
where we have substituted $F^{\mathcal T}(P, L) \underset{g \to 0}{\longrightarrow} \xi S(P, L) \xi^\dagger$, as follows directly from eq.~\eqref{eq:FT_def}.

We next relate $\xi S(P, L) \xi^\dagger$ to a version of the L\"uscher finite-volume function, denoted here by $\widetilde F(P,L)$.%
\footnote{As above, the tilde here indicates rescaling by powers of $k^\star_{\sf os}$. We do not use the tilde for $S(P,L)$ since this is a non-standard quantity anyway, as has been defined immediately with the rescaling.}
The key distinction between $\xi S(P, L) \xi^\dagger$ and $\widetilde F(P,L)$ is simply that the latter is defined with a sum-integral difference while $\xi S(P, L) \xi^\dagger$ is defined only with a sum in isolation [see also eq.~\eqref{eq:sum_over_s}]. Studying the definitions carefully, one finds that the relation takes the form
\begin{align}
\xi S(P, L) \xi^\dagger & = \widetilde F(P,L) + \widetilde I(P) \,,
\label{eq:xiSxi_FI}
\end{align}
where each of the quantities carries two set of angular-momentum indices, and we have introduced
\begin{equation}
\widetilde I_{\ell m, \ell' m'}(P) = {\sf p.v.} \frac12 \int \frac{d^3 \boldsymbol k}{(2\pi)^3} \, \frac{4\pi \, Y_{\ell m}(\hat{\boldsymbol k}^\star) \, Y^*_{\ell' m'}(\hat{\boldsymbol k}^\star) \, \vert \boldsymbol k^\star \vert^{\ell + \ell'} \, H(\boldsymbol k^\star)}{4 \omega_N(\boldsymbol k) \, \big[ (k_{\sf os}^\star)^2 - (\boldsymbol k^\star)^2 \big]} \,.
\label{eq:Ipv_def}
\end{equation}
This implicit definition of $\widetilde F(P,L)$ exactly matches the quantity denoted $\widetilde F(E, \boldsymbol P, L)$ in ref.~\cite{Peterken:2023zwu}. See e.g.~eq.~(A1) of that work. Equivalently, it is related to the finite-volume function defined by Kim, Sachrajda, and Sharpe in ref.~\cite{Kim:2005gf} according to $\widetilde F_{\ell m, \ell' m'} = (k^\star_{\sf os})^{\ell + \ell'} \text{Re}(i F^{\sf KSS}_{\ell m, \ell' m'}/2)$, where we have added the superscript ${\sf KSS}$ to distinguish the earlier definition.

The final missing ingredient in recovering the standard scattering formalism from eq.~\eqref{eq:qcnew_gzero_alt} is the relation between $\overline{\mathcal K}^{\sf os}(P)$ and the scattering amplitude. This can mostly easily be recovered from the integral equation, eq.~\eqref{eq:int_eq1}, which simplifies in three ways for $g = 0$: First, the breaking of exchange symmetry in $\mathcal M^{\sf aux}$ is removed so that this just directly becomes the scattering amplitude. Second, the off-shell dependence of $\mathcal M^{\sf aux}$ on $\vert \boldsymbol{k}^\star \vert$, which is inherited from $\mathcal{T}$, is removed when the coupling vanishes. Finally, we can replace $\mathcal K^{\mathcal T} \to \overline{\mathcal{K}}^{\sf os}$ everywhere, since the difference between the two is proportional to $g^2$. We reach
\begin{multline}
\widetilde {\mathcal M}_{\ell m, \ell' m'} (P) = \overline{\mathcal K}^{\sf os}_{\ell m, \ell' m'}(P)
\\
- \widetilde {\mathcal M}_{\ell m, \ell'' m''} (P) \frac12 \int_0^\infty \! \frac{d \vert \boldsymbol k^\star \vert}{(2\pi)^3} \frac{4\pi \vert \boldsymbol k^\star \vert^{2\ell'' + 2} H(\boldsymbol k^\star)}{4\omega_N(\boldsymbol k^\star) \big[ (k_{\sf os}^\star)^2 - (\boldsymbol k^\star)^2 + i\epsilon \big]} \overline{\mathcal K}^{\sf os}_{\ell'' m'', \ell' m'} (P ) \,.
\end{multline}
Switching to a matrix notation to drop indices and multiplying both sides by $\mathcal M(P)^{-1} $ on the right and $\overline{\mathcal K}^{\sf os}(P)^{-1}$ on the left, we find
\begin{equation}
\overline{\mathcal K}^{\sf os}(P)^{-1} = \widetilde{\mathcal M}(P)^{-1} - \widetilde I^{i \epsilon}(P) \,,
\label{eq:Kinvos_Minv}
\end{equation}
where we have introduced
\begin{equation}
\widetilde I^{i \epsilon}_{\ell m, \ell' m'}(P) = \delta_{\ell \ell'} \delta_{mm'} \frac12 \int_0^\infty \! \frac{d \vert \boldsymbol k^\star \vert}{(2\pi)^3} \frac{4\pi \vert \boldsymbol k^\star \vert^{2\ell + 2} H(\boldsymbol k^\star)}{4\omega_N(\boldsymbol k^\star) \big[ (k_{\sf os}^\star)^2 - (\boldsymbol k^\star)^2 + i\epsilon \big]} \,.
\label{eq:Iieps_def}
\end{equation}
The notation for this quantity is well-chosen. As can be seen by evaluating the angular integrals in eq.~\eqref{eq:Ipv_def}, $\widetilde I(P)$ and $\widetilde I^{i \epsilon}(P)$ differ only in the pole prescription.

In particular, since the principal value defining $\widetilde I(P)$ is equivalent to the real part of the $i \epsilon$ prescription, the difference in integrals is just the imaginary part of $\widetilde I^{i \epsilon}(P)$
\begin{align}
\widetilde I^{i \epsilon}_{\ell m, \ell' m'}(P) - \widetilde I_{\ell m, \ell' m'}(P) & = \delta_{\ell \ell'} \delta_{mm'} i \, \text{Im} \frac12 \int_0^\infty \! \frac{d \vert \boldsymbol k^\star \vert}{(2\pi)^3} \frac{4\pi \vert \boldsymbol k^\star \vert^{2\ell + 2} H(\boldsymbol k^\star)}{4\omega_N(\boldsymbol k^\star) \big[ (k_{\sf os}^\star)^2 - (\boldsymbol k^\star)^2 + i\epsilon \big]} \,, \\[5pt]
& = - \delta_{\ell \ell'} \delta_{mm'} i \rho(s) (k_{\sf os}^\star)^{2\ell}
\label{eq:Iieps_I_diff}
= - i \widetilde \rho_{\ell m, \ell' m'}(s) \,,
\end{align}
where $\rho(s)$
is the usual phase space, defined in eq.~\eqref{eq:rhoN_pN_def} above, as is $ \widetilde \rho_{\ell m, \ell' m'}(s)$ in eq.~\eqref{eq:rho_tilde_def}.

It is exactly this phase space factor that appears in the relation between the scattering amplitude $\widetilde {\mathcal M}$ and the standard two-particle K-matrix
\begin{equation}
\widetilde {\mathcal K}(P)^{-1} = \widetilde {\mathcal M}(P)^{-1} + i \widetilde \rho(P) \,.
\label{eq:Kinv_Minv}
\end{equation}
For convenience we also repeat the relation to the scattering phase shift given in eq.~\eqref{eq:Ktilde_delta_relation}
\begin{equation}
\widetilde {\mathcal K}_{\ell m, \ell' m'}(P) = \frac{1}{(k_{\sf os}^\star)^{2\ell}} \delta_{\ell \ell'} \delta_{mm'} \frac{16 \pi \sqrt{s}}{k_{\sf os}^\star} \tan \delta^{(\ell)}(k_{\sf os}^\star) \,.
\end{equation}

Finally, combining eqs.~\eqref{eq:Kinvos_Minv}, \eqref{eq:Iieps_I_diff} and \eqref{eq:Kinv_Minv}, we deduce
\begin{align}
\overline{\mathcal K}^{\sf os}(P)^{-1} & = \widetilde {\mathcal K}(P)^{-1} - \widetilde I(P) \,.
\end{align}
Together with eq.~\eqref{eq:xiSxi_FI}, this is the second key identity of this subsection.
Substituting the identities into eq.~\eqref{eq:qcnew_gzero_alt}, one finally obtains
\begin{equation}
\det_{\ell m} \left[ \widetilde {\mathcal K}(P_j)^{-1} - \widetilde I(P_j) + \widetilde F(P_j,L) + \widetilde I(P_j) \right] = \det_{\ell m} \left[ \widetilde {\mathcal K}(P_j)^{-1} + \widetilde F(P_j,L) \right] = 0 \,.
\end{equation}
This is the standard quantization condition of L\"uscher, generalized to nonzero momentum $\boldsymbol P$ in the finite-volume frame \cite{Luscher:1986pf,Rummukainen:1995vs,Kim:2005gf}. We have thus achieved the aim of this section.

Before concluding, we briefly return to the case of $g \neq 0$. If we restrict our attention to energies above the left-channel cut, i.e.~with $s > (2M_N)^2 - M_\pi^2$, then we can ignore the presence of the latter in the on-shell Bethe-Salpeter kernel, up to exponentially suppressed $L$-dependence, albeit dependence that can be enhanced if we are too close to saturating the inequality. In this case, one can formally group the $g$ dependent term into $\overline B^{\mathfrak T}$ and apply an effective $g = 0$ analysis to again deduce the usual formulas. It would be instructive to analytically recover the standard formalism from our results in the above threshold regime, and to quantify the neglected exponentially suppressed corrections. We leave this for future work.

As mentioned above, our perspective is that, in a numerical calculation, one should ensure that there are no statistically significant differences between the scattering predictions arising from the standard and the improved formalisms and, if there are, to use the latter.

\subsection{\texorpdfstring{$S$}{S}-wave dominance}
\label{sec:s_wave_dominance}

In this subsection, we consider the form of the quantization condition and the integral equations in the case that only the $S$-wave component of $\overline{\mathcal K}^{\sf os}$ is non-zero. Beginning with the quantization condition, eq.~\eqref{eq:new_ellm_qc} becomes
\begin{equation}
\overline{\mathcal K}^{{\sf os}(0)}(P_j)^{-1} + F_0^{\mathcal T}(P_j, L) = 0 \,,
\end{equation}
where we have introduced the $S$-wave quantities
\begin{align}
\overline{\mathcal K}^{{\sf os}(0)}(P_j) & = \overline{\mathcal K}_{00,00}^{\sf os}(P_j) \,, \\
F^{\mathcal T}_0(P, L) & = Z(P,L) - V^\dagger (P,L) \frac{1}{1 + 2 g^2 \mathcal T(P) S(P, L)} 2 g^2 \mathcal T(P) V(P,L) \,,
\end{align}
with the latter expressed in terms of two new building blocks, a scalar function $Z(P,L)$ and a vector function $V(P,L)$:
\begin{align}
Z(P,L) & \equiv [\xi S(P, L) \xi^\dagger]_{00,00} = \frac{1}{2L^3} \sum_{\boldsymbol k} \frac{e^{- \alpha [ (\boldsymbol k^\star)^2 - (k_{\sf os}^\star)^2]}}{4 \omega_N(\boldsymbol k) \, \big [ (k_{\sf os}^\star)^2 - (\boldsymbol k^\star)^2\big ]} \,, \\
V^\dagger_{\boldsymbol k'^\star \ell' m'}(P,L) & \equiv \xi_{\boldsymbol k^\star} S_{\boldsymbol k^\star 0 0 , \boldsymbol k'^\star \ell' m'} (P, L) = \frac{1}{2L^3} \, \frac{\sqrt{4\pi} \, Y^*_{\ell' m'}(\hat{\boldsymbol k'}^\star) \, \vert \boldsymbol k'^\star \vert^{\ell'} \, e^{- \alpha [ (\boldsymbol k'^\star)^2 - (k_{\sf os}^\star)^2]}}{4 \omega_N(\boldsymbol k') \, \big [ (k_{\sf os}^\star)^2 - (\boldsymbol k'^\star)^2\big ]} \,.
\end{align}
The key observation here is that $V(P,L)$ is still populated with all angular momentum components, even though $F^{\mathcal T}(P, L) $ has been truncated to the $S$-wave. This is because the exchanges, encoded in $\mathcal T(P)$, depend on all partial waves, and the truncation of this object is logically separate from the truncation of the quantization condition. In practice, the approach is to evaluate $F^{\mathcal T}_0(P, L)$ for various truncations of $\mathcal T(P)$ and look for saturation. The convergence will depend on the values of $g, P, L$. Further investigation is needed to understand this in detail.

To give one explicit example, we consider the case of zero total momentum and the trivial irrep of the finite-volume symmetry group, called $A_1^+$. Then the lowest-lying contaminating partial wave is $\ell = 4$ and the $m$ component is removed by irrep projection such that one can define a two-dimensional angular momentum space. The resulting building blocks for $F^{\mathcal T}_0(P, L) $ are then given by
\begin{align}
V^\dagger_{\boldsymbol k}(P,L) & = \frac{1}{2L^3} \, \frac{\, e^{- \alpha [ (\boldsymbol k)^2 - (k_{\sf os})^2]}}{4 \omega_N(\boldsymbol k) \, \big [ (k_{\sf os})^2 - (\boldsymbol k)^2\big ]} \Big (\begin{matrix} 1 \ \ & \ \ \mathcal Y_4^{A_1^+}(\boldsymbol k) \end{matrix} \Big ) \,, \\[5pt]
\mathcal T_{\boldsymbol k, \boldsymbol k'}(P) & =
\begin{pmatrix}
\mathcal T^{(\ell=0), A_1^+}_{\boldsymbol k, \boldsymbol k'}(P) & 0 \\ 0 & \mathcal T^{(\ell=4), A_1^+}_{\boldsymbol k, \boldsymbol k'}(P)
\end{pmatrix} \,, \\[5pt]
S_{\boldsymbol k , \boldsymbol k'} (P, L) & = \frac{1}{2L^3} \, \frac{\delta_{\boldsymbol k \boldsymbol k'} \, e^{- \alpha [ (\boldsymbol k)^2 - (k_{\sf os})^2]}}{4 \omega_N(\boldsymbol k) \, \big [ (k_{\sf os})^2 - (\boldsymbol k)^2\big ]} \begin{pmatrix} 1 \\[5pt] \mathcal Y_4^{A_1^+}(\boldsymbol k) \end{pmatrix} \Big (\begin{matrix} 1 \ \ & \ \ \mathcal Y_4^{A_1^+}(\boldsymbol k) \end{matrix} \Big ) \,,
\end{align}
where we have introduced the $A_1^+$-projected $G$-wave spherical harmonic:
\begin{equation}
\mathcal Y_4^{A_1^+}(\boldsymbol k) = \frac{\sqrt{21}}{4} \Big(5 (k_x^{4} + k_y^{4} + k_z^{4}) - 3 (\boldsymbol k^2)^2 \Big) \,.
\end{equation}
This is also used to define the $G$-wave component of $\mathcal T(P)$, which is given explicitly by
\begin{multline}
\mathcal T^{(\ell=4), A^+_1}_{\boldsymbol k ,\boldsymbol k'}(P) = \frac{21}{16 (4 \pi )^2 \vert \boldsymbol k \vert^4 \vert \boldsymbol k' \vert^4} \int \! d \Omega_{\hat {\boldsymbol n}} \int \! d \Omega_{\hat {\boldsymbol n}}' \,
\\ \times
\frac{\big (5 ( \hat {n}_x^4 + \hat {n}_y^4 + \hat {n}_z^4 ) - 3 \big ) \big (5 ( \hat {n}'^4_x + \hat {n}'^4_y + \hat {n}'^4_z ) - 3 \big )}{2\omega_N(\boldsymbol k) \omega_N(\boldsymbol k') - 2 \vert \boldsymbol k \vert \vert \boldsymbol k' \vert \hat {\boldsymbol n} \cdot \hat {\boldsymbol n}' - 2M_N^2 + M_\pi^2 - i\epsilon}
\,,
\end{multline}
where $\hat {\boldsymbol n} = (\sin \theta \cos \phi, \sin \theta \sin \phi, \cos \theta)$ and similar with primed coordinates. One can evaluate the integral to find
\begin{align}
\mathcal T^{(\ell=4), A^+_1}_{\boldsymbol k ,\boldsymbol k'}(P) = \frac{\mathcal G \Big ( \vert \boldsymbol k \vert/M_N, \vert \boldsymbol k' \vert/M_N \Big )}{M_N^{10}} \,,
\end{align}
where we have introduced
\begin{align}
\begin{split}
\mathcal G(x,y) & = \frac{5 F(x, y) (-21 F(x, y)^2 + 44 x^2 y^2)}{384 x^8 y^8} \\
& \hspace{20pt} + \frac{35 F(x, y)^4 - 120 x^2 y^2 F(x, y)^2 + 48x^4 y^4}{512 x^9 y^9} \log \left[ \frac{F(x, y) - x y}{F(x, y) + x y} \right]
\,, \label{eq:G_def}
\end{split}\\[5pt]
F(x, y) & = -2 \sqrt{1 + x^2} \sqrt{1 + y^2} + 2 - (M_\pi/M_N)^2 + i\epsilon \,.
\end{align}
The second function also allows us to express the $S$-wave component in eq.~\eqref{eq:tchan_swave} in a more compact form:
\begin{equation}
\mathcal T^{(\ell=0), A^+_1}_{\boldsymbol k ,\boldsymbol k'}(P) = \frac{1}{M_N^2} \frac{1}{4 x y} \log \left [ \frac{F(x, y) - x y}{F(x, y) + x y} \right ] \,.
\end{equation}

These expressions can be readily extended to higher dimensions, simply by working out the trivial irrep harmonics and the integrals of the latter entering $\mathcal T(P)$. In this way one can reliably estimate $F^{\mathcal T}_0(P, L)$ and thereby determine the relation between $\overline{\mathcal K}_0^{\sf os}(P)$, in a given parametrization, and the finite-volume energies on the left-hand cut.

The second step of the procedure is then using the $S$-wave determined $\overline{\mathcal K}_0^{\sf os}(P)$ together with $g$ to solve for scattering amplitude. The strategies presented in refs.~\cite{\intEquationStrats} will clearly be useful here. We leave a detailed numerical exploration to future work.

\subsection{Comparison to previous work}
\label{sec:previous_work}

In this subsection, we briefly compare our results with those of refs.~\cite{Sato:2007ms,Meng:2021uhz}.

We begin with ref.~\cite{Sato:2007ms}, one of the first papers to emphasize the importance of single-pion exchange in exponentially suppressed volume corrections to the L\"uscher quantization condition. The aim of the publication is to point out that such effects may be large enough to bias the extracted phase shifts for sufficiently small volumes, and to use non-relativistic nuclear chiral effective field theory (EFT) to numerically estimate the size of the corrections. In contrast to our formalism, ref.~\cite{Sato:2007ms} uses a version of the Bethe-Salpeter kernel that is given solely by the single-pion exchange. This is referred to as a potential and denoted by $V$. Comparing eq.~(32) of the earlier work with our eq.~\eqref{eq:tchan_swave}, one can see that our two approaches are brought closer by setting $\overline B = 0$ in our derivation and taking the non-relativistic expansion.

The additional volume effects identified in ref.~\cite{Sato:2007ms} then correspond to the sum-integral-difference acting on a difference between the on-shell and off-shell versions of $V$, contained in a quantity called $\mathbb F$, defined in eq.~(20) of ref.~\cite{Sato:2007ms}. The authors assume that the finite-volume energies are not on the left-hand cut, so $\mathbb F$ is real-valued, and this leads to real-valued exponentially suppressed corrections to the standard L\"uscher scattering formalism. Other differences from our approach include the neglect of partial-wave mixing in the finite volume, the restriction to interactions described by the specific EFT, and the fact that the former work restricts attention to (and provides numerical estimates for) the two-nucleon system with physical masses.

The more recent work of ref.~\cite{Meng:2021uhz} advocates using a plane-wave basis, as opposed to the partial-wave basis typically used for two-particle quantization conditions. Also here, the authors suggest using an EFT-based framework to describe the one-pion exchange, and show that applying this in the plane-wave basis circumvents the partial-wave mixing effects by providing an alternative approach to treat the broken rotational symmetry of the finite volume.

The main difference between our approach and that of ref.~\cite{Meng:2021uhz} is that our quantization condition [given in eq.~\eqref{eq:qcnew}] is defined over a combined plane-wave (indexed by $\boldsymbol k^\star, \boldsymbol k'^\star$) and angular momentum (indexed by $\ell m, \ell' m'$) basis. Our alternative, equivalent form of the quantization condition [given in eq.~\eqref{eq:new_ellm_qc}] is defined by a determinant only over the angular momentum space. Importantly, both versions of our main result depend on the same K-matrix analog, the quantity we call $\overline {\mathcal K}^{\sf os(\ell)}(s)$, and this does not depend on the magnitudes or directions of the plane-wave basis indices $\boldsymbol k^\star$ and $\boldsymbol k'^\star$. Implementing our method in practice will require the use of various models or parameterizations of $\overline {\mathcal K}^{\sf os(\ell)}(s)$, which will include tests to ensure that a sufficient set of partial wave components are included so that the neglected higher partial waves are numerically irrelevant. Such tests have already been performed in ref.~\cite{Meng:2021uhz}, and it will be instructive to quantitatively compare the results.

The above summary concerns qualitative similarities and differences between the methods, and a direct numerical test and quantitative comparison is still outstanding and urgently needed. This is left to future work, as is a more detailed theoretical exploration of regimes in which the formulas are equivalent.

\section{Conclusions}
\label{sec:conclusions}

In this work, we have addressed the breakdown of the L\"uscher scattering formalism \cite{Luscher:1986pf} and extensions \cite{\twoScatter} for finite-volume energies that arise far enough below threshold to approach (or overlap) the left-hand branch cut from single meson exchange. The initial motivation for this investigation was a lattice calculation \cite{Green:2021qol} of baryon-baryon scattering in the $H$-dibaryon sector with quark masses at the $SU(3)$-flavor-symmetric point. The authors of that study found finite-volume energies on the left-hand cut and highlighted the need for an extension of finite-volume methods in order to make use of these energies. More recently, the same issue was pointed out in lattice calculations of $DD^*$ scattering, investigating the doubly charmed tetraquark $T_{cc}(3875)^+$ \cite{\TccCollection}.

To address this issue, we have proposed an alternative approach to the standard procedure, allowing one to extract the physical amplitude in the region of the cut. The detailed relations presented here cover the case of a single channel of identical scattering particles, denoted $N$, exchanging a lighter particle, denoted $\pi$, in the $t$- and $u$-channels.

In particular, in section~\ref{sec:FVformalism}, we have derived an alternative quantization condition, summarized in eq.~\eqref{eq:qcnew}, which is valid both in the elastic regime, and for subthreshold energies down to the first two-particle-exchange cut. The quantization condition allows one to extract an intermediate infinite-volume quantity, denoted $\overline{\mathcal K}^{\sf os}$, together with the coupling governing the cut, denoted $g$. We have additionally derived integral equations in section~\ref{sec:int_eqs} that allow one to relate $\overline{\mathcal K}^{\sf os}$ to the physical scattering amplitude. This gives a complete workflow to extract the scattering amplitude from the finite-volume spectrum in an extended kinematic region.

As with much of the extensive literature relating finite-volume data to amplitudes, our derivations are based on a generic all-orders EFT of nucleons and pions with interactions only constrained by symmetries. Expressions are given for any spin of the nucleon, with emphasis on the toy spin-zero and the physical spin-half cases. We have made an effort to detail the steps that fail in the presence of subthreshold cuts, and to illustrate and motivate the adaptations required to solve the issue. To summarize, two key ingredients underpin our approach: the separation of the Bethe-Salpeter kernel into dangerous pion exchanges and a safe remainder, and the use of partial off-shellness as a tool for correctly representing the finite-volume correlator, in particular the fact that it does not contain left-hand cuts.

Further developments, already underway, include the generalization of the method to non-identical particles, non-degenerate masses and coupled channels. The relation to the three-particle finite-volume formalism is also being explored, e.g.~whether this approach can be recovered as a limiting case of the three-particle finite-volume relations. $DD^*$ scattering provides a good testing ground for these extensions, with studies from the three-particle perspective already ongoing.

More ambitious future work could involve the treatment of lower left-hand cuts arising from multi-particle exchanges, such as that shown in figure~\ref{fig:various_examples}(b), thereby extending the range of validity of the approach to lower energies, should these be extracted in a given system. Applications to real lattice data are also planned. In particular, an analysis to the baryon-baryon data of ref.~\cite{Green:2021qol} is possible immediately given the results of this work.

\acknowledgments

We thank
Ra{\'u}l Brice{\~n}o,
John Bulava,
Zohreh Davoudi,
Will Detmold,
Evgeny Eppelbaum,
Andrew Jackura,
Fernando Romero-L\'opez,
Srijit Paul,
and
Steve Sharpe
for useful discussions. We especially thank Steve Sharpe for a detailed reading of an earlier version of this manuscript. Both ABR and MTH and supported by UK STFC grant ST/P000630/1. MTH is additionally supported by UKRI Future Leader Fellowship MR/T019956/1.

\appendix

\section{Analyticity of the Bethe-Salpeter kernel}
\label{app:BS_kernel_sing}

In this appendix we give additional details concerning the conditions under which the Bethe-Salpeter kernel is an analytic function of $E$ with exponentially suppressed volume effects in its finite-volume analog.

The discussion is broken into two subsections: First we analyze the diagram shown in figure~\ref{fig:single_u_channel_loop}(a) from various perspectives to illustrate an important and subtle point about diagrams with $u$-channel-like momentum routing. We show that, in our energy range of interest: $(2 M_N)^2 - (2 M_\pi)^2 < s < (2 M_N + M_\pi)^2$, this diagram has no singularities besides that associated with the two-particle $s$-channel cut. At the same time, we find that applying the analysis of section~\ref{sec:loop_cont} to this diagram does lead to neglected singularities inside the term denoted $\mathcal I^{[2]}(P)$, together with spurious singularities in the terms that are kept explicit. This is an artifact of the partially off-shell kinematics and is removed once all external legs of the Bethe-Salpeter kernel are set on shell. The issue must nonetheless be addressed, since our derivation uses properties of the partially off-shell Bethe-Salpeter kernel to reach the final result.

After demonstrating the issue in detail for the diagram of figure~\ref{fig:single_u_channel_loop}(a), we explain how the result generalizes to all contributions to the Bethe-Salpeter kernel. The basic insight is that a diagram with a $u$-channel subdiagram is invariant under the replacement $k \to P - k$, which converts it to a $t$-channel subdiagram. In this sense, the distinction between $u$- and $t$-channel is artificial and relies on the particular choice of internal momentum routing. This ultimately leads to the definition of $B^{\mathfrak T}$, an intermediate quantity that resolves the issue. When the partially off-shell kinematics are projected fully on shell, $B^{\mathfrak T}$ can be safely symmetrized to the usual kernel. As a result, our final expressions are not affected by these issues.

\subsection{The \texorpdfstring{$u$}{u}-channel loop}

\begin{figure}
\centering
\includegraphics[width=\textwidth]{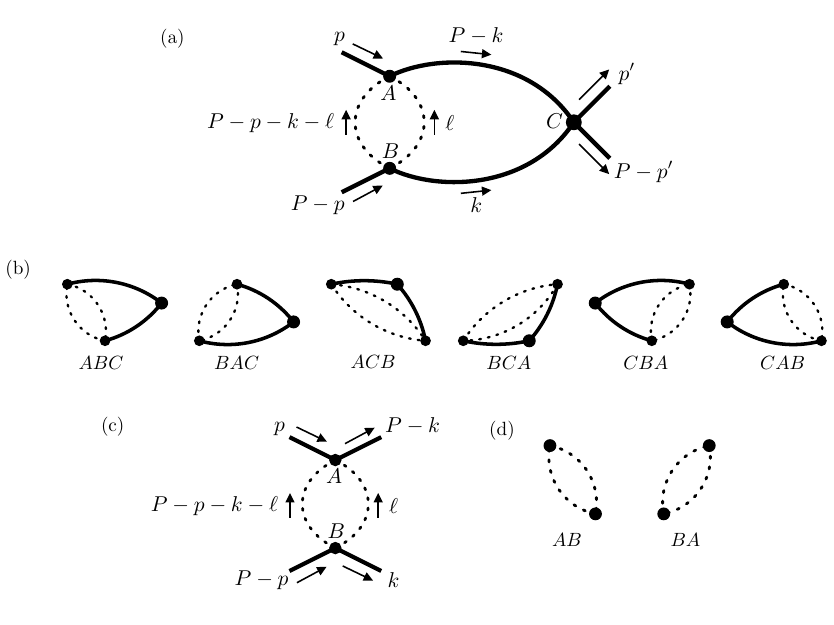}
\caption{(a) $u$-channel-loop diagram with momentum routing that causes problems in the standard analysis of section~\ref{sec:loop_cont}; (b) the time-ordered perturbation theory (TOPT) contributions to the diagram of (b); (c) the $u$-channel contribution to the Bethe-Salpeter kernel in isolation; (d) TOPT contributions corresponding to (c).
\label{fig:single_u_channel_loop}}
\end{figure}

We begin by using time-ordered perturbation theory (TOPT) to directly identify the singularities that arise in the full diagram shown in figure~\ref{fig:single_u_channel_loop}(a). For concreteness, we envision evaluating the diagram in a finite volume, such that all spatial loop momenta are summed. The corresponding infinite-volume behavior can be inferred by replacing the sums at any stage with an integral, together with a pole prescription. The external kinematics of the diagram are evaluated with $p$ on shell: $p^0 = \omega_N(\boldsymbol p)$, but with $P - p = (E - \omega_N(\boldsymbol p), \boldsymbol P - \boldsymbol p)$ generally off shell.

As is shown in figure~\ref{fig:single_u_channel_loop}(b), six time orderings contribute. The corresponding singularities are given by multiplying factors of $[E_{\sf cut} - \sum_i \omega_i]^{-1}$, one for each vertical cut, before summing over time orderings. Here $E_{\sf cut}$ is the energy flowing across the cut and $\sum_i \omega_i$ the sum of on-shell energies for the internal propagators that intersect the cut. Following ref.~\cite{sterman_1993}, $E_{\sf{cut}}$ is defined as the sum of inflowing $p^0$ components into all vertices appearing to the left of the cut. So, for example, when only vertex $B$ appears to the left, then $E_{\sf{cut}} = E - \omega_N(\boldsymbol p)$.

Studying \ref{fig:single_u_channel_loop}(b), we note that only two of the six time orderings have the $E - \omega_N(\boldsymbol k) - \omega_N(\boldsymbol P - \boldsymbol k)$ singularity leading to the relevant two-particle pole. These are the first two diagrams, labeled $ABC$ and $BAC$, respectively. The full singularity structure of the two orderings is
\begin{align}
\!\!\!\![ABC] & =
\frac{1}{\omega_N(\boldsymbol p) - \omega_N(\boldsymbol P - \boldsymbol k) - \omega_\pi(\boldsymbol P - \boldsymbol p - \boldsymbol k - \boldsymbol \ell) - \omega_\pi(\boldsymbol \ell)}
\frac{1}{E - \omega_N(\boldsymbol k) - \omega_N(\boldsymbol P - \boldsymbol k)} \,, \\
\!\!\!\![BAC] & =
\frac{1}{E - \omega_N(\boldsymbol p) - \omega_N( \boldsymbol k) - \omega_\pi(\boldsymbol P - \boldsymbol p - \boldsymbol k - \boldsymbol \ell) - \omega_\pi(\boldsymbol \ell)}
\frac{1}{E - \omega_N(\boldsymbol k) - \omega_N(\boldsymbol P - \boldsymbol k)} \,,
\end{align}
where we have introduced $[ABC]$, etc.~as a shorthand, representing the poles generated by a given vertex ordering.

Crucially, both the $[ABC]$ and $[BAC]$ have no singularities in the relevant kinematic range, besides that at $E = \omega_N(\boldsymbol k) + \omega_N(\boldsymbol P - \boldsymbol k)$. However, if we consider the simple $u$-channel loop of figure~\ref{fig:single_u_channel_loop}(c) and the corresponding time orderings of figure~\ref{fig:single_u_channel_loop}(d) in isolation, we instead get the following contributions:
\begin{align}
[AB] & = \frac{1}{\omega_N(\boldsymbol p) - E + \omega_N(\boldsymbol k) - \omega_\pi(\boldsymbol P - \boldsymbol p - \boldsymbol k - \boldsymbol \ell) - \omega_\pi(\boldsymbol \ell)} \,, \\
[BA] & = \frac{1}{E - \omega_N(\boldsymbol p) - \omega_N(\boldsymbol k) - \omega_\pi(\boldsymbol P - \boldsymbol p - \boldsymbol k - \boldsymbol \ell) - \omega_\pi(\boldsymbol \ell)} \,.
\end{align}
In these one-loop diagrams, the notion of external momentum has changed. For example, vertex $A$ now has the external momenta $p - P +k$ flowing into it. As a result, the cut in $[AB]$ has $E_{\sf cut} =\omega_N(\boldsymbol p) - E + \omega_N(\boldsymbol k) $.
We see that, while $[BA]$ exactly matches the factor appearing within $[BAC]$, the $[AB]$ contribution differs. The reason this matters is that the $[AB]$ form is what appears after $k^0$ integration is applied to the general two-particle loop in section \ref{sec:loop_cont}. See, in particular, eq.~\eqref{eq:Cloop_intm}, which is repeated in the paragraph after next.

The first factor in $[ABC]$ can be converted to $[AB]$ via the replacement $\omega_N(\boldsymbol P - \boldsymbol k) \to E - \omega_N(\boldsymbol k)$. Thus, the difference between the two is something that vanishes whenever $E = \omega_N(\boldsymbol k) + \omega_N(\boldsymbol P - \boldsymbol k)$. Such replacements are often valid in deriving finite-volume relations, since they amount to neglecting a difference with exponentially suppressed $L$ dependence.
In this case the replacement is invalid, since the form of $[AB]$ in isolation has unphysical singularities. For example, in the case where $\boldsymbol P = \boldsymbol 0$ and $\boldsymbol p = - \boldsymbol k$, the $[AB]$ factor has a pole at $E = 2 \omega_N(\boldsymbol p) - 2 \omega_\pi(\boldsymbol \ell)$. There is no pole in the original diagram at this location; the singularity is only an artifact of setting the Bethe-Salpeter kernel to partially off-shell kinematics.

An alternative way to identify the same complication is to track what goes wrong in the derivation of section~\ref{sec:loop_cont}, for this type of diagram. The problem is the step going from eq.~\eqref{eq:loop_cont2},
\begin{equation*}
\mathcal L \, \circ_{\sf fv} \, \mathcal R^\dagger = \mathcal I^{[1]}(P) + \frac{1}{2} \int \frac{d k^0}{2 \pi} \frac{1}{L^3} \sum_{\boldsymbol k} \frac{\mathcal L(P, k) \, i^2 \, \mathcal R^*(P, k)}{[k^2 - M_N^2 + i\epsilon][(P-k)^2 - M_N^2 + i\epsilon]} \,,
\end{equation*}
to eq.~\eqref{eq:Cloop_intm},
\begin{equation*}
\mathcal L \, \circ_{\sf fv} \, \mathcal R^\dagger = \mathcal I^{[2]}(P) + \sum_{\boldsymbol k} \frac{\mathcal L (P,k) \, i \, \mathcal R^* (P,k)}{2L^3 \cdot 2 \omega_N(\boldsymbol k) \big[ (E - \omega_N(\boldsymbol k) )^2 - \omega_N (\boldsymbol P - \boldsymbol k)^2 \big]} \, \Bigg \vert_{k^0 = \omega_N(\boldsymbol k)} \,,
\end{equation*}
(both repeated here for convenience). As we now show, the $k^0$ integration has contributions other than the $k^0 = \omega_N(\boldsymbol k)$ part and these generate singularities in $\mathcal I^{[2]}(P)$, which thus carries power-like $L$ dependence. Simultaneously, singularities and associated power-like $L$-dependence are hidden within $\mathcal L(P,k)$ in eq.~\eqref{eq:Cloop_intm}. The poles cancel and are also both removed when all external legs are set on shell.

To avoid clutter of notation, we demonstrate the problem explicitly for slightly simplified three-vector kinematics. In particular, we take $\boldsymbol P = \boldsymbol 0$ and $\boldsymbol p = \boldsymbol k$ (aligned rather than anti-aligned as above). For these values, the pion energies are distinct: $\omega_{\pi}(\boldsymbol \ell)$ and $\omega_{\pi}(2 \boldsymbol p + \boldsymbol \ell)$ while all on-shell nucleons carry $\omega_N(\boldsymbol p)$. The relevant singularities are then captured by the function
\begin{align}
\begin{split}
\mathcal F(\boldsymbol p, \boldsymbol \ell) & \equiv \int \! \frac{d k^0}{2 \pi} \int \! \frac{d \ell^0}{2 \pi} \frac{1}{\ell^0 - \omega_\pi(\boldsymbol \ell) + i\epsilon} \frac{1}{\ell^0 + \omega_\pi(\boldsymbol \ell) - i\epsilon} \\
&
\frac{1}{k^0 + \ell^0 + \omega_N(\boldsymbol p) - E - \omega_{\pi}(2 \boldsymbol p + \boldsymbol \ell) + i \epsilon} \frac{1}{k^0 + \ell^0 + \omega_N(\boldsymbol p) - E + \omega_{\pi}(2 \boldsymbol p + \boldsymbol \ell) - i \epsilon} \\
& \frac{1}{k^0 - \omega_N(\boldsymbol p) + i\epsilon} \frac{1}{k^0 + \omega_N(\boldsymbol p) - i\epsilon} \frac{1}{k^0 - E - \omega_N(\boldsymbol p) + i\epsilon} \frac{1}{k^0 - E + \omega_N(\boldsymbol p) - i\epsilon} \,,
\end{split}
\end{align}
which follows from simply factorizing the four covariant propagators.

Beginning with the $\ell^0$ integral (which is internal to $\mathcal L(P,k)$ in the general construction), and closing the contour in the lower half of the complex $\ell^0$ plane, we note that two terms arise from encircling the poles at $\ell^0 = \omega_\pi(\boldsymbol \ell) - i \epsilon$ and
$\ell^0 = - k^0 - \omega_N(\boldsymbol p) + E + \omega_\pi(2 \boldsymbol p + \boldsymbol \ell) - i \epsilon$. The result reads
\begin{align}
\begin{split}
\mathcal F(\boldsymbol p, \boldsymbol \ell) & \equiv (-i) \frac{1}{2 \omega_\pi(\boldsymbol \ell)} \int \! \frac{d k^0}{2 \pi} \frac{1}{k^0 + \omega_\pi(\boldsymbol \ell) + \omega_N(\boldsymbol p) - E - \omega_{\pi}(2 \boldsymbol p + \boldsymbol \ell) + i \epsilon} \\
& \hspace{100pt} \frac{1}{k^0 + \omega_\pi(\boldsymbol \ell) + \omega_N(\boldsymbol p) - E + \omega_{\pi}(2 \boldsymbol p + \boldsymbol \ell) - i \epsilon} \\
& \hspace{100pt} \frac{1}{k^0 - \omega_N(\boldsymbol p) + i\epsilon} \frac{1}{k^0 + \omega_N(\boldsymbol p) - i\epsilon} \\
& \hspace{100pt} \frac{1}{k^0 - E - \omega_N(\boldsymbol p) + i\epsilon} \frac{1}{k^0 - E + \omega_N(\boldsymbol p) - i\epsilon} \\[10pt]
& + (-i) \frac{1}{2 \omega_{\pi}(2 \boldsymbol p + \boldsymbol \ell)} \int \! \frac{d k^0}{2 \pi} \frac{1}{-k^0 - \omega_N(\boldsymbol p) + E + \omega_{\pi}(2 \boldsymbol p + \boldsymbol \ell) - \omega_\pi(\boldsymbol \ell) + i\epsilon} \\
& \hspace{100pt} \frac{1}{-k^0 - \omega_N(\boldsymbol p) + E + \omega_{\pi}(2 \boldsymbol p + \boldsymbol \ell) + \omega_\pi(\boldsymbol \ell) - i\epsilon} \\
& \hspace{100pt} \frac{1}{k^0 - \omega_N(\boldsymbol p) + i\epsilon} \frac{1}{k^0 + \omega_N(\boldsymbol p) - i\epsilon} \\
& \hspace{100pt} \frac{1}{k^0 - E - \omega_N(\boldsymbol p) + i\epsilon} \frac{1}{k^0 - E + \omega_N(\boldsymbol p) - i\epsilon} \,.
\end{split}
\end{align}

When we then subsequently evaluate the $k^0$ integral, each of these two terms generates three, for a total of six terms in the final result. But the paradigm reviewed in the main text only keeps the $k^0 = \omega_N(\boldsymbol k)$ contributions explicit with all others buried inside $\mathcal I^{[2]}(P)$. However, if we consider the $k^0= - \omega_N(\boldsymbol p) + E + \omega_{\pi}(2 \boldsymbol p + \boldsymbol \ell) + \omega_\pi(\boldsymbol \ell) - i\epsilon$ contribution within the second term above, we identify a contribution of the form
\begin{align}
\nonumber
\mathcal F(\boldsymbol p, \boldsymbol \ell) & \supset - (-i)^2 \frac{1}{2 \omega_\pi(\boldsymbol \ell)} \frac{1}{2 \omega_{\pi}(2 \boldsymbol p + \boldsymbol \ell)} \bigg [
\frac{1}{E -2 \omega_N(\boldsymbol p) + \omega_{\pi}(2 \boldsymbol p + \boldsymbol \ell) + \omega_\pi(\boldsymbol \ell)} \\
& \hspace{150pt} \frac{1}{E + \omega_{\pi}(2 \boldsymbol p + \boldsymbol \ell) + \omega_\pi(\boldsymbol \ell)} \\
& \hspace{150pt} \frac{1}{\omega_{\pi}(2 \boldsymbol p + \boldsymbol \ell) + \omega_\pi(\boldsymbol \ell) - 2 \omega_N(\boldsymbol p)}
\frac{1}{\omega_{\pi}(2 \boldsymbol p + \boldsymbol \ell) + \omega_\pi(\boldsymbol \ell)} \bigg ] \,, \nonumber
\end{align}
which includes a singularity at $E = 2 \omega_N(\boldsymbol p)- \omega_\pi(\boldsymbol \ell) - \omega_\pi(2 \boldsymbol p + \boldsymbol \ell)$.

In the construction of section~\ref{sec:loop_cont}, this singularity is included in $\mathcal I^{[2]}(P)$. Similarly, keeping the $k^0 = \omega_N(\boldsymbol p)$ result leads to a contribution to $\mathcal L(P,k)$ with the same pole, the one identified in the TOPT analysis of figure~\ref{fig:single_u_channel_loop}(c) above. These poles cancel in the full expression for the diagram, but they formally disrupt the smoothness assumptions for both $\mathcal I^{[2]}(P)$ and $\mathcal L(P,k)$.

\subsection{Shuffling \texorpdfstring{$t$}{t}- and \texorpdfstring{$u$}{u}-type subdiagrams}

Having discussed the problem of $u$-channel-like Bethe-Salpeter contributions in the context of a particular example, we now turn to the solution, first for the specific case considered and then for all contributions to the Bethe-Salpeter kernel.

For the example of the single $u$-channel loop, the problem is solved by simply replacing $\mathcal L(P,k) \to \mathcal L(P,P-k)$ before $k^0$ integration. Since the expression is invariant under $k \to P-k$, this does not change the value of the full diagram, but it does change the definition (and indeed the basic properties) of particular contributions such as $\mathcal I^{[2]}(P)$.
In particular, the TOPT contributions identified above are transformed to
\begin{align}
[ABC] & =
\frac{1}{\omega_N(\boldsymbol p) - \omega_N(\boldsymbol k) - \omega_\pi(\boldsymbol k - \boldsymbol p - \boldsymbol \ell) - \omega_\pi(\boldsymbol \ell)}
\frac{1}{E - \omega_N(\boldsymbol k) - \omega_N(\boldsymbol P - \boldsymbol k)} \,, \\ \nonumber
[BAC] & =
\frac{1}{E - \omega_N(\boldsymbol p) - \omega_N(\boldsymbol P - \boldsymbol k) - \omega_\pi(\boldsymbol k - \boldsymbol p - \boldsymbol \ell) - \omega_\pi(\boldsymbol \ell)}
\frac{1}{E - \omega_N(\boldsymbol k) - \omega_N(\boldsymbol P - \boldsymbol k)} \,, \\
[AB] & = \frac{1}{\omega_N(\boldsymbol p) - \omega_N(\boldsymbol k) - \omega_\pi(\boldsymbol k - \boldsymbol p - \boldsymbol \ell) - \omega_\pi(\boldsymbol \ell)} \,, \\
[BA] & = \frac{1}{- \omega_N(\boldsymbol p) + \omega_N(\boldsymbol k) - \omega_\pi(\boldsymbol k - \boldsymbol p - \boldsymbol \ell) - \omega_\pi(\boldsymbol \ell)} \,.
\end{align}
In this case we find that the $[AB]$ and $[BA]$ terms faithfully represent the true singularities, without adding any spurious poles. For $[AB]$, the expression for the sub-diagram matches the full diagram exactly, while for $[BA]$ it is related by the replacement $E - \omega_N(\boldsymbol P - k) \to \omega_N(\boldsymbol k)$. In contrast to the situation of the preceding subsection, here the replacement does not generate spurious poles and in fact leads to an $E$-independent expression for the sub-diagram. (To see that $[AB]$ and $[BA]$ are safe note that either denominator vanishing would correspond to the kinematics of an on-shell $N \to N \pi \pi$ decay, which is not possible.)

\begin{figure}
\centering
\includegraphics[width=\textwidth]{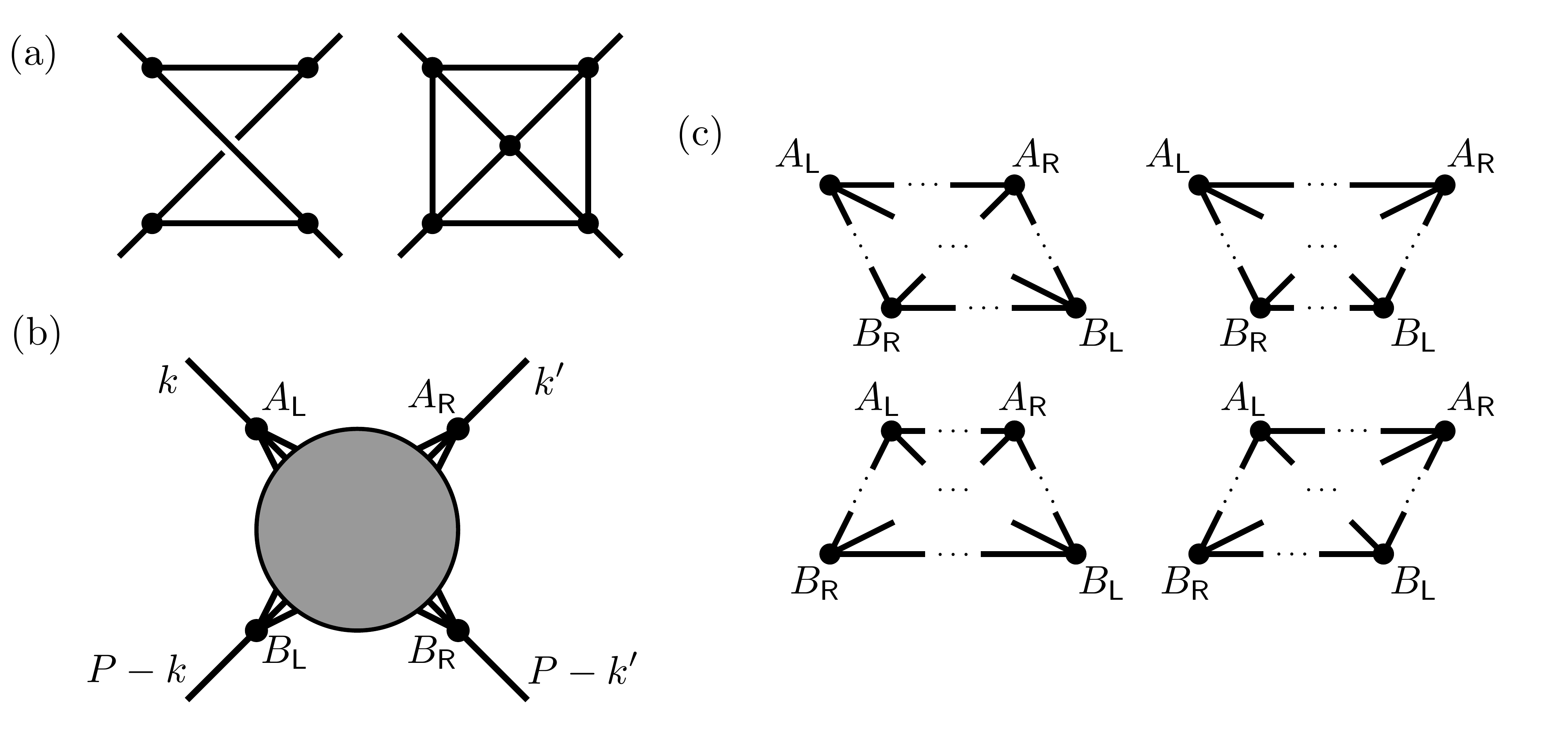}
\caption{(a) Examples of diagrams for which the notion of $u$-channel-like and $t$-channel like is ambiguous. (b) Scheme for labeling vertices in a generic diagram. (c) Summary of orderings for which a momentum rerouting is required.}
\label{fig:routing_procedure}
\end{figure}

It remains to show how the replacement $\mathcal L(P,k) \to \mathcal L(P,P-k)$ is generalized to all diagrams. The guiding principle is to replace $u$-channel-like momentum routing with $t$-channel-like routing. However, this is ambiguous since certain diagrams, like those shown in figure~\ref{fig:routing_procedure}(a), have both $u$- and $t$-channel like behavior. Such ambiguities are resolved by performing all internal $\ell^0$ integrals in a given Bethe-Salpeter contribution before applying the momentum reassignments.

This leads to a diagrammatic definition of the modified Bethe-Salpeter kernel $B^{\mathfrak{T}}(P,p,p')$:
\begin{enumerate}
\item For a generic diagram contributing to the Bethe-Salpeter kernel, label the four vertices connected to propagators in the adjacent two-particle loops as $A_{\sf L}$, $B_{\sf L}$, $A_{\sf R}$, and $B_{\sf R}$, where ${\sf L}$ and ${\sf R}$ stand for left and right, respectively. [See figure~\ref{fig:routing_procedure}(b).]
\item If $A_{\sf L}$ and $B_{\sf L}$ or $A_{\sf R}$ and $B_{\sf R}$ are the same vertex, (if [$(A_{\sf L} = B_{\sf L})$ or $(A_{\sf R} = B_{\sf R})$] is true), then the diagram is a contribution to $B^{\mathfrak{T}}(P,p,p')$ without any further modification.
\item If $A_{\sf L}$ and $B_{\sf L}$ are distinct from each other, and also $A_{\sf R}$ and $B_{\sf R}$ are distinct, then evaluate the TOPT contributions to the diagram by enumerating all orderings of $A_{\sf L}$, $B_{\sf L}$, $A_{\sf R}$, and $B_{\sf R}$, as well as the set of all vertices not attached to neighboring loops. Note, it can still be the case that $A_{\sf L} = A_{\sf R}$ or $B_{\sf L} = B_{\sf R}$ (or else $A_{\sf L} = B_{\sf R}$ or $B_{\sf L} = A_{\sf R}$) as with the single $u$-channel loop.
\item For concreteness, define the momentum flowing into each vertex from the neighboring loops as follows: $k$ into $A_{\sf L}$, $P-k$ into $B_{\sf L}$, $k'$ into $A_{\sf R}$, $P-k'$ into $B_{\sf R}$. [Again see figure~\ref{fig:routing_procedure}(b).] Then all time orderings contribute to $B^{\mathfrak{T}}(P,p,p')$ without modification, except for the following orderings of the external vertices, also shown in figure~\ref{fig:routing_procedure}(c):
\subitem $A_{\sf L} \cdots B_{\sf R} \cdots A_{\sf R} \cdots B_{\sf L}$,
\subitem $A_{\sf L} \cdots B_{\sf R} \cdots B_{\sf L} \cdots A_{\sf R}$,
\subitem $B_{\sf R} \cdots A_{\sf L} \cdots A_{\sf R} \cdots B_{\sf L}$,
\subitem $B_{\sf R} \cdots A_{\sf L} \cdots B_{\sf L} \cdots A_{\sf R}$,

where the ellipses represent other internal vertices and cuts that can appear in the ordering. (For our $u$-channel loop, both orderings are of this type.)
\item For all orderings given above, dangerous cuts appear between the middle two vertices (i.e.~in the central ellipses). The energy flowing across these cuts is $u$-channel like, of the form $\omega_N(\boldsymbol k) - E + \omega_N(\boldsymbol k')$, and this leads to the spurious singularities discussed above. Therefore, for all such orderings we replace $\boldsymbol k$ with $\boldsymbol P - \boldsymbol k$ and $\omega_N(\boldsymbol k)$ with $E - \omega_N(\boldsymbol k)$. This replaced definition yields the contribution of this diagram to $B^{\mathfrak{T}}(P,p,p')$. For all other time orderings the contribution to $B^{\mathfrak{T}}(P,p,p')$ is unmodified. This completes our construction of the modified Bethe-Salpeter kernel.
\end{enumerate}

This modified kernel is useful because it allows us to follow the procedure of section \ref{sec:loop_cont} without encountering spurious singularities. In particular, the $k^0$ integration of eq.~\eqref{eq:loop_cont2} is now valid. The ordering of vertices guarantees that the only energies that can flow across the cuts in the TOPT approach are $E$ and $\pm[\omega_N(\boldsymbol k) - \omega_N(\boldsymbol k')]$. Thus, in the case where no single meson exchange occurs, $B^{\mathfrak{T}}(P,p,p')$ is analytic in $s$ and has exponentially suppressed volume effects for $4 M_N^2 - 4 M_\pi^2 <s < (2 M_N + M_\pi)^2$. This is the same range of validity holds for the subtracted kernel $\overline B^{\mathfrak{T}}(P,p,p')$ in theories with a single meson exchange.

\section{Manipulating the finite-volume \texorpdfstring{$S$}{S} function}
\label{sec:manipulating_S}

In this appendix, we describe the steps required to go from eq.~\eqref{eq:Cloop_intm} to eqs.~\eqref{eq:Cloop_compact} and \eqref{eq:S_def} of the main text. Recalling the definitions, the task is to show that
\begin{equation*}
\mathcal L \, \circ_{\sf fv} \, \mathcal R^\dagger = \mathcal I^{[2]}(P) + \sum_{\boldsymbol k} \frac{\mathcal L (P,k) \, i \, \mathcal R^* (P,k)}{2L^3 \cdot 2 \omega_N(\boldsymbol k) \big[ (E - \omega_N(\boldsymbol k) )^2 - \omega_N (\boldsymbol P - \boldsymbol k)^2 \big]} \, \Bigg \vert_{k^0 = \omega_N(\boldsymbol k)} \,,
\end{equation*}
can be rewritten as
\begin{multline}
\mathcal L \, \circ_{\sf fv} \, \mathcal R^\dagger = \mathcal I^{[3]}(P) \\ + \widetilde{\mathcal L}_{\boldsymbol k^\star \ell m}(P) \, \frac{1}{2L^3} \, \frac{i 4\pi \, Y_{\ell m}(\hat{\boldsymbol k}^\star) \, Y^*_{\ell' m'}(\hat{\boldsymbol k}^\star) \, \delta_{\boldsymbol k^\star \boldsymbol k'^\star} \, \vert \boldsymbol k^\star \vert^{\ell + \ell'} \, H(\boldsymbol k^\star)}{4 \omega_N(\boldsymbol k) \, \big [ (k_{\sf os}^\star)^2 - (\boldsymbol k^\star)^2\big ]} \, \widetilde{\mathcal R}^*_{\boldsymbol k'^\star \ell' m'}(P) \,,
\label{eq:combining_L_R_S}
\end{multline}
where we have combined eqs.~\eqref{eq:Cloop_compact} and \eqref{eq:S_def} to reach eq.~\eqref{eq:combining_L_R_S}.

Taking the difference of these two results and using the definitions of $ \widetilde{\mathcal L}_{\ell m} (P, \vert \boldsymbol k^\star \vert ) $ and $ \widetilde{\mathcal R}_{\ell m} (P, \vert \boldsymbol k^\star \vert ) $, eqs.~\eqref{eq:harm_proj_L} and \eqref{eq:harm_proj_R}, it remains to show
\begin{multline}
\mathcal I^{[2]}(P) - \mathcal I^{[3]}(P) = \frac{i}{2 L^3} \sum_{\boldsymbol k} \frac{1}{2 \omega_N(\boldsymbol k)} \mathcal L (P,k) \, \mathcal R^* (P,k) \\
\times \bigg [ \frac{1}{(E - \omega_N(\boldsymbol k) )^2 - \omega_N (\boldsymbol P - \boldsymbol k)^2} - \frac{H(\boldsymbol k^\star)/2}{(k_{\sf os}^\star)^2 - (\boldsymbol k^\star)^2} \bigg ] \bigg \vert_{k^0 = \omega_N(\boldsymbol k)}\,,
\end{multline}
with the key claim being that $\mathcal I^{[2]}(P) - \mathcal I^{[3]}(P)$ only has exponentially suppressed $L$-dependence. This follows from the Poisson summation formula, provided that the summand on the right-hand side has a horizontal strip of analyticity in the complex-$\vert \boldsymbol k \vert$ plane, which includes the real axis. Here we abuse notation by thinking of $z = \vert \boldsymbol k \vert$ as a complex variable.

For $P^2 = s < (2 M_N + M_\pi)^2$, the analyticity holds for $\mathcal L (P,k) \, \mathcal R^* (P,k)$ by construction, and also for $\omega_N(\boldsymbol k)$, which is analytic for $\vert \text{Im}[\vert \boldsymbol k \vert] \vert < M_N$. The non-trivial step is to show the same for the difference of the two terms in square brackets. To demonstrate this we define the quantity in question via
\begin{equation}
Q(P, \boldsymbol k) = \frac{1}{(E - \omega_N(\boldsymbol k) )^2 - \omega_N (\boldsymbol P - \boldsymbol k)^2} - \frac{H(\boldsymbol k^\star)/2}{(k_{\sf os}^\star)^2 - (\boldsymbol k^\star)^2} \,.
\end{equation}
We then observe that
\begin{equation}
(E - \omega_N(\boldsymbol k) )^2 - \omega_N (\boldsymbol P - \boldsymbol k)^2 = (P - k)^2 - M_N^2 = (E^\star - \omega_N(\boldsymbol k^\star))^2 - \omega_N(\boldsymbol k^\star)^2 \,,
\end{equation}
where we have used $P^\mu = (E, \boldsymbol P)^\mu$ and $k^\mu = (\omega_N(\boldsymbol k), \boldsymbol k)^\mu$ in the middle step. This makes manifest that the combination is a Lorentz scalar and can be written in the CM frame, as we have done.

Further algebraic manipulations then yield
\begin{equation}
(E - \omega_N(\boldsymbol k) )^2 - \omega_N (\boldsymbol P - \boldsymbol k)^2 =\frac{E^\star}{E^\star + 2 \omega_N(\boldsymbol k^\star)} \big [E^{\star 2} - 4 \omega_N(\boldsymbol k^\star)^2 \big ]\,.
\end{equation}
Substituting $(k_{\sf os}^\star)^2 = E^{\star 2}/4 - M_N^2$ and then substituting back for $Q(P, \boldsymbol k)$, we finally deduce
\begin{equation}
Q(P, \boldsymbol k) = \bigg [ \frac{E^\star + 2 \omega_N(\boldsymbol k^\star)}{4 E^\star} - \frac{H(\boldsymbol k^\star)}{2} \bigg ] \, \frac{1}{(k_{\sf os}^\star)^2 - (\boldsymbol k^\star)^2} \,.
\end{equation}
Note that the factor in square brackets scales as $[(k_{\sf os}^\star)^2 - (\boldsymbol k^\star)^2]$ as $( \boldsymbol k^\star)^2 \to (k _{\sf os}^\star)^2$, and therefore cancels the pole. This holds because
\begin{equation}
\bigg [ \frac{E^\star + 2 \omega_N(\boldsymbol k^\star)}{4 E^\star} - \frac{H(\boldsymbol k^\star)}{2} \bigg ] \bigg \vert_{( \boldsymbol k^\star)^2 = (k _{\sf os}^\star)^2} = 0 \,,
\end{equation}
together with the fact that all functions are analytic for $( \boldsymbol k^\star)^2 > - M_N^2$.
We thus conclude a non-zero strip of analyticity for $Q(P, \boldsymbol k)$

\section{Details of the derivation}
\label{app:derivation_details}

In this appendix we discuss the manipulations required to go from eq.~\eqref{eq:CL_decomposed_three}:
\begin{equation}
C_L(P) = \mathcal I^{[1]}_C(P) + \sum_{n=0}^\infty A^{[1]}(P) \, iS(P, L) \, \Big [ \big( i\overline{\mathcal K}^{[1]}(P) + 2 ig^2\mathcal T(P) \big) \, iS(P, L) \Big ]^n \, {A}^{[1]\dagger}(P) \,,
\nonumber
\end{equation}
to eq.~\eqref{eq:final_geometric_series}:
\begin{align}
C_L(P) & = \mathcal I_C(P) + \sum_{n = 0}^\infty A^{\sf os}(P) \, \xi \, iS(P,L) \left[ \left( \xi^\dagger\, i\overline{\mathcal K}^{\sf os}(P) \, \xi + 2 i g^2 \mathcal T(P) \right) iS(P,L) \right]^n \xi^\dagger \, A^{\sf os}(P)^{\dagger} \,,
\nonumber
\end{align}
where we have repeated both results here for convenience.

The basic idea is that the initial four infinite-volume quantities ($\mathcal I^{[1]}_C(P)$, $\overline{\mathcal K}^{[1]}(P)$, $A^{[1]}(P)$, and $A^{[1]\dagger}(P)$) are modified to alternatives ($\mathcal I_C(P)$, $\overline{\mathcal K}^{\sf os}(P)$, $A^{\sf os}(P)$, and $A^{\sf os}(P)^\dagger$) with $\ell m$ indices in place of $\boldsymbol{k}^\star \ell m$ everywhere, i.e.~the $\boldsymbol{k}^\star$ index is removed in quantities labelled by ${\sf os}$, which stands for \emph{on shell}. This is possible because, if both nucleons in a given two-nucleon state are on shell, then the state is completely specified by energy and angular momentum and the $\boldsymbol{k}^\star$ index is redundant. This is not true for the quantities labeled by $[1]$ as these are off-shell and the magnitude of $\boldsymbol{k}^\star$ can be freely varied.

The simplest instance of replacing off-shell with on-shell quantities is given by the relation
\begin{equation}
\mathcal I^{[1]}_C(P) + A^{[1]}(P) iS(P, L) A^{[1]\dagger}(P) = \widetilde {\mathcal I}^{[1]}_C(P) + A^{[1]{\sf os}}(P) \, \xi \, iS(P, L) \, \xi^\dagger \, A^{[1]{\sf os}}(P)^\dagger \,,
\label{eq:first_A1_on_shell_example}
\end{equation}
which trivially follows from definition
\begin{align}
\widetilde {\mathcal I}^{[1]}_C(P) & = \mathcal I^{[1]}_C(P) + A^{[1]}(P) iS(P, L) A^{[1]\dagger}(P) - A^{[1]{\sf os}}(P) \, \xi \, iS(P, L) \, \xi^\dagger \, A^{[1]{\sf os}}(P)^\dagger \,,
\label{eq:IC1_single_S_cancel}
\end{align}
where $A_{\ell m}^{[1]{\sf os}}(P) = A_{\boldsymbol k^\star \ell m}^{[1]}(P) \Big \vert_{\vert \boldsymbol k^\star \vert = k^\star_{\sf os}}$ and similar for the conjugated factor. This replacement is valid because $A_{\ell m}^{[1]{\sf os}}(P)$ does not have a left-hand cut and the singularity of $S(P,L)$ cancels between the last two terms of eq.~\eqref{eq:IC1_single_S_cancel}. Thus the sum, implicit in the second two terms, can be replaced with an integral up to exponentially suppressed terms that we neglect.

In this appendix we will describe a series of such replacements that have the effect of converting eq.~\eqref{eq:CL_decomposed_three} to eq.~\eqref{eq:final_geometric_series}. The guiding principles of the replacements are as follows:
\begin{enumerate}
\item In quantities without a left-hand cut in the region that we control ($4 M_N^2 - 4 M_\pi^2 < s$), one can safely set $\vert \boldsymbol{k}^\star \vert$ to $k_{\sf os}^\star$ up to differences that cancel neighboring $S(P,L)$ poles.
\item The cancellation of $S(P,L)$ poles, in turn, allows one to replace sums with integrals and the integrated terms are absorbed in redefinitions of infinite-volume quantities, such as in eq.~\eqref{eq:IC1_single_S_cancel}.
\item An iterative process then emerges. This is because the resulting integrals create new quantities for which the replacement $\vert \boldsymbol{k}^\star \vert \to k_{\sf os}^\star$ can safely be performed.
\end{enumerate}

To explain in detail, it will be useful to introduce a more compact notation. We use bold symbols to represent the building blocks compactly and suppress all arguments, factors of $i$ and $g^2$, and overlines. We then define the $(n+1)$th or $(n'+1)$th term in our starting and target sums, respectively, as
\begin{align}
C^{[1], (n)}_L(P) & = {\textbf A}^{\![1]} \, {\textbf S} \, \Big [ \big( {\textbf K}^{[1]} + {\textbf T} \big) \, {\textbf S} \Big ]^{n-1} \, {{\textbf A}}^{[1]\dagger} \,,
\\
C^{[{\sf os}],(n')}_L(P) & = {\textbf A}^{\! \sf os} \, \xi \, {\textbf S} \left[ \left( \xi^\dagger\, {\textbf K}^{\sf os} \, \xi + {\textbf T} \right) {\textbf S} \right]^{n'-1} \xi^\dagger \, {\textbf A}^{\! \sf os \dagger} \,,
\end{align}
with $n = 0$ corresponding to the ${\textbf S}$-independent quantity in each case, i.e.~$\mathcal I^{[1]}_C(P)$ or $\mathcal I_C(P)$.

The key claim is that, for any non-negative integer $n$, $C_L^{[1], (n)}$ generates a series of contributions to $C^{[{\sf os}],(n')}_L(P)$ for all $n' \leq n$. We now sketch the argument for $n=0,1,2,3$, since new features emerge at each of these orders. We then give the all orders construction.

For $n = 0$ it is straightforward; we simply note that $\mathcal I^{[1]}_C(P)$ is one of the terms entering the definition of $\mathcal I_C(P)$. In addition, we have already addressed $n = 1$ above in our motivating example. In our more compact notation we begin with
\begin{equation}
C^{[1], (1)}_L(P) = {\textbf A}^{\![1]} \, {\textbf S} \, {{\textbf A}}^{[1]\dagger} \,.
\end{equation}
The on-shell projection is then exactly shown in eq.~\eqref{eq:first_A1_on_shell_example}. We generate contributions to $n'=0$ and $n'=1$, i.e. to
\begin{equation}
C^{[{\sf os}],(0)}_L(P) = \mathcal I_C(P) \,, \qquad C^{[{\sf os}],(1)}_L(P) = {\textbf A}^{\! \sf os} \, \xi \, {\textbf S} \, \xi^\dagger \, {\textbf A}^{\! \sf os \dagger} \,.
\end{equation}

Before moving to higher $n$, we introduce some additional notation and show how this applies to the $n=1$ case. In particular, we define
\begin{equation}
\delta^\dagger {{\textbf A}}^{\! [1] \, {\sf os} \, \dagger} = {{\textbf A}}^{\! [1] \, \dagger} - \xi^\dagger {\textbf A}^{\! [1] \, \sf os \, \dagger} \,, \qquad {\textbf A}^{\![1] \, {\sf os}} \delta = {\textbf A}^{\![1]} - {\textbf A}^{\! [1] \, \sf os} \, \xi \,,
\end{equation}
and write
\begin{equation}
C^{[1], (1)}_L(P) = {\textbf A}^{\![1]} \, {\textbf S} \, {{\textbf A}}^{[1]\dagger} = {\textbf A}^{\! [1] \, {\sf os}} \, (\xi + \delta) \, {\textbf S} \, (\xi^\dagger + \delta^\dagger) \, {\textbf A}^{\! [1] \, {\sf os} \, \dagger} \,.
\end{equation}
Note that the quantities with a $\delta$ are not truly on shell as the label implies. We nonetheless use this notation so that $\xi$ and $\delta$ can be compactly grouped as shown. We next use the rule that, whenever one or more $\delta$ factors appears next to an $\textbf S$, the singularity cancels and the sum is replaced with an integral.%
\footnote{It should be emphasized that similar tricks were used heavily in addressing singularities and redefinitions for the three-to-three finite-volume formalism \cite{Hansen:2014eka} and in the study of two-to-two processes with an external current \cite{Briceno:2015tza}.}
In this way, of the four terms generated by multiplying out the $(\xi + \delta)$ and $(\xi^\dagger + \delta^\dagger)$ binomials, three contribute to $C^{[{\sf os}],(0)}_L(P) $, as claimed.

We are now ready to turn to $n=2$, where a new and subtle aspect of the derivation arises. We write out the expression as
\begin{align}
C^{[1], (2)}_L(P) & = {\textbf A}^{\![1]} \, {\textbf S} \, {\textbf K}^{[1]} \, {\textbf S} \, {{\textbf A}}^{[1]\dagger} + {\textbf A}^{\![1]} \, {\textbf S} \, {\textbf T} \, {\textbf S} \, {{\textbf A}}^{[1]\dagger} \,, \\
\begin{split}
& = {\textbf A}^{\![1] \, {\sf os}} \, (\xi + \delta) \, {\textbf S} \, (\xi^\dagger + \delta^\dagger) \, {\textbf K}^{[1] \, {\sf os}} \, (\xi + \delta) \, {\textbf S} \, (\xi^\dagger + \delta^\dagger) \, {{\textbf A}}^{\![1] \, {\sf os} \,\dagger}
\\
& \hspace{150pt} + {\textbf A}^{\![1] \, {\sf os}} \, (\xi + \delta) \, {\textbf S} \, {\textbf T} \, {\textbf S} \, (\xi^\dagger + \delta^\dagger) \, {{\textbf A}}^{\![1] \, {\sf os} \,\dagger} \,,
\label{eq:C12_xi_delta_decom}
\end{split}
\end{align}
where in the second equality we have applied the $\xi + \delta$ decomposition to both $\textbf{A}^{\![1]}$ and $\textbf K^{[1]}$.

First consider the $2^4 = 16$ terms generated by expanding all binomials in the first term of eq.~\eqref{eq:C12_xi_delta_decom}. Using the usual rule that a $\delta$ always annihilates a neighboring $\textbf S$ we see that all terms can be identified as contributions to $C_L^{[\sf os], (n')}(P)$ with $n' \leq 2$.

More effort is required for the second term of \eqref{eq:C12_xi_delta_decom}. Of the $2^2 = 4$ terms, the term with $\xi$ and $\xi^\dagger$ has a straightforward match in the ${\textbf A}^{\! \sf os} \, \xi \, {\textbf S} \, {\textbf T} \, {\textbf S} \, \xi^\dagger \, {\textbf A}^{\! \sf os \dagger}$ part of $C^{[{\sf os}],(1)}_L(P)$. Also, the term with $\delta$ and $\delta^\dagger$ is clear, since all singularities are cancelled and one reaches a fully integrated contribution to $C^{[{\sf os}],(0)}_L(P)$. The challenge is the remaining contributions with a single factor of either $\delta$ or $\delta^\dagger$.

Consider, for concreteness, $ {\textbf A}^{\![1] \, {\sf os}} \, \delta \, {\textbf S} \, {\textbf T} \, {\textbf S} \, \xi^\dagger \, {{\textbf A}}^{\![1] \, {\sf os} \,\dagger}$. The issue is that $\textbf T$ cannot be placed on shell but, at the same time, the $\delta$ cancelling the pole in $\textbf S$ will lead to an integral combining the off-shell $\textbf T$ into the left endcap. We require a final piece of notation to reflect this:
\begin{equation}
{\textbf A}^{\![k+1]} = {\textbf A}^{\![k] \, {\sf os}} \, \delta \, {\textbf S} \, {\textbf T} \,.
\label{eq:A_kplus1_def}
\end{equation}
Of which the special case relevant here is simply ${\textbf A}^{\![2]} = {\textbf A}^{\![1] \, {\sf os}} \, \delta \, {\textbf S} \, {\textbf T} $. So we find that $\textbf T$ can convert an on-shell contribution back to an off-shell contribution. In other words, ${\textbf A}^{\![2]} $ has dependence on $\boldsymbol k^\star$ within $\boldsymbol k^\star \ell m$, and is thus not defined on the desired index space.

Having articulated the extra complication, the resolution follows from an imitation of what we have already done. The terms that we have to address are of the form $ {\textbf A}^{\![2]} \, {\textbf S} \, \xi^\dagger \, {{\textbf A}}^{\![1] \, {\sf os} \,\dagger}$ and $ {\textbf A}^{\![1] \,{\sf os}} \, \xi \, {\textbf S} \, {{\textbf A}}^{\![2] \,\dagger}$. But, now that the $\textbf T$ factors within ${\textbf A}^{\![2]}$ have integrated momenta, these no longer have cuts when set on shell.
It follows that the on-shell projections of $ {\textbf A}^{\![2]}$ and ${{\textbf A}}^{\![2] \,\dagger}$ are safe, and we can write
\begin{align}
{\textbf A}^{\![2]} \, {\textbf S} \, \xi^\dagger \, {{\textbf A}}^{\![1] \, {\sf os} \,\dagger} & = {\textbf A}^{\![2] \, {\sf os}} \, (\xi + \delta) \, {\textbf S} \, \xi^\dagger \, {{\textbf A}}^{\![1] \, {\sf os} \,\dagger} \,, \\
{\textbf A}^{\![1] \,{\sf os}} \, \xi \, {\textbf S} \, {{\textbf A}}^{\![2] \,\dagger} & = {\textbf A}^{\![1] \,{\sf os}} \, \xi \, {\textbf S} \, (\xi^\dagger + \delta^\dagger) \, {{\textbf A}}^{\![2] \, {\sf os} \,\dagger}\,.
\end{align}
At this stage, the $\delta$-dependent terms are absorbed into $C_L^{[\sf os], (0)}(P)$ and the remaining terms are absorbed into the part of $C_L^{[\sf os], (1)}(P)$ that is linear in $\textbf S$.

The last claim above relies on the definition for the full on-shell matrix element
\begin{equation}
{\textbf A}^{\! \sf os} = \sum_{k=1}^\infty {\textbf A}^{\![k] \, \sf os} \,, \qquad {\textbf A}^{\! {\sf os} \, \dagger} = \sum_{k=1}^\infty {\textbf A}^{\![k] \, {\sf os} \, \dagger} \,.
\end{equation}
These definitions indicate that the final endcap factors are built from an infinite set of $\textbf T$ factors, attached with integrals to a neighboring factor of $\textbf S$ in which the pole has been cancelled. In all such cases, the factor of $\textbf T$ is not evaluated at $\vert \boldsymbol k^\star \vert = k^\star_{\sf os}$ until after integration. In this way, one avoids introducing spurious left-hand cuts in the finite-volume correlator.

A final new feature arises within $C_L^{[1], (3)}$, defined as
\begin{equation}
C^{[1], (3)}_L(P) = {\textbf A}^{\![1]} \, {\textbf S} \, \big( {\textbf K}^{[1]} + {\textbf T} \big) \, {\textbf S} \, \big( {\textbf K}^{[1]} + {\textbf T} \big) \, {\textbf S} \, {{\textbf A}}^{[1]\dagger} \,,
\label{eq:C13_definition}
\end{equation}
namely that factors of $\textbf T$ must also be absorbed into a redefinition of the K matrix.

At this stage, we think it more pedagogical to explain this partly in words, without introducing additional heavy notation to represent the absorptions. When one inserts the $(\xi + \delta)$ decomposition for all ${\textbf A}^{\![1]} $, ${{\textbf A}}^{[1]\dagger}$ and ${\textbf K}^{[1]} $ in eq.~\eqref{eq:C13_definition}, a term arises with the left ${\textbf K}^{[1]}$ replaced by $\xi^\dagger {\textbf K}^{[1]\, {\sf os}} \delta $, leading to combinations such as the following:
\begin{equation}
C^{[1], (3)}_L(P) \supset {\textbf A}^{\![1]\, {\sf os}} \, \xi \, {\textbf S} \, \xi^\dagger \Big ( {\textbf K}^{[1] \, {\sf os}} \, \delta \, {\textbf S} \, {\textbf T} \Big ) {\textbf S} \, \xi^\dagger \, {{\textbf A}}^{\! [1] \,{\sf os} \, \dagger} \,,
\end{equation}
where the $\supset$ symbol indicates we have only kept one term. Observe that the factor in parentheses here is analogous to ${\textbf A}^{\![2]}$ defined in eq.~\eqref{eq:A_kplus1_def} above.
Any number of $\textbf T$ factors can be attached on either side of $\textbf K$ in this way. After such an attachment, the function no longer has a cut in the region of interest and can be set on shell. The sum over all such on-shell, $\textbf T$-absorbing factors defines the final K-matrix denoted by ${\textbf K}^{\sf os}$.

Fortunately we do not have to make use of this definition, since the relation of ${\textbf K}^{\sf os}$ to the finite-volume correlation function provides an alternative definition in terms of the two-to-two scattering amplitude. Here the situation is analogous to the three-particle formalism of refs.~\cite{Hansen:2014eka}. Also in that work a very ugly iterative definition of a K-matrix was provided but never used, favoring instead an alternative derivation of a more direct relation to the scattering amplitude \cite{Hansen:2015zga}.

At this stage we have presented all features that arise in expressing $C_L^{[1], (n)}(P)$ in terms of\linebreak$C_L^{[{\sf os}], (n')}(P)$ for $n' \leq n$. We summarize the general construction as follows:
\begin{enumerate}
\item For a given $C_L^{[1], (n)}(P)$ insert all allowed factors of $\xi + \delta$ and $\xi^\dagger + \delta^\dagger$. A total of $2 n$ such binomials arise, 2 from the endcaps and the remaining $(2 n -2)$ from the $n - 1$ insertions of $\textbf K^{[1]}$.
\item Multiplying out the terms yields $2^{2n}$ combinations, with various sequences of $\xi$, $\xi^\dagger$, $\delta$ and $\delta^\dagger$.
\item All contributions are immediately identified within $C_L^{[{\sf os}], (n')}(P)$, except for those where an off-shell $\textbf T$ is attached to either an endcap or a K-matrix.
\item These remaining terms correspond to a contribution within $C_L^{[1], (n-j)}(P)$ for some $j \geq 1$. That is, they match a factor with fewer explicit $\textbf T$s, since some are absorbed into $\textbf A^{[k]}$, $\textbf A^{[k]\dagger}$ or the corresponding K matrix.
\item For these remaining terms, new $\xi + \delta$ insertions arise and are processed as above.
\item The iterative procedure always either generates terms that match those in $C_L^{[{\sf os}], (n')}(P)$ or else reduces the explicit $\textbf{T}$ factors in the remaining, unmatched terms.
\item It follows that, after $n$ such iterations on $C_L^{[1], (n)}(P)$, the process terminates and the matching to the desired expression is complete.
\end{enumerate}

Thus, we have achieved our aim to show the equivalence of eqs.~\eqref{eq:CL_decomposed_three} and \eqref{eq:final_geometric_series} (the first two equations of this appendix). We emphasize that the result requires a sequence of redefinitions that is analogous to the three-particle finite-volume scattering formalism of refs.~\cite{Hansen:2014eka,Hansen:2015zga}. These parallels are expected since the same underlying $N N \pi$ state is leading to the left-hand cut addressed here.

We close this appendix with two comments. First, we consider the analogous equivalence arising for $\mathcal M_L^{\sf aux}$. For this quantity, the starting point is eq.~\eqref{eq:aux_amp}, which can be recast as
\begin{equation}
i \mathcal M_L^{\sf aux}(P) = \sum_{n=0}^\infty \Big [ \big( i\overline{B}^{\mathfrak T} + 2 ig^2\mathcal T \big) \, \big [ \! \circ_{\sf rm} +\, iS(P, L) \big ] \, \Big ]^n \big( i\overline{B}^{\mathfrak T} + 2 ig^2\mathcal T \big) + i \Delta \mathcal M^{\sf aux}_L \, .
\end{equation}
Then, following the same steps as with $C_L(P)$, this can first be rewritten as
\begin{equation}
i \mathcal M_L^{\sf aux}(P) = \sum_{n=0}^\infty
\Big [ \big( i\overline{\mathcal K}^{[1]}(P) + 2 ig^2\mathcal T(P) \big) \, iS(P, L) \Big ]^n \, \big( i\overline{\mathcal K}^{[1]}(P) + 2 ig^2\mathcal T(P) \big)
+ i \Delta \mathcal M^{\sf aux}_L
\,,
\end{equation}
where $\overline{\mathcal K}^{[1]}(P)$ is defined in eq.~\eqref{eq:CL_decomposed_Kbar}. The aim is then to show that this can be rewritten as eq.~\eqref{eq:aux_amp2}, repeated here for convenience
\begin{equation}
i \mathcal M^{\sf aux}_L (P) \equiv
\sum_{n = 0}^\infty \Big[ \left( \xi^\dagger\, i\overline{\mathcal K}^{\sf os}(P) \, \xi + 2 i g^2 \mathcal T(P) \right) iS(P, L) \, \Big]^n \left( \xi^\dagger\, i\overline{\mathcal K}^{\sf os}(P) \, \xi + 2i g^2 \mathcal T(P) \right) \, . \nonumber
\end{equation}
The argument follows the pattern used for $C_L(P)$ but with the additional feature that any quantities that vanish when external legs are on shell are absorbed into $\Delta \mathcal M_L^{\sf aux}$. For example, the $n=1$ term includes contributions with a difference between on- and off-shell $\overline{\mathcal K}^{[1]}(P)$. This cancels the singularity in $S(P,L)$ and can be absorbed as an $n=0$ contribution. Also, order by order, contributions with off-shell external kinematics can be set on-shell with the difference absorbed in $\Delta \mathcal M_L^{\sf aux}$. We do not spell out the steps in more detail here as they are very repetitive to those given above.

Finally, we briefly address the exchange symmetry and Lorentz invariance of $\overline {\mathcal K}^{\sf os}(P)$. Note that $\overline {\mathcal K}^{[1]}(P)$ is neither exchange nor Lorentz invariant, since it is defined with $\overline B^{\mathfrak T}$, for which both of these properties are broken due to the separation of $t$- and $u$-channel-like diagrams at the level of TOPT. However, invariance is recovered for $\overline {\mathcal K}^{\sf os}(P)$ when its external momenta are set on-shell, also due to the fact that any non-symmetric part does not contribute to $C_L(P)$. For example, in any given diagram, when the outermost factors of $\overline {\mathcal K}^{\sf os}(P)$ are combined with the exchange-symmetric endcaps, any antisymmetric component is annihilated in the sum. This effectively symmetrizes the outermost insertions and a recursive argument can be used to see that all factors of $\overline {\mathcal K}^{\sf os}(P)$ are symmetrized. In practice, this means that only even partial waves of $\overline {\mathcal K}^{\sf os}(P)$ are nonzero for indistinguishable particles. The symmetrization of $\overline {\mathcal K}^{\sf os}(P)$ also leads to the same for $\overline B^{\mathfrak T}$ and this has the consequence that Lorentz invariance is recovered for $\overline {\mathcal K}^{\sf os}(P)$. The Lorentz invariance can also be seen from the relation between $\overline {\mathcal K}^{\sf os}(P)$ and the scattering amplitude, which is manifestly Lorentz invariant.

\bibliographystyle{JHEP}
\bibliography{refs.bib}
\end{document}